\documentclass[twocolumn,twocolappendix]{aastex631}

\newcommand\cheops{\textit{CHEOPS}}
\newcommand\tess{\textit{TESS}}

\usepackage[T1]{fontenc}
\usepackage{xurl}
\usepackage{multirow}
\usepackage{xcolor}
\usepackage{rotating}
\usepackage{soul}

\usepackage{graphicx}	% Including figure files
\usepackage{amsmath}	% Advanced maths commands

\begin{document}

\title{Characterization of a set of small planets with TESS and CHEOPS and an analysis of photometric performance}

\correspondingauthor{Dominic Oddo}
\email{doddo@unm.edu}

\author[0000-0002-2702-7700]{Dominic Oddo}
\affiliation{Department of Physics and Astronomy, University of New Mexico, 210 Yale Blvd NE, Albuquerque, NM, 87106, USA}

\author[0000-0003-2313-467X]{Diana Dragomir}
\affiliation{Department of Physics and Astronomy, University of New Mexico, 210 Yale Blvd NE, Albuquerque, NM, 87106, USA}

\author[0000-0002-7201-7536]{Alexis Brandeker}
\affiliation{Department of Astronomy, Stockholm University, AlbaNova University Center, 10691 Stockholm, Sweden}

\author[0000-0002-4047-4724]{Hugh~P.~Osborn}
\affiliation{Department of Physics and Kavli Institute for Astrophysics and Space Research, Massachusetts Institute of Technology, Cambridge, MA 02139, USA}
\affiliation{NCCR/Planet-S, Universität Bern, Gesellschaftsstrasse 6, 3012 Bern, Switzerland}

\author[0000-0001-6588-9574]{Karen Collins}
\affiliation{Center for Astrophysics $\vert$ Harvard \& Smithsonian, 60 Garden Street, Cambridge, MA 02138, USA}

\author[0000-0002-3481-9052]{Keivan G. Stassun}
\affiliation{Department of Physics and Astronomy, Vanderbilt University, Nashville, TN 37235, USA}

\author[0000-0002-8462-515X]{Nicola Astudillo-Defru}
\affiliation{Departamento de Matem\'{a}tica y F\'{i}sica Aplicadas, Universidad Cat\'{o}lica de la Sant\'{i}sima Concepci\'{o}n, Alonso de Rivera 2850, Concepci\'{o}n, Chile}

\author[0000-0001-6637-5401]{Allyson Bieryla}
\affiliation{Center for Astrophysics $\vert$ Harvard \& Smithsonian, 60 Garden Street, Cambridge, MA 02138, USA}

\author[0000-0002-2532-2853]{Steve~B.~Howell}
\affiliation{NASA Ames Research Center, Moffett Field, CA 94035, USA}

\author[0000-0002-5741-3047]{David~R.~Ciardi}
\affiliation{NASA Exoplanet Science Institute, Caltech/IPAC, Pasadena, CA 91125, USA}

\author[0000-0002-8964-8377]{Samuel Quinn}
\affiliation{Center for Astrophysics $\vert$ Harvard \& Smithsonian, 60 Garden Street, Cambridge, MA 02138, USA}

\author[0000-0003-3208-9815]{Jose M. Almenara}
\affiliation{Univ. Grenoble Alpes, CNRS, IPAG, F-38000 Grenoble, France}

\author[0000-0001-7124-4094]{C\'{e}sar Brice\~{n}o}
\affiliation{Cerro Tololo Inter-American Observatory, Casilla 603, La Serena, Chile}

\author[0000-0003-2781-3207]{Kevin I.\ Collins}
\affiliation{George Mason University, 4400 University Drive, Fairfax, VA, 22030 USA}

\author[0000-0001-8020-7121]{Knicole D. Col\'{o}n}
\affiliation{NASA Goddard Space Flight Center, Greenbelt, MD 20771, USA} 

\author[0000-0003-2239-0567]{Dennis M.\ Conti}
\affiliation{American Association of Variable Star Observers, 185 Alewife Brook Parkway, Suite 410, Cambridge, MA 02138, USA}

\author[0000-0001-7866-8738]{Nicolas Crouzet}
\affiliation{Leiden Observatory, Leiden University, Postbus 9513, 2300 RA, Leiden, The Netherlands}

\author[0000-0001-9800-6248]{Elise Furlan}
\affiliation{NASA Exoplanet Science Institute, Caltech/IPAC, Pasadena, CA 91125, USA}

\author[0000-0002-4503-9705]{Tianjun Gan}
\affiliation{Department of Astronomy and Tsinghua Centre for Astrophysics, Tsinghua University, Beijing 100084, China}

\author[0000-0003-2519-6161]{Crystal L. Gnilka}
\affiliation{NASA Ames Research Center, Moffett Field, CA 94035, USA}

\author{Robert~F.~Goeke}
\affiliation{Department of Physics and Kavli Institute for Astrophysics and Space Research, Massachusetts Institute of Technology, Cambridge, MA 02139, USA}

\author{Erica Gonzales}
\affiliation{Department of Astronomy and Astrophysics, University of California, Santa Cruz, Santa Cruz, CA, United States}

\author{Mallory Harris}
\affiliation{Department of Physics and Astronomy, University of New Mexico, 210 Yale Blvd NE, Albuquerque, NM, 87106, USA}

\author[0000-0002-4715-9460]{Jon M. Jenkins}
\affiliation{NASA Ames Research Center, Moffett Field, CA 94035, USA}

\author[0000-0002-4625-7333]{Eric L.\ N.\ Jensen}
\affiliation{Department of Physics \& Astronomy, Swarthmore College, Swarthmore PA 19081, USA}

\author[0000-0001-9911-7388]{David Latham}
\affiliation{Center for Astrophysics $\vert$ Harvard \& Smithsonian, 60 Garden Street, Cambridge, MA 02138, USA}

\author{Nicholas Law}
\affiliation{Department of Physics and Astronomy, The University of North Carolina at Chapel Hill, Chapel Hill, NC 27599-3255, USA}

\author[0000-0003-2527-1598]{Michael B. Lund}
\affiliation{NASA Exoplanet Science Institute, Caltech/IPAC, Pasadena, CA 91125, USA}

\author[0000-0003-3654-1602]{Andrew W. Mann}
\affiliation{Department of Physics and Astronomy, The University of North Carolina at Chapel Hill, Chapel Hill, NC 27599-3255, USA}

\author[0000-0001-8879-7138]{Bob Massey}
\affil{Villa '39 Observatory, Landers, CA 92285, USA}

\author{Felipe Murgas}  %(LCOGT 1-m time contribution)
\affiliation{Instituto de Astrof\'isica de Canarias (IAC), E-38205 La Laguna, Tenerife, Spain}
\affiliation{Departamento de Astrof\'isica, Universidad de La Laguna (ULL), E-38206 La Laguna, Tenerife, Spain}

\author[0000-0003-2058-6662]{George Ricker}
\affiliation{MIT Kavli Institute for Astrophysics and Space Research, Massachusetts Institute of Technology, Cambridge, MA 02139, USA}
\affiliation{MIT Department of Physics, Massachusetts Institute of Technology, Cambridge, MA 02139, USA}

\author{Howard M. Relles}
\affiliation{Center for Astrophysics $\vert$ Harvard \& Smithsonian, 60 Garden Street, Cambridge, MA 02138, USA}

\author[0000-0002-4829-7101]{Pamela~Rowden}
\affiliation{Royal Astronomical Society, Burlington House, Piccadilly, London W1J 0BQ, UK}

\author[0000-0001-8227-1020]{Richard P. Schwarz}
\affiliation{Center for Astrophysics $\vert$ Harvard \& Smithsonian, 60 Garden Street, Cambridge, MA 02138, USA}

\author{Joshua Schlieder}
\affiliation{Goddard Space Flight Center, National Aeronautics and Space Administration, Greenbelt, United States}

\author[0000-0002-1836-3120]{Avi~Shporer}
\affiliation{Department of Physics and Kavli Institute for Astrophysics and Space Research, Massachusetts Institute of Technology, Cambridge, MA 02139, USA}

\author[0000-0002-6892-6948]{Sara Seager}
\affiliation{MIT Kavli Institute for Astrophysics and Space Research, Massachusetts Institute of Technology, Cambridge, MA 02139, USA}
\affiliation{Earth and Planetary Sciences, Massachusetts Institute of Technology, 77 Massachusetts Avenue, Cambridge, MA 02139, USA}
\affiliation{Department of Aeronautics and Astronautics, MIT, 77 Massachusetts Avenue, Cambridge, MA 02139, USA}

\author{Gregor Srdoc}
\affiliation{Kotizarovci Observatory, Sarsoni 90, 51216 Viskovo, Croatia}

\author[0000-0002-5286-0251]{Guillermo~Torres}
\affiliation{Center for Astrophysics $\vert$ Harvard \& Smithsonian, 60 Garden Street, Cambridge, MA 02138, USA}

\author[0000-0002-6778-7552]{Joseph~D.~Twicken}
\affiliation{SETI Institute, Mountain View, CA 94043, USA}
\affiliation{NASA Ames Research Center, Moffett Field, CA 94035, USA}

\author[0000-0001-6763-6562]{Roland~Vanderspek}
\affiliation{Department of Physics and Kavli Institute for Astrophysics and Space Research, Massachusetts Institute of Technology, Cambridge, MA 02139, USA}

\author[0000-0002-4265-047X]{Joshua~N.~Winn}
\affiliation{Department of Astrophysical Sciences, Princeton University, 4 Ivy Lane, Princeton, NJ 08544, USA}

\author{Carl Ziegler}
\affiliation{Department of Physics, Engineering and Astronomy, Stephen F. Austin State University, 1936 North St, Nacogdoches, TX 75962, USA}

\begin{abstract}

The radius valley carries implications for how the atmospheres of small planets form and evolve, but this feature is visible only with highly precise characterizations of many small planets. We present the characterization of nine planets and one planet candidate with both NASA TESS and ESA CHEOPS observations, which adds to the overall population of planets bordering the radius valley. While four of our planets—TOI 118 b, TOI 455 b, TOI 560 b, and TOI 562 b—have already been published, we vet and validate transit signals as planetary using follow-up observations for five new TESS planets, including TOI 198 b, TOI 244 b, TOI 262 b, TOI 444 b, and TOI 470 b. While a three times increase in primary mirror size should mean that one CHEOPS transit yields an equivalent model uncertainty in transit depth as about nine TESS transits in the case that the star is equally as bright in both bands, we find that our CHEOPS transits typically yield uncertainties equivalent to between two and 12 TESS transits, averaging 5.9 equivalent transits. Therefore, we find that while our fits to CHEOPS transits provide overall lower uncertainties on transit depth and better precision relative to fits to TESS transits, our uncertainties for these fits do not always match expected predictions given photon-limited noise. We find no correlations between number of equivalent transits and any physical parameters, indicating that this behavior is not strictly systematic, but rather might be due to other factors such as in-transit gaps during CHEOPS visits or nonhomogeneous detrending of
CHEOPS light curves.

\end{abstract}

\section{Introduction} \label{sec:intro}

The number of officially confirmed and validated exoplanets has now exceeded 5200\footnote{From NASA Exoplanet Archive, \url{https://exoplanetarchive.ipac.caltech.edu/} \citep{Exoplanet_Archive}}. Due to this substantial sample size, our knowledge of exoplanetary systems, their properties, and formation and evolutionary processes has greatly expanded in the past three decades. It is now known that the distribution of planet radii between the size of Earth and Neptune is bifurcated in two distinct populations: super-Earths and sub-Neptunes \citep{berger2020gaia,fulton2018california,Parviainen2015LDTK,petigura2022california}. Super-Earths are those planets which have radii between 1 and 1.5 times that of Earth and have densities indicative of rocky compositions \citep{dressing2015kepler93}, whereas sub-Neptunes have radii between 2 and 3.5 times that of Earth and have relatively lower densities \citep{chouqar2020properties,bean2021subneptunes}, suggesting different formation/evolution pathways for these different populations \citep{swain2019terrestrial,Luque_2022}. Therefore, "the radius valley", as it is called, is the result of physical processes which shape this feature in the distribution of planet radii. 

 Large populations of precisely characterized planets are required to resolve the valley \citep{macdonald2019examining,petigura2022california}. The NASA \emph{Transiting Exoplanet Survey Satellite} (\tess; \citealt{ricker2015TESS}) mission is poised to deliver on this requirement via its full-sky observations of transiting exoplanets, making it the largest survey for transiting exoplanets to date. Meanwhile, the ESA \emph{CHaracterizing ExOPlanets Satellite} (\cheops; \citealt{Broeg2013CHEOPS,Benz2021CHEOPS}) is a larger space telescope, launched in December 2019, with a 32 cm aperture for the purpose of precision followup of known planetary systems. \cheops\ has the capability to improve radius measurements and orbital properties of planets it observes. Thus, when these two photometric telescopes are used in tandem, further insights into important physical phenomena which govern the characteristics of known planets may be gleaned.

There is a growing sample of systems which have been observed by both \tess\ and \cheops, for which ultra-high precision measurements are important. For example, these observations have illuminated properties of planets in multi-planet systems \citep{Bonfanti2021CHEOPS,Hoyer_2022HD108236,Serrano_2022HD93963,Wilson_2022subNeptunes}, young planetary systems \citep{zhou2022minineptune}, phase curves of KELT-1b \citep{Parviainen_2022ArXiv}, and spin-orbit misalignment of planets orbiting rapidly-rotating stars \citep{Garai_2022}, among others. These observations require precision on the $\sim$ tens of ppm scales in order to draw meaningful conclusions. Given the different sizes of the primary apertures of these telescopes, it is reasonable to expect varying photometric performances from these telescopes, but quantifying these differences has not yet been explored fully. Previous work comparing these telescopes has found that for a V $\approx 9$ mag solar-like star and a transit signal of $\approx 500$ ppm that one \cheops\ transit is equivalent in photometric precision to eight \tess\ transits combined \citep{Bonfanti2021CHEOPS}. We expand on this by writing formalism for comparison between the two, which is presented in section \ref{sec:comparison}.

We present the characterization of 10 systems which were initially detected by \tess\ and subsequently observed by \cheops, including the validation of four new \tess\ planets. These are systems which host small planets that may border the radius valley, but whose properties were poorly constrained prior to their observations with \cheops. In this work, we present the largest single sample to date of planets which were observed by both of these telescopes and analyzed homogeneously. Importantly, our sample size gave us the opportunity to compare the relative photometric performances of \tess\ and \cheops, in addition to measuring the properties of these planets. We investigated the possibility that system properties are not influenced by our analysis by modeling and fitting with three different methods. We compare system properties and uncertainties on these values as a metric for the photometric performance of these telescopes.  

Out of our ten systems, four of them were already published as validated/confirmed planets. We attempted to validate six new planets in this work, but were only able to do so for five out of those six. Given the evidence presented in section \ref{subsec:val_518}, we were not able to conclusively validate the planet candidate TOI 518.01. In our vetting process, we make use of follow-up observations for each of these new planets, including high-resolution imaging, ground-based photometric followup, and reconnaissance spectroscopic radial velocity (RV) characterization.

We describe photometric observations with \tess\ and \cheops\ in section \ref{sec:obs}, along with other follow-up observations which were performed in order to validate new planets. We present our sample of systems in section \ref{sec:sample}, including host star properties and how these values were derived. Next, we validate the systems which have not yet been validated in section \ref{sec:val}. We present our fitting and modeling of \tess\ and \cheops\ photometry in section \ref{sec:meth}. We then discuss results in section \ref{sec:results} and present our discussion in section \ref{sec:disc}, concluding and summarizing in section \ref{sec:conc}. 

\subsection{Target Selection}

Here we describe how we selected the sample of TOIs for which we obtained \cheops\ observations. The main science goal of the \cheops\ proposals was to better constrain the density and bulk composition of small \tess\ planets, so we only selected TOIs smaller than 5 $R_{Earth}$, and for which one or two \cheops\ transits were expected to substantially improve the precision on the planet radius measurement available at the time of proposal submission. We also ensured the TOIs had already undergone reconnaissance spectroscopic and photometric follow-up to rule out eclipsing binaries as the source of the transit signal. 

We then applied two cuts based on brightness (G mag $<$ 12 as recommended by the \cheops\ AO policies and procedures) and \cheops\ observability at a minimum of 50\% efficiency. 
Lastly, we removed any targets found on the \cheops\ guaranteed time observing (GTO) program reserved target list at the time of proposal submission. The final sample for which we obtained \cheops\ observations consists of ten TOIs.

\section{Observations}\label{sec:obs}

\begin{table*}
 \caption{\tess\ observations of the systems presented here.}
 \label{tab:tess_obs}
 \centering
 \begin{tabular*}{\linewidth}{llllllll}
  \hline
  TOI ID & TIC ID & \tess\ Sectors & PM Cadence & EM Cadence & Camera-CCD & RA & Dec \\
  \hline
  TOI 118 & TIC 266980320 & [1,28] & 2 min & 2 min & 2-2 & 23:18:14.22 & -56:54:14.35 \\[2 pt]
  TOI 198 & TIC 12421862 & [2,29] & 2 min & 2 min & 1-2 & 00:09:05.16 & -27:07:18.28 \\[2 pt]
  TOI 244 & TIC 118327550 & [2,29] & 2 min & 20 s & 2-3 & 00:42:16.74 & -36:43:04.71 \\[2 pt] 
  TOI 262 & TIC 70513361 & [3,30] & 2 min & 2 min & 2-3 & 02:10:08.32 & -31:04:14.26 \\[2 pt] 
  TOI 444 & TIC 179034327 & [4,5,31,32] & 2 min & 2 min & 2-1 \& 2-2 & 04:16:44.16 & -26:45:59.07 \\[2 pt] 
  TOI 455 & TIC 98796344 & [4,31] & 2 min & 20 s & 2-4 & 03:01:50.99 & -16:35:40.18 \\[2 pt]
  TOI 470 & TIC 37770169 & [6,33] & 2 min & 2 min & 2-2 \& 2-1 & 06:16:02.38 & -25:01:53.08 \\[2 pt]
  TOI 518 & TIC 264979636 & [7,34\footnote{Poor data quality, not used}] & 2 min & 2 min & 1-4 & 07:42:52.03 & +08:52:00.86 \\[2 pt]
  TOI 560 & TIC 101011575 & [8,34] & 2 min & 2 min & 2-3 \& 2-4 & 08:38:45.19 & -13:15:23.50 \\[2 pt] 
  TOI 562 & TIC 413248763 & [8,35] & 2 min & 2 min & 2-3 & 09:36:01.79 & -21:39:54.23 \\[2 pt] 
  \hline
 \end{tabular*}
\end{table*}

In this section, we outline the observations of our targets, including with \tess\ and \cheops\ for all targets. For those targets which have not yet been validated, we briefly outline additional observations in subsection  \ref{subsec:followup}.

\subsection{\tess}

\tess\ is a spacecraft with four telescopes conducting an all-sky survey of nearby bright stars in search of transiting exoplanets \citep{ricker2015TESS}. It was launched in April of 2018 and systematically surveys 24 deg $\times$ 96 deg portions of the sky called sectors for approximately 27 days at a time. During its two-year Primary Mission (PM), it observed both the southern and northern ecliptic skies in 13 sectors each, for a total of 26 sectors. It followed a similar path during its 27 month First Extended Mission (henceforth EM1), part of whose purpose is to provide additional followup and shorter cadence observations of targets which were observed in the PM (its sky pattern in EM1 was slightly different, encompassing more of the ecliptic than in the PM). As such, each of our targets were observed in at least two \tess\ sectors including the PM and the 1EM. As of September 2022, the Second Extended Mission (EM2) has begun, but this work relies only on the PM and EM1. 

In order to maximize sky coverage, visibility, and stability, \tess\ is on a 13.7 d lunar-resonant eccentric orbit \citep{tess_orbit}. In the PM and EM1, it remains pointed at the same part of the sky for two orbits at a time, with data downlink between the two orbits, leading to regular data gaps in the middle of \tess\ sectors. The $2048\times2048$ imaging area on each CCD has a pixel scale of about 21"/pix. Pixel readout occurs continuously at 2 second cadence, which is then stacked to either 2-minute Postage Stamps for 20,000 pre-selected targets\footnote{Lists of 2 min and 20 s cadence targets: \url{https://tess.mit.edu/observations/target-lists/}} for each sector or 30-min Full Frame Images (FFIs) for the full field in the PM. During the EM1, FFI cadence is reduced from 30-min to 10-min, and there are an additional 1,000 pre-selected targets at 20 s cadence, along with 20,000 pre-selected 2 min targets. The number of pre-selected targets observed at 20 s cadence increased to 1,300 by the end of EM1.

All of the systems which are the subject of this work were first observed by \tess\ during Year 1 of its PM. All of our targets in the \tess\ Input Catalog (TIC; \citealt{Stassun_2018,Stassun_2019}) were observed at 2 min cadence in the PM. \tess\ observations for these systems were processed by both the Science Processing Operations Center (SPOC; \citealt{Jenkins2016SPOC}) at NASA Ames Research Center and the Massachusetts Institute of Technology (MIT) Quick-Look Pipeline (QLP; \citealt{Huang_2020_QLP1,Huang_2020_QLP2}). SPOC and QLP are both pipelines for extracting light curves, except QLP extracts light curves exclusively from FFIs. These systems were flagged as potential candidates by either SPOC or QLP, vetted by the \tess\ vetting team, and were designated as \tess\ Objects of Interest (TOIs; \citealt{guerrero2021tess}). Each of these TOIs were then re-observed by \tess\ in its EM1, providing a longer baseline of photometric observations, which refined uncertainties in the periods of our targets, as well as other parameters such as transit depth and planet radius. Eight of our systems were observed at 2 min cadence again, but TOI 244 and TOI 455 were observed at 20 s cadence in the EM1, as shown in Table \ref{tab:tess_obs}. We used the shortest available cadence for all of our analysis, meaning we used 2 min cadence light curves for a majority of our \tess\ light curves, but used 20 s cadence light curves for TOI 244 and TOI 455. 

Since all of our targets were selected for observation in short cadence, we analyzed short cadence SPOC light curves for all of our systems. Additionally, SPOC applies Presearch Data Conditioning to its light curves which were extracted via Simple Aperture Photometry (PDCSAP), a procedure initially developed for the \emph{Kepler} mission \citep{Smith_2012,Stumpe_2012,Stumpe_2014}. We chose to work with PDCSAP light curves for our analysis, which is assumed to be corrected for instrumental effects.

%new addition!
A DOI for these \tess\ observations has been created and is hosted by MAST at the following web address: \url{http://dx.doi.org/10.17909/dshz-jz09}.

\subsection{\cheops}\label{subsec:CHEOPS}

\begin{table*}
    \centering
    \caption{Details of \cheops\ visits for all targets, including month of observation, visit duration in hours, number of frames, observing efficiency as reported by the DRP, and Median Absolute Difference (MAD).}
    \begin{tabular}{cccccc}
    \hline
        TOI ID & Obs. Date & Visit Duration & Number of Frames & Efficiency & MAD \\
             &           &  [hrs]           &                  & [\%]         & [ppm] \\ \hline
         118 & Aug 2021  & 7.4            & 305              & 68.5\%     & 257 \\
         198 & Sept 2021 & 8.4            & 472              & 93.4\%     & 611 \\
         244 & Oct 2021  & 10.9 \& 10.6   & 484 \& 488   & 73.7\% \& 76.3\% & 741 \& 602 \\
         262 & Oct 2020  & 16.2            & 1441          & 83.2\%     & 205 \\
         444 & Dec 2020  & 14.9            & 815            & 90.9\%     & 252 \\
         455 & Oct 2020  & 6.5            & 357              & 91.7\%     & 278 \\
         470 & Dec 2020  & 6.3            & 365              & 96.5\%     & 588 \\
         518 & Dec 2021 \& Mar 2022  & 8.1 \& 7.5  & 299 \& 291 & 61.4\% \& 64.6\%     & 406 \& 443 \\
         560 & Jan 2021  & 5.0            & 224              & 72.7\%     & 215 \\
         562 & Mar 2022  & 6.5            & 359              & 92.2\%     & 358 \\ \hline
    \end{tabular}
    \label{tab:cheops_obs}
\end{table*}

%Diana to add how CHEOPS targets were selected and the proposal IDs
The \emph{CHaracterising ExOPlanets Satellite (CHEOPS)} mission is a European Space Agency small-class mission dedicated to studying bright, nearby exoplanet host stars for the purpose of making high-precision observations of transiting super-Earth and sub-Neptune planets. It was launched in December of 2019 and is currently in a sun-synchronous orbit $\sim700$ km above Earth. There are two consequences of \cheops's orbital configuration which manifest themselves in our observations. First, the low-Earth orbit of the spacecraft renders certain parts of the sky unobservable by \cheops, but importantly this also means that stray light from the Earth will sometimes surpass acceptable levels during observation, leading to gaps in the data at these times. As such, there is an associated observing efficiency associated with each \cheops\ observation, which is the fraction of the observation which is successfully retained. Second, given its nadir-locked orbit (i.e. the Z-axis of the spacecraft is antiparallel to the nadir direction), the field of view rotates about the central optical axis with the same period as the spacecraft orbit. %This means the location of the target star on the detector may change, whose point-spread function (PSF) varies on the CCD. This is an effect which we addressed by applying PSF photometry to generate our light curves, which we discuss below.
 
%Seven targets were proposed as part of the AO-1, but we got data for 5 of them (I think). Then the rest were submitted as part of a similar campaign for AO-2
We proposed five of these targets for observation in \cheops's first Announcement of Opportunity (AO-1), which spanned the period from March 2020 to March 2021. We proposed a similar campaign for the remaining targets on our list in AO-2, which spanned the following year from March 2021 to March 2022. 

%this paragraph may be removed if we end up going with PIPE analysis
Our observations were processed by the \cheops\ Data Reduction Pipeline (DRP; \citealt{hoyer2020expected}), which calibrates and corrects for instrumental and environmental effects, such as bias, gain, and flat-fielding. The DRP performs three main functions, which are calibration, correction, and photometry. The calibration phase corrects for instrumental response, the correction phase accounts for environmental effects (such as stray light or cosmic rays), and the photometry phase transforms the calibrated and corrected images into light curves. The calibration phase consists of standard CCD reduction, including corrections for bias, gain, dark current, and flat fielding. The correction phase accounts for smearing, bad pixels, depointing, pixel to sky mapping, and background and stray light. Finally, the DRP produces aperture photometry for radii from 15 up to 40 pixels. %Finally, the photometry phase performs aperture photometry on the images to produce light curves of the observations. Four different aperture radii are available to the user by default, including DEFAULT (25 pix), RINF (DEFAULT * 0.9 = 22.5 pix), RSUP (DEFAULT * 1.2 = 30 pix), and OPTIMAL. The OPTIMAL radius is not a fixed value, but rather is determined by calculation of the maximum of the signal-to-noise ratio of the light curve, which is computed continuously from 15 to 40 pixels. 

In addition to DRP aperture photometry, we extracted light curves from our \cheops\ visits with the point-spread function (PSF) photometry package \texttt{PIPE}\footnote{\url{https://github.com/alphapsa/PIPE}} developed specifically for \cheops\ (Brandeker et al.\ in prep.; see also descriptions in \citealt{sza21} and \citealt{mor21}) that has been proven to produce light curves consistent with aperture photometry but less affected by background stars \citep{Serrano_2022HD93963}. Briefly, \texttt{PIPE} derives the PSF from the imagettes returned by the telescope, and then fits the PSF to each image cutout, i.e. imagette, to yield a light curve. PSF photometry has the advantage of reducing background noise from nearby stars relative to aperture photometry. Light curves generated with \texttt{PIPE} exhibited lower Median Absolute Differences (MADs) than those generated by the DRP. Therefore, we chose to use \texttt{PIPE} light curves for all of our \cheops\ visits, with the exception of TOI 455 since it is part of a highly blended stellar system. Our observations are described in Table  \ref{tab:cheops_obs}, and our \cheops\ light curves are shown in Fig. \ref{fig:CHEOPS_LCs}. 
Additionally, all of our \cheops\ detrended and raw extraction light curves are publicly available online on ExoFOP-TESS\footnote{\url{https://exofop.ipac.caltech.edu/tess/}}.

\subsection{Followup Observations}\label{subsec:followup}
Here, we detail our followup observations which were acquired to vet and validate the planetary signals for new systems.

\subsubsection{Precise RVs of TOI 198 with VLT-ESPRESSO}

From July 4th to September 12th, 2019, we acquired 23 spectra of TOI-198 with ESPRESSO under ESO program-id 0103.C-0849(A). The spectrograph is stabilized in pressure and temperature for precise radial velocity measurements, operating in the wavelength domain from 380 nm to 788 nm with a resolving power of 140,000 in the high-resolution mode used here \citep{Pepe_2013}. The raw data were reduced with the ESO dedicated pipeline and radial velocities were computed by a template-matching approach following \citet{Astudillo-Defru_2017}.  That is, we obtained a enhanced signal-to-noise spectrum from all available spectra that is Doppler shifted in radial velocity where tellurics were neglected. Then we maximize the likelihood between individual spectra and the template to obtain the radial velocity, whose uncertainty is derived following \citet{Bouchy_2001}. These RVs are given in Table \ref{tab:TOI198_RVs}.

\subsubsection{Reconnaissance Spectra with FLWO-TRES \& CTIO/SMARTS-CHIRON}

Reconnaissance spectra were obtained with the Tillinghast Reflector Echelle Spectrograph (TRES; \citealt{gaborthesis}) which is mounted on the 1.5m Tillinghast Reflector telescope at the Fred Lawrence Whipple Observatory (FLWO) atop Mount Hopkins, Arizona. TRES is an optical, fiber-fed echelle spectrograph with a wavelength range of 390-910\,nm and a resolving power of R$\sim44\,000$. The TRES spectra were extracted as described in \citet{buchhave2010} and a multi-order relative velocity analysis was performed for TOI-262 and TOI-444 by cross-correlating the strongest observed spectrum as a template, order by order, against the remaining spectra, for each target. We used methods designed for M-dwarf stars (TRES41; \citealt{irwin2018}) to derive the rotational velocities for TOI-198 and TOI-244. TRES41 uses the wavelength range 707-717\,nm which is dominated by TiO to cross-correlate an observed spectrum against Barnard’s star to estimate the rotational velocity of the star. Stellar parameters were derived for TOI-444, TOI-470, and TOI-518 using the Stellar Parameter Classification (SPC; \citealt{buchhave2012}) tool. SPC cross correlates an observed spectrum against a grid of synthetic spectra based on Kurucz atmospheric models \citep{kurucz1992} to derive effective temperature, surface gravity, metallicity, and rotational velocity of the star. 

Reconnaissance spectra for each of the systems we validate were also  obtained with the CHIRON fiber-fed cross-dispersed echelle spectrometer \citep{Tokovinin_2013} at the Cerro Tololo Inter-American Observatory (CTIO)/Small and Moderate Aperture Research Telescope System (SMARTS) 1.5-m telescope. CHIRON has a spectral resolving power of R = 80,000 over the wavelength range of 410–870 nm. Spectra from CHIRON were reduced as per \citet{Paredes_2021}. The radial velocities were measured following the procedure from \citet{Zhou_2020} via a least-squares deconvolution of each observation against a synthetic nonrotating template generated from the ATLAS9 model atmospheres \citep{Castelli_Kurucz_2004}. 

\subsubsection{High-contrast imaging with Gemini-Zorro/'Alopeke, Keck2-NIRC2, Palomar-PHARO, \& SOAR-HRCam}

In an effort to measure the impact of possible contamination from nearby stars and rule out the possibility of stellar companions, we obtained high-contrast imaging with multiple large ground-based telescopes. Bound stellar companions, in addition to diluting transit signals and leading to the underestimation of planet radii \citep{ciardi2015}, can create false-positive transit signals if they are eclipsing binaries (EBs).

We obtained high-contrast images from Gemini-Zorro/'Alopeke \citep{Scott_2021}, Keck2-NIRC2 \citep{2000SPIE.4007....2W}, Palomar-PHARO \citep{hayward2001}, and SOAR-HRCam \citep{Tokovinin_2018}. Our Gemini observations are described in Table \ref{tab:Gemini}. These observations are more precisely described in section \ref{sec:val}.

\subsubsection{Ground-based photometry with LCOGT}\label{subsec:LCO}
%Add other LCO observations

The \tess\ pixel scale is $\sim 21\arcsec$ pixel$^{-1}$, and photometric apertures typically extend out to roughly 1 arcminute, which generally results in multiple stars blending in the TESS aperture. We acquired ground-based time-series follow-up photometry of our planet candidates as part of the \tess\ Follow-up Observing Program Sub Group 1 \citep[TFOP SG1;][]{collins_2019}\footnote{\url{https://tess.mit.edu/followup} \citep{EXOFOP}} to attempt detect the transit-like events on target and to rule out or identify nearby eclipsing binaries (NEBs) as the potential sources of the \tess\ detections. 

We observed full predicted transit windows of TOI-198.01, TOI-244.01, TOI-444.01, and TOI-470.01 using the Las Cumbres Observatory Global Telescope \citep[LCOGT;][]{Brown_2013} 1.0\,m network. We observed TOI-198.01 on UT 2022 September 14 and TOI-244.01 on UT 2019 August 01 from the South Africa Astronomical Observatory (SAAO) node in Pan-STARRS $z$-short (zs) band. We observed TOI-444.01 on UT 2020 October 31 from the Siding Spring Observatory node in zs-band, and TOI-470.01 on UT 2021 October 23 from the Cerro Tololo Inter-American Observatory (CTIO) node in both Sloan $g'$ and zs bands. We used the {\tt TESS Transit Finder}, which is a customized version of the {\tt Tapir} software package \citep{Jensen_2013}, to schedule our transit observations. The 1\,m telescopes are equipped with $4096\times4096$ SINISTRO cameras having an image scale of $0\farcs389$ per pixel, resulting in a $26\arcmin\times26\arcmin$ field of view. The images were calibrated by the standard LCOGT {\tt BANZAI} pipeline \citep{Mccully_2018}. Differential photometric data were extracted with {\tt AstroImageJ} \citep{Collins_2017} using target star circular photometric apertures which exclude all flux from the nearest known Gaia DR3 stars that are bright enough to be capable of causing the TESS detection. Transit-like events that are consistent with the depths, durations, and ephemerides measured by \textit{TESS} were detected in the follow-up apertures, confirming that the TOI-198.01, TOI-244.01, TOI-444.01, and TOI-470.01 signals occur on-target relative to known Gaia DR3 stars. These observations are further discussed in section \ref{sec:val}.

\section{Stellar Properties}\label{sec:sample}

\subsection{Published system parameters}\label{sec:stell_pars}

\begin{sidewaystable*}
\centering
\caption{Stellar parameters for previously published targets, TOI 118, TOI 455, TOI 560, and TOI 562.}
\label{tab:pub_stell_props}
\begin{tabular}{|c|c|c|c|c|c|}
\hline
Parameter   & Unit       & TOI 118         & TOI 455         & TOI 560        & TOI 562 \\ \hline
G mag.      & mag        & 9.650           & 10.058          & 9.270          & 9.880 \\
TESS mag.   & mag        & 9.179           & 8.840           & 8.592          & 8.741 \\
Spect. type &            & G5V             & M3.0            & K4V            & M2.5V \\
T$_{\rm eff}$        & K          & $5527\pm65$       & $3340\pm150$    & $4511\pm110$   & $3505\pm51$ \\
{[}Fe/H{]}  & dex        & $0.04\pm0.04$  & $-0.34\pm0.09$  & $0.00\pm0.09$  & $-0.12\pm0.16$ \\
log(g)      &            & $4.40\pm0.11$    & $4.62\pm0.12$  & $4.94\pm0.07$  \\
R$_{\star}$  & R$_{\sun}$ & $1.03\pm0.03$  & $0.265\pm0.011$ & $0.65\pm0.02$  & $0.337\pm0.015$ \\
M$_{\star}$  & M$_{\sun}$ & $0.92\pm0.03$  & $0.257\pm0.014$ & $0.73\pm0.02$  & $0.342\pm0.011$ \\
log(R'HK)   &            & $-5.07\pm0.03$ &                  & $-4.47\pm0.02$ & $-5.37$ \\
P$_{rot}$   & days       &                &                  & $12.2\pm0.2$   & $77.8\pm2.1$  \\
Age         & Gyr        & $10.0\pm2.0$   &                  & $0.48\pm0.19$  &        \\ 
Source    &  & \citet{Esposito2019HD219666} & \citet{Winters_2022} & \citet{Barragan2022TOI560} & \citet{Luque2019GJ357} \\ \hline
\end{tabular}
\end{sidewaystable*}

Some of these planets have been published in previous works. We give stellar parameters as computed by these authors in Table \ref{tab:pub_stell_props}. Given that these are parameters for systems which are already well-characterized, we do not perform any further stellar analysis, and use these parameters to calculate planet properties later. We include the reference from which these parameters were taken for each star at the bottom of the tables.

\subsection{Spectral energy distribution (SED) analysis of new systems}

As an independent determination of the basic stellar parameters for previously unpublished systems, we performed an analysis of the broadband spectral energy distribution (SED) of each star together with the {\it Gaia\/} EDR3 parallax \citep[with no systematic offset applied; see, e.g.,][]{StassunTorres:2021}. This was in order to determine an empirical measurement of the stellar radius, following the procedures described in \citet{Stassun:2016,Stassun:2017,Stassun:2018}. Depending on the photometry available for each source, we pulled the $B_T,V_T$ magnitudes from {\it Tycho-2} \citep{TGAS}, the $BV$ $g,r,i$ magnitudes from {\it APASS} \citep{APASS}, the $JHK_S$ magnitudes from {\it 2MASS} \citep{2MASS}, the W1--W4 magnitudes from {\it WISE}, the $G$, $G_{\rm BP}$, $G_{\rm RP}$ magnitudes from {\it Gaia} \citep{Gaia}, and the FUV and/or NUV fluxes from {\it GALEX} \citep{GALEX}. Together, the available photometry generally spans the stellar SED over the approximate wavelength range 0.2--22~$\mu$m (see Appendix \ref{app:SED}).  

We performed fits to the photometry using Kurucz stellar atmosphere models, with the principal parameters being the effective temperature ($T_{\rm eff}$), metallicity ([Fe/H]), and surface gravity ($\log g$), for which we adopted the spectroscopically determined values when available. We included the extinction, $A_V$, as a free parameter but limited to the full line-of-sight value from the Galactic dust maps of \citet{Schlegel:1998}; for systems that have small distances according to {\it Gaia}, we fixed $A_V \equiv 0$. The resulting fits shown in Appendix \ref{app:SED} have a reduced $\chi^2$ ranging from 0.7 to 1.6 (in some cases excluding the GALEX photometry, if a UV excess indicative of chromospheric activity is present; see below), and the best-fit parameters are summarized in Table \ref{tab:SED}.

Integrating the model SED gives the bolometric flux at Earth, $F_{\rm bol}$. Taking the $F_{\rm bol}$ together with the {\it Gaia\/} parallax directly gives the luminosity, $L_{\rm bol}$. Similarly, the $F_{\rm bol}$ together with the $T_{\rm eff}$ and the parallax gives the stellar radius, $R_\star$. Moreover, the stellar mass, $M_\star$, can be estimated from the empirical eclipsing-binary based relations of \citet{Torres:2010} or the $M_K$-based relationships of \citet{Mann:2019} for the cooler stars, and the (projected) rotation period can be calculated from $R_\star$ together with the spectroscopically measured $v\sin i$. The mean stellar density, $\rho_\star$, follows from the mass and radius. When available, the {\it GALEX} photometry allows the activity index, $\log R'_{\rm HK}$ to be estimated from the empirical relations of \citet{Findeisen:2011}. All quantities are summarized in Table~\ref{tab:SED}. 

Where possible, we have also estimated the system ages from the $R'_{\rm HK}$ activity and/or the stellar rotation, using the activity-age and/or rotation-age empirical relations of \citet{Mamajek:2008} which are applicable for $T_{\rm eff} \lesssim 6500$~K, or the relations of \citet{Engle:2018} for the cooler stars. These quantities are also summarized in Table~\ref{tab:SED}. 

\begin{sidewaystable*}
\centering
\caption{Derived stellar parameters for the TOIs we validate from SED analysis. We compare some parameters, including T$_{\rm eff}$ and log(g) to values from the TIC.}
\label{tab:SED}
\begin{tabular}{|c|cc|cccccc|}
\hline\hline
Parameter   & Unit                   & Source         & TOI 198  & TOI 244 & TOI 262 & TOI 444    & TOI 470   & TOI 518 \\ \hline
G mag       & mag                    & Gaia DR2       & 10.915   & 11.549  & 8.678 & 9.612      & 11.245    & 10.564 \\
TESS mag    & mag                    & TIC           & 9.928     & 10.347  & 8.134 & 9.056      & 10.700    & 10.143  \\
Spect. type &                        & SIMBAD         & M0V      & M2      & K0V & K1/2V      & late G    & early G \\
Av          & mag                    &                & 0.0      & 0.0     & $0.12\pm0.02$ & $0.07\pm0.02$  & $0.07\pm0.03$   & $0.05\pm0.05$ \\
T$_{\rm eff}$        & \multirow{2}{*}{K}     & SED   & $3650\pm75$        & $3450\pm75$ & $5310\pm124$  & $5225\pm70$  & $5190\pm90$  & $5845\pm70$ \\
            &                        & TIC            & $3782\pm157$       & $3407\pm157$ & $5302\pm135$ & $5091\pm124$ & $5112\pm125$ & $5891\pm122$ \\
{[}Fe/H{]}  & dex                    & SED/TRES       & $-0.7\pm0.5$       & $0.0\pm0.3$  & $0.28\pm0.08$ & $0.08\pm0.08$ & $0.10\pm0.08$ & $0.00\pm0.08$ \\
log(g)      & \multirow{2}{*}{cgs}   & SED/TRES       & $4.75\pm0.25$      & $4.75\pm0.25$ & $4.56\pm0.10$  & $4.64\pm0.10$ & $4.54\pm0.10$  & $4.46\pm0.10$ \\
            &                        & TIC            & $4.783\pm0.005$    & $4.820\pm0.004$ & $4.53\pm0.08$ & $4.567\pm0.082$ & $4.524\pm0.086$ & $4.714\pm0.101$  \\
F$_{bol}$   & erg s$^{-1}$ cm$^{-2}$ & SED            & $1.765\pm0.041$    & $1.336\pm0.047$ & $8.82 \pm 0.21$ & $3.895\pm0.014$ & $0.830\pm0.001$ & $1.44\pm0.051$ \\
            & $\times10^{-9}$        &                &                    &     &  &  &    \\
R$_{star}$  & R$_{\sun}$             & SED            & $0.441\pm0.019$    & $0.399\pm0.019$ & $0.861 \pm 0.019$ & $0.779\pm0.053$ & $0.831\pm0.053$  & $1.027\pm0.025$ \\
M$_{star}$  & M$_{\sun}$             & SED            & $0.467\pm0.023$    & $0.424\pm0.021$ & $0.96 \pm 0.06$ & $0.96\pm0.13$ & $0.87\pm0.09$ & $1.07\pm0.06$  \\
log(R'HK)   &                        &                & $-5.05\pm0.08$     &  & $-5.08 \pm 0.05$ & $-4.46\pm0.05$  &   & $-4.76\pm0.05$  \\
P$_{rot}$   & days                   & pred,R'HK      & $45.4\pm5.3$       &  & $21.79 \pm 5.47$ & $13.0\pm3.0$ & $18.28\pm3.99$  & $15.9\pm1.6$  \\
Age         & Gyr                    & pred,P$_{rot}$ & $5.3\pm1.0$ &      & $8.1 \pm 1.0$ & $0.57\pm0.14$  & $1.4\pm0.6$ & $2.7\pm0.6$  \\ \hline
\end{tabular}
\end{sidewaystable*}

\section{Validation of new \tess\ planets}\label{sec:val}

\begin{figure*}
\centering
 \includegraphics[width=0.49\linewidth]{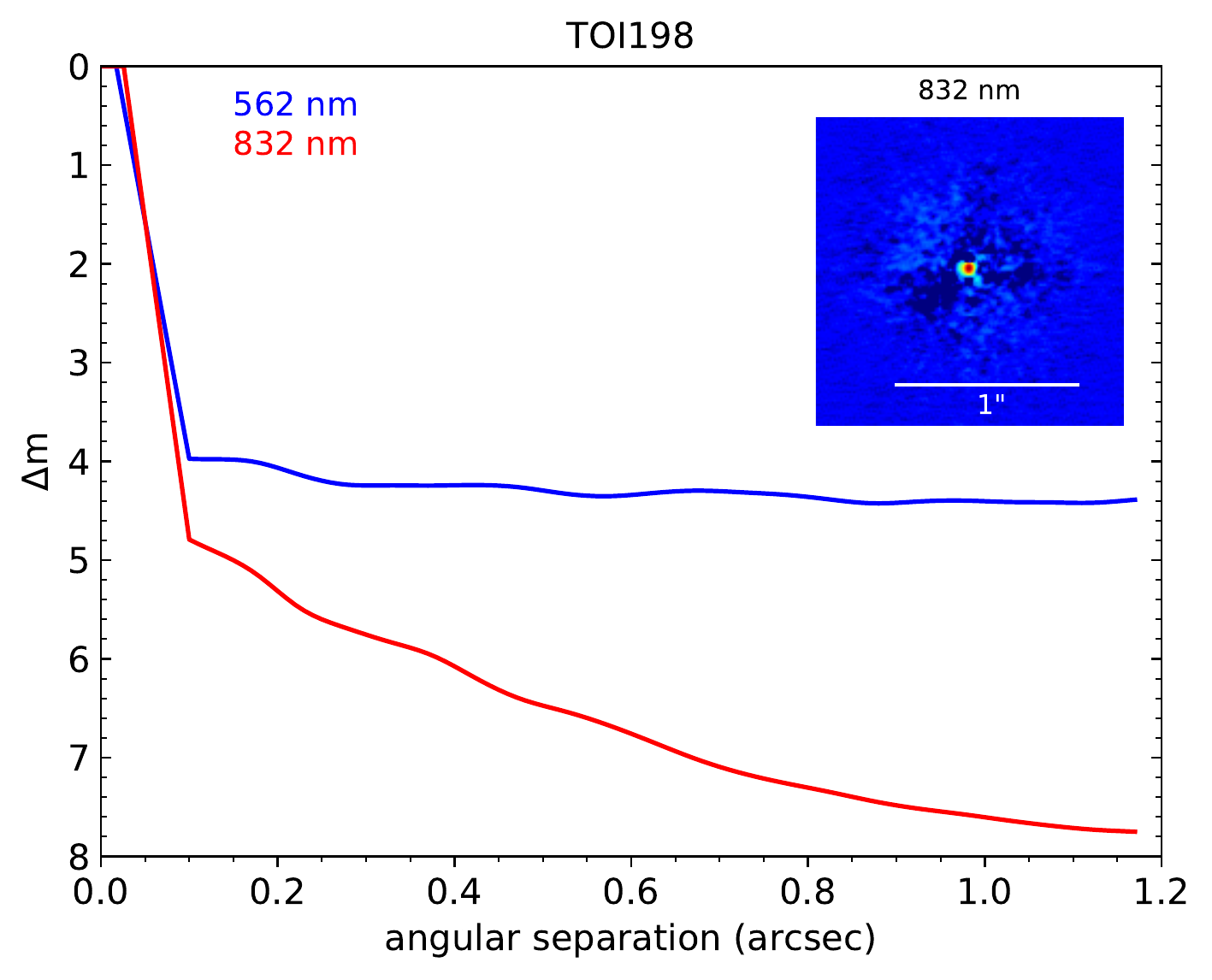}
 \includegraphics[width=0.49\linewidth]{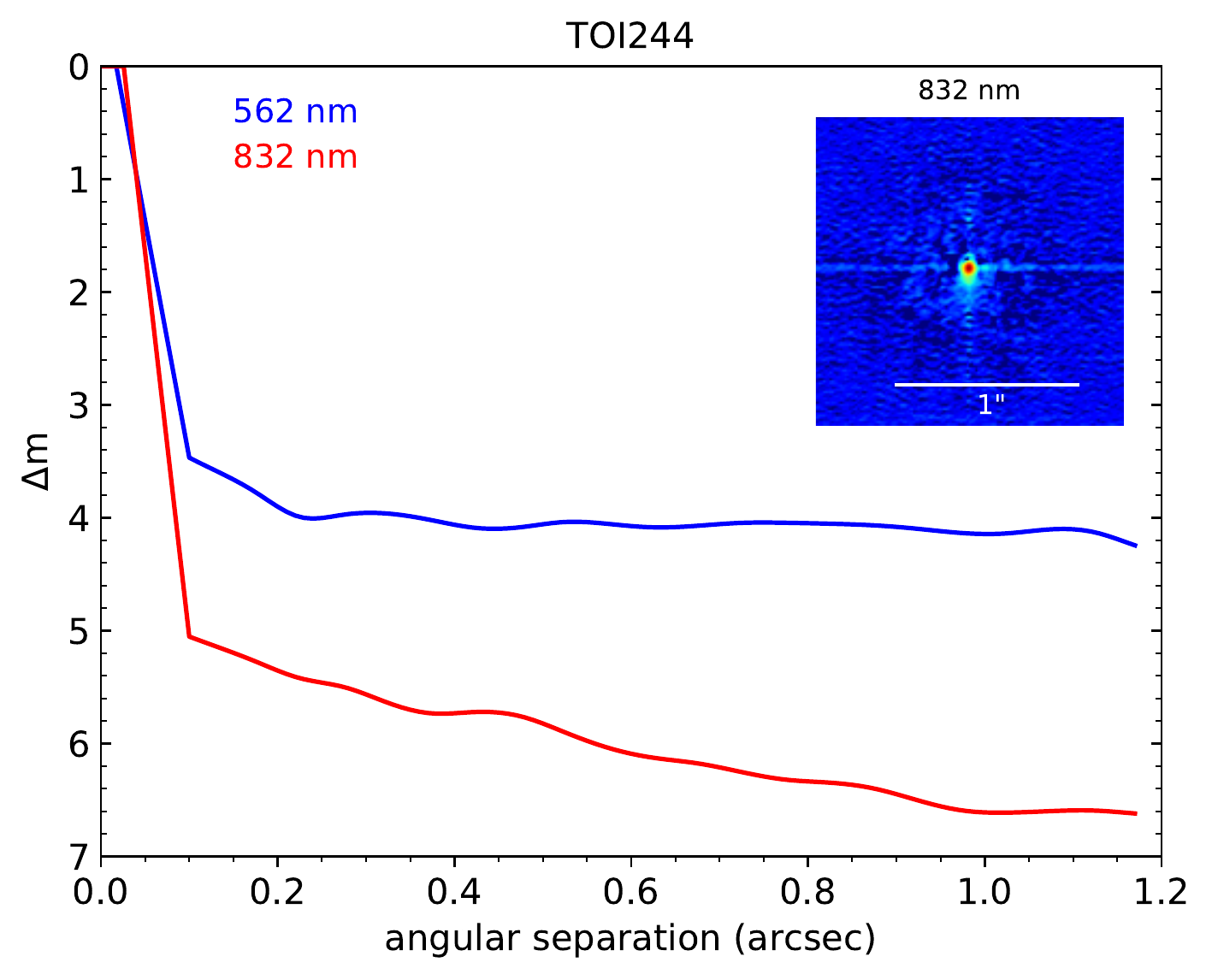}
 \includegraphics[width=0.49\linewidth]{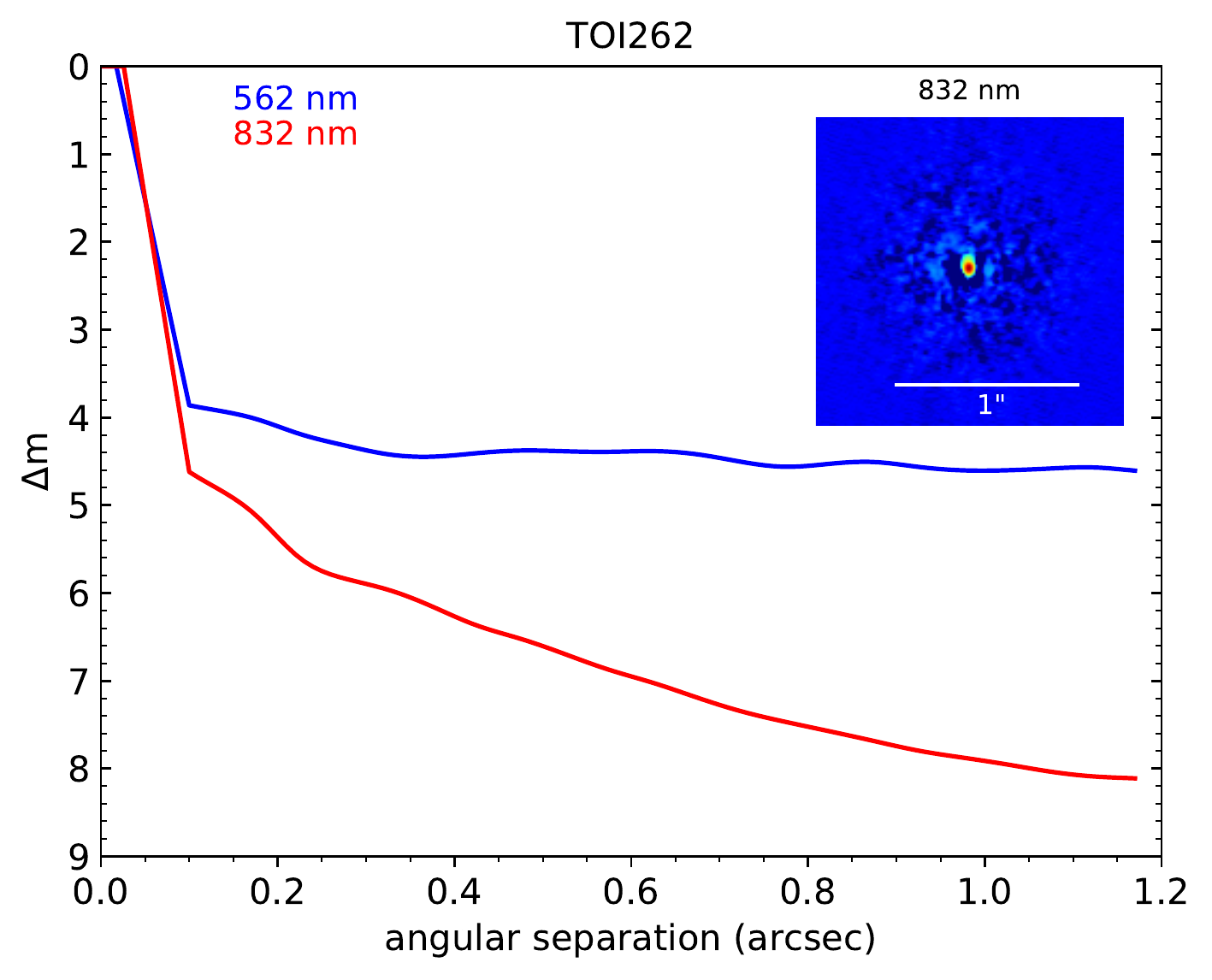}
 \includegraphics[width=0.49\linewidth]{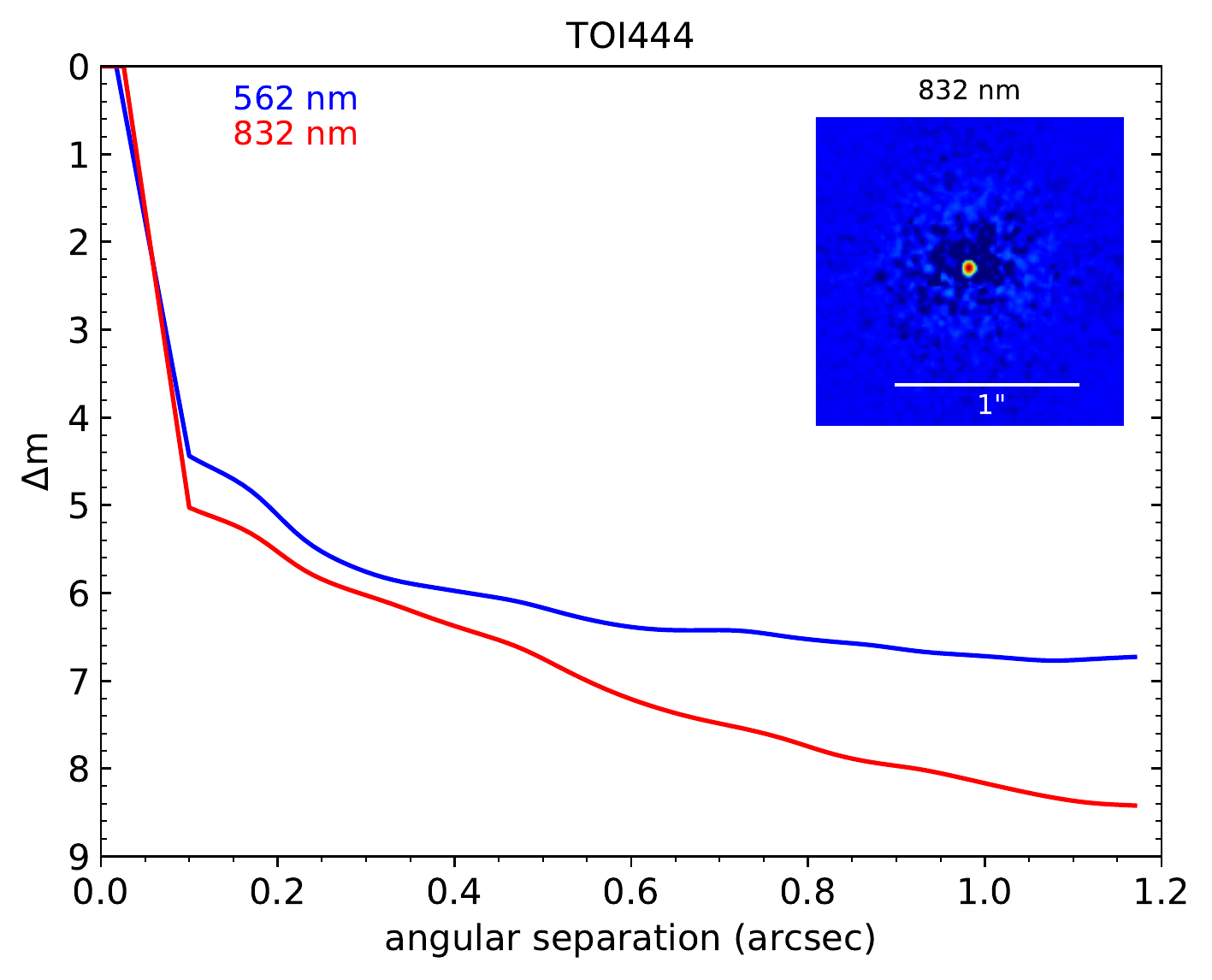}
 \includegraphics[width=0.49\linewidth]{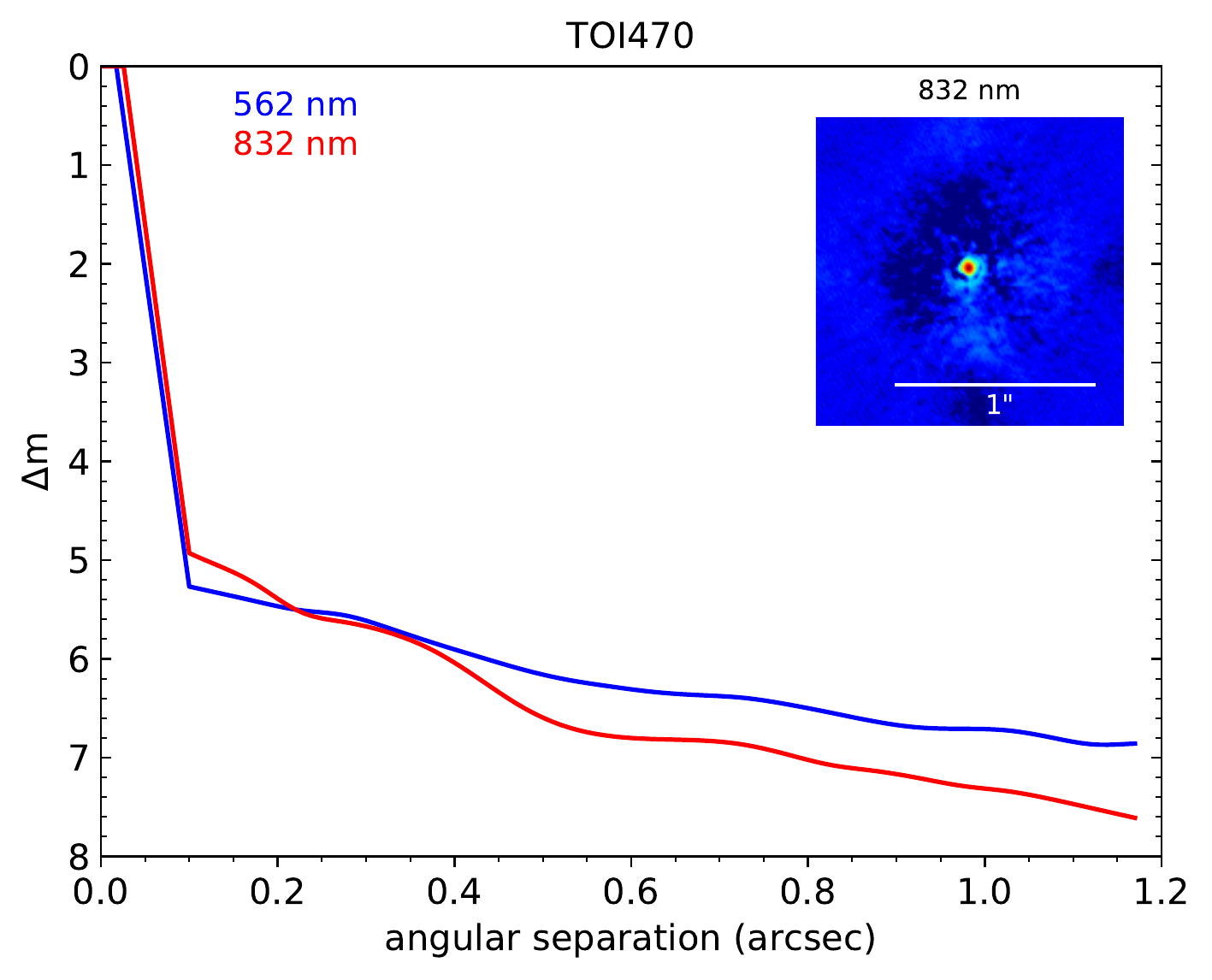}
 \includegraphics[width=0.49\linewidth]{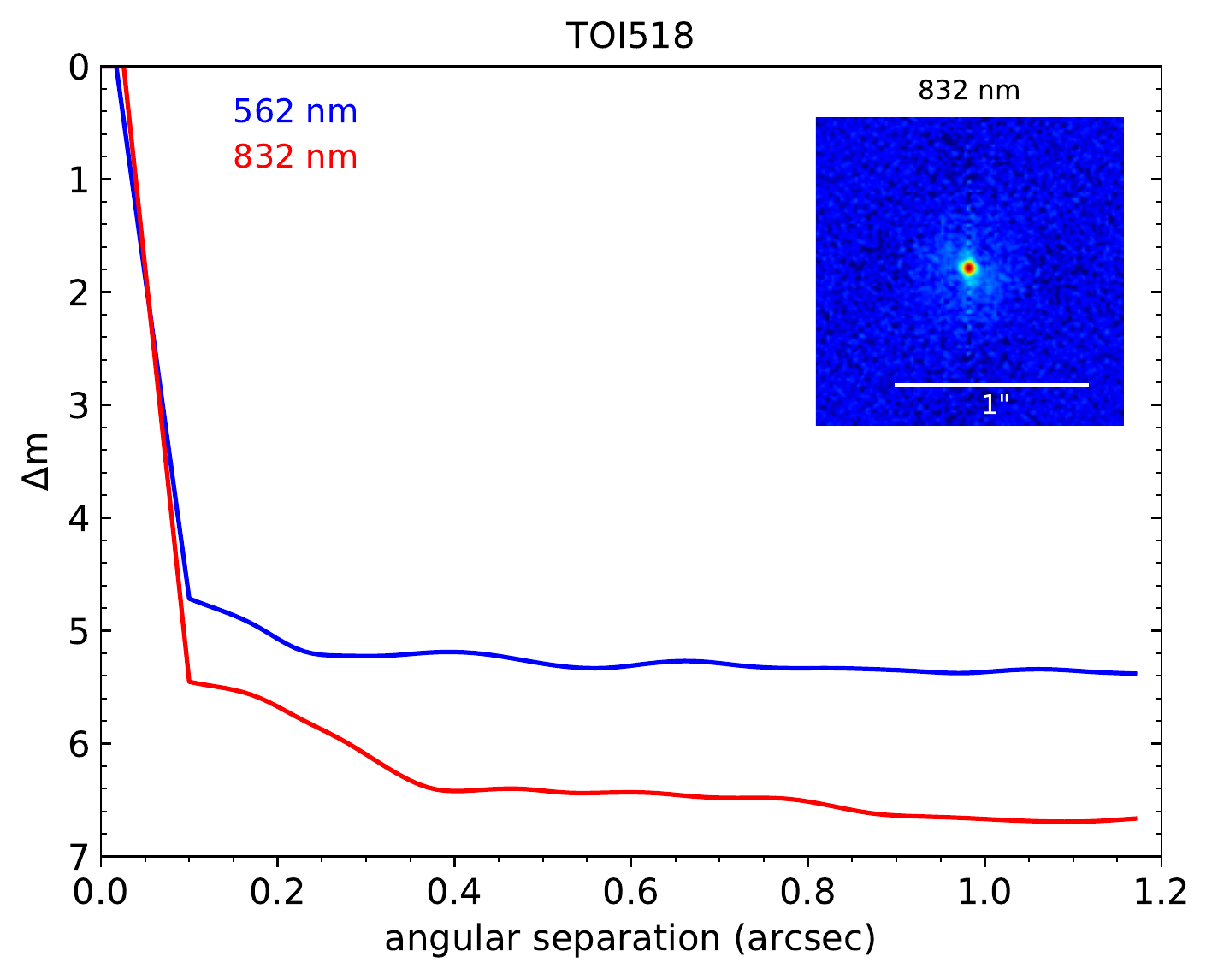}
 \caption{Contrast curves in 562 nm and 832 nm (blue and red, respectively) for TOI 198, TOI 244, TOI 262, TOI 444, TOI 470, and TOI 518 from Gemini observations with insets showing the central star at 832 nm.}
 \label{fig:contrast_curves}
\end{figure*}

In this section, we vet and validate five new planets, which is the process of identifying potential sources of the transit signal and calculating the probability that the given signal is likely to be due to a planet transiting the host star in question. We vet and validate transit signals for the following planets: TOI 198 b, TOI 244 b, TOI 262 b, TOI 444 b, and TOI 470 b. Our observations of TOI 518.01 to date were not sufficient to validate it as a planet. We also discuss our observations of TOI 518 and the reasons why it could not be validated in section \ref{subsec:val_518}. Our validation process included analyzing our photometric observations of these targets with \tess\ and \cheops, as well as followup observations with high-resolution imaging and reconnaissance spectroscopy, which serve multiple purposes. These followup observations may either reveal or rule out the possibility of nearby stellar-mass companions to the target star, provide stellar parameters for our target stars, or, if they are precise enough, reveal the mass of the orbiting companion, or limits on that value. As part of our statistical analysis of our photometric observations with \tess, we use the Bayesian statistical validation analysis code \texttt{TRICERATOPS} (Tool for Rating Interesting Candidate Exoplanets and Reliability Analysis of Transits Originating from Proximate Stars; paper: \citealt{Giacalone2021TRICERATOPS_paper}; code: \citealt{Giacalone2020TRICERATOPS_code}). Among the sources of astrophysical false positives, \texttt{TRICERATOPS} calculates the probability that the transit-like signal is due to an Eclipsing Binary (EB), including the probability that a planet exists but it orbits either the Primary or Secondary component of the binary (PEB and SEB, respectively). This code also calculates the probability that the signal is the result of a foreground star or EB that dilutes or mimics the possible transit signature, or a background star or EB which may have the same effect. Finally, \texttt{TRICERATOPS} calculates the probability that the given signal is off target and due to a nearby star or EB (NEB). 

For each of our validated targets, we show the star field around the target star with \tess\ pixels overlaid, as well as the SPOC report difference image centroid offset (as in Fig. \ref{fig:TOI198_S2}, \citealt{Twicken_2018,Li_2019}). For the star field, the yellow star in the center of the figure indicates the target star, and other dots represent other stars in the field according to their measured positions in Gaia DR2, scaled in color by their \tess\ magnitudes. Additionally, we include directional arrows pointing North and East. The grey dashed line represents an equidistant circle of radius 200 arcseconds from the target star in each star field image. In all cases our centroid offset plots show the in-transit centroid location from multiple sectors. The red asterisk denotes the location of the target, the pink cross denotes the $1\sigma$ centroid offset, and the blue circle represents the $3\sigma$ centroid offset. In each case except TOI 518 (for which we only analyze one \tess\ sector), the green crosses represent the centroid location for each sector. 

\begin{figure*}
\centering
 \includegraphics[width=.8\linewidth]{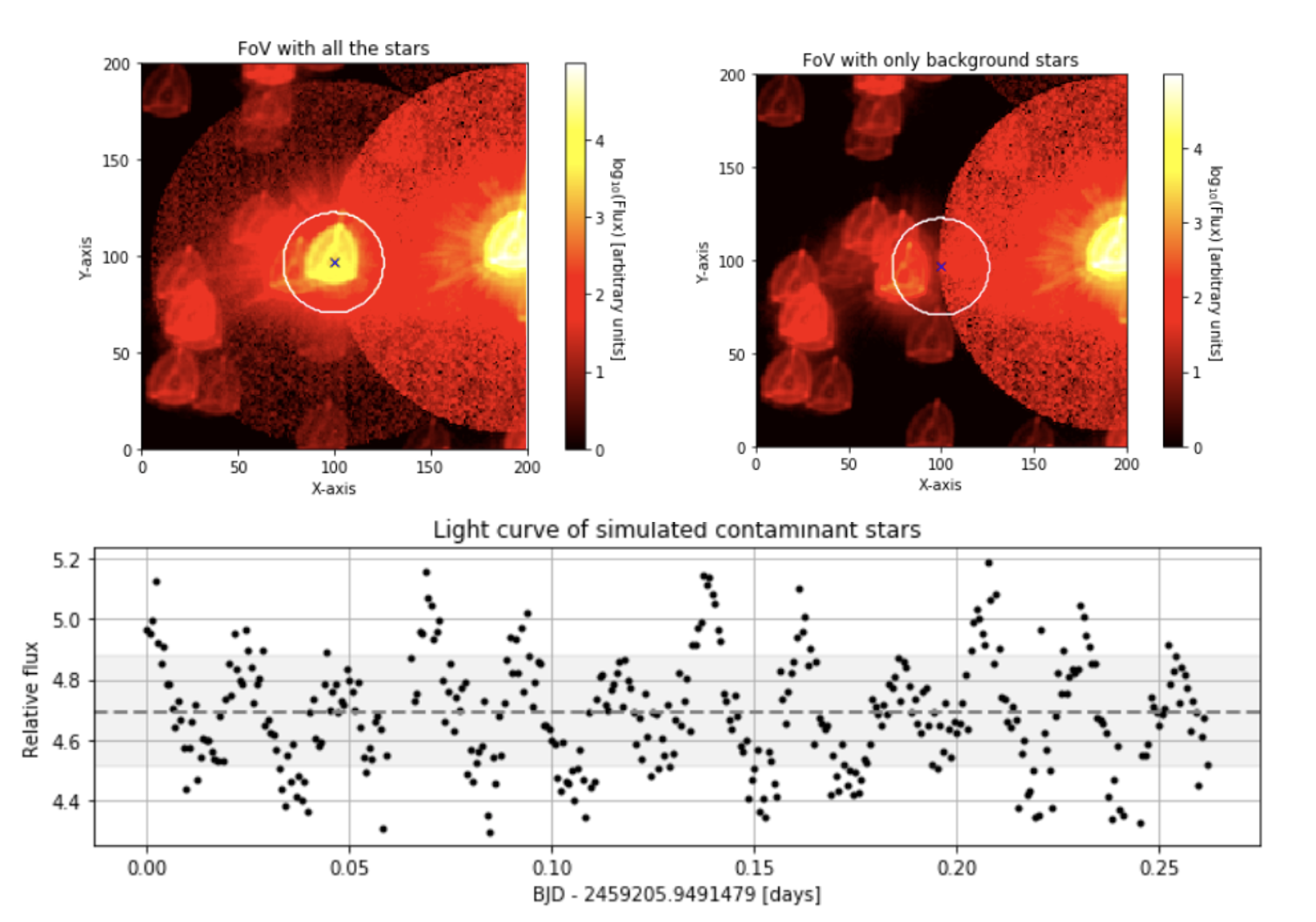}
 \caption{Diagnostics of contamination from nearby stars in the \cheops\ light curve for TOI 470. Top left: \cheops\ field of view, including both our target star (centered on black 'X') and field stars. Top right: DRP simulated image of only field stars without target star, exhibiting relative fluxes of non-target stars. Note difference in color scales between field stars with and without target star. Bottom: Simulated contributions to \cheops\ light curve from field stars, which is subtracted from light curve. Average contribution from off-axis stars is 4.7\% for this light curve.}
 \label{fig:TOI470_contam}
\end{figure*}

We also use diagnostics from our \cheops\ photometry (which has a pixel scale of $\sim1$ arcsec) to validate these planets. For example, we estimate the contributions of background flux in our \cheops\ light curves and analyze the centroid position of the flux both in-transit and out-of-transit. Although we did not use the DRP light curves, DRP diagnostics are still useful. The \cheops\ DRP \citep{hoyer2020expected} uses field star properties derived from Gaia DR2 to simulate the brightnesses of nearby stars in \cheops\ light curves, where magnitudes are converted to \cheops\ magnitudes. Then, based on the relative position of the background stars in or near the chosen aperture and their brightnesses in the \cheops\ band, the DRP estimates the flux contribution from background stars in the light curve and reports this as a percentage for each frame. This contribution is then divided out from the light curve in each \cheops\ frame such that the light curve is de-biased from background sources. Therefore, noting potential sources of background contamination with \cheops's extremely high-precision photometry is useful for characterizing whether the transit signal is on target. We demonstrate an example of this in Fig. \ref{fig:TOI470_contam}.

In addition to light curve photometry, we employ high-resolution imaging, which is part of the standard process for validating transiting exoplanets to assess the possible contamination of bound or unbound companions on the derived planetary radii \citep{ciardi2015}. Close stellar companions (bound or line of sight) can confound exoplanet discoveries in a number of ways.  The detected transit signal might be a false positive due to a background eclipsing binary and even real planet discoveries will yield incorrect stellar and exoplanet parameters if a close companion exists and is unaccounted for \citep{Furlan_2020}. Additionally, the presence of a close companion star may mask the detection of small planets \citep{Lester_2021}. Given that nearly one-half of solar-like stars are in binary or multiple star systems \citep{Matson_2018},  high-resolution imaging provides crucial information toward our understanding of exoplanetary formation, dynamics and evolution \citep{Howell_2021FrASS,Howell_2021AJ}.

\begin{table*}
\caption{Our speckle observations with Gemini of the six TOIs we validate, including target observed, date of observation, distance to star (in pc), whether a luminous companion was detected, the achieved contrast in $\Delta$mag, and the inner and outer visibility limits.}
\begin{tabular}{llllllll}\label{tab:Gemini}
TOI & TIC       & UT Date   & Dist (pc) & Companion? & Contrast ($\Delta$mag) & Inner (AU) & Outer (AU) \\ \hline
198 & 12421862  & 4-Aug-20  & 23.7   & N          & 5.0-8.0            & 0.5        & 28   \\
244 & 118327550  & 4-Aug-20  & 22     & N          & 5-6.5              & 0.44       & 26   \\
262 & 70513361  & 15-Oct-19  & 43.9   & N         & 5-6.6               & 0.88      & 52    \\
444 & 179034327 & 9-Jan-20  & 57.4   & N          & 5.0-8.0            & 1.15       & 69   \\
470 & 37770169 & 24-Feb-21 & 130.5  & N          & 5-7.5              & 2.6        & 157  \\
518 & 264979636 & 9-Feb-21  & 159.8  & N          & 5-6.8              & 3.2        & 192  \\   
\hline
\end{tabular}
\end{table*}

Six TOIs were observed using the ‘Alopeke and Zorro speckle instruments on the Gemini North/South 8-m telescopes \citep{Scott_2021,Howell_2022FrASS}.  Both speckle instruments provide simultaneous speckle imaging in two bands (562 nm and 832 nm) with output data products that include reconstructed images and robust contrast limits on companion detections. A number of different sets of speckle observations were obtained for each star and processed in our standard reduction pipeline (see \citealt{Howell_2011}). Fig. \ref{fig:contrast_curves} shows our final $5\sigma$ contrast curves and the 832 nm reconstructed speckle image, and the details of our observations are shown in Table \ref{tab:Gemini}. These contrast curves are also used by \texttt{TRICERATOPS} to aid in our statistical vetting. 

Using \tess\ photometry and these high-resolution contrast curves, \texttt{TRICERATOPS} reports a False Positive Probability (FPP) and Nearby FPP (NFPP) for each target. FPP represents the aggregate probability that the observed transit is due to something other than a transiting planet around the target star, and NFPP is the same except that it suggests the origin of the signal is a nearby known TICv8 star. In order for the planet candidate to be considered statistically validated, we require FPP $< 0.015$ and NFPP $< 10^{-3}$, as recommended by \citet{Giacalone2021TRICERATOPS_paper}. In order to account for intrinsic scatter in the statistical calculation, we ran 20 trials of calculating the FPP and NFPP, and report the mean and standard deviation of these values for each of our validated systems. 20 trials allows us to explore the possibility that our result is not sensitive to intrinsic scatter in the calculation.

Further, we employ reconnaissance spectroscopic observations of our validated systems. We employed spectroscopic observations from both FLWO-TRES and SMARTS-CHIRON, whose observations of specific targets are delineated in the following subsections. We use spectroscopic observations in this paper for the purpose of ruling out a stellar-mass companion, with the exception of TOI 198 b, for which we have high-precision radial velocities from VLT-ESPRESSO and are thus able to place constraints on the planet mass.

\begin{table}
\caption{SMARTS-CHIRON RVs for the TOIs we validate. The implications of these RVs are delineated in the respective subsection for each validated planet.}
\centering
\begin{tabular*}{130 pt}{cll}\label{tab:CHIRON_RVs}

BJD\_UTC                          & \multicolumn{1}{c}{vrad}   & \multicolumn{1}{c}{svrad} \\ \hline
\multicolumn{3}{c}{TOI 198}                                                                \\ \hline
2458650.88886                     & \multicolumn{1}{c}{20.551} & \multicolumn{1}{c}{0.085} \\
2458653.90559                     & \multicolumn{1}{c}{20.477} & \multicolumn{1}{c}{0.165} \\
2459200.60414                     & \multicolumn{1}{c}{20.387} & \multicolumn{1}{c}{0.068} \\ \hline
\multicolumn{3}{c}{TOI 244}                                                                \\ \hline
2458824.62649                     & \multicolumn{1}{c}{15.254} & \multicolumn{1}{c}{0.219} \\
2459209.62434                     & \multicolumn{1}{c}{15.205} & \multicolumn{1}{c}{0.179} \\ \hline
\multicolumn{3}{c}{TOI 262}                                                                \\ \hline
2458527.55528                     & \multicolumn{1}{c}{33.222} & \multicolumn{1}{c}{0.025} \\
2458666.91538                     & \multicolumn{1}{c}{33.172} & \multicolumn{1}{c}{0.035} \\
\multicolumn{1}{l}{2459206.56806} & 33.234                     & 0.023                     \\
\multicolumn{1}{l}{2459454.80242} & 33.227                     & 0.024                     \\ \hline
\multicolumn{3}{c}{TOI 444}                                                                \\ \hline
\multicolumn{1}{l}{2458531.59407} & 1.106                      & 0.029                     \\
\multicolumn{1}{l}{2459218.64491} & 1.130                      & 0.027                     \\
\multicolumn{1}{l}{2459454.85929} & 1.095                      & 0.024                     \\ \hline
\multicolumn{3}{c}{TOI 470}                                                                \\ \hline
\multicolumn{1}{l}{2458545.63013} & 30.296                     & 0.044                     \\
\multicolumn{1}{l}{2459349.44464} & 30.390                     & 0.022                     \\ \hline
\multicolumn{3}{c}{TOI 518}                                                                \\ \hline
\multicolumn{1}{l}{2458570.58672} & 45.460                     & 0.039                     \\
\multicolumn{1}{l}{2458626.45015} & 45.507                     & 0.029                     \\ \hline
\end{tabular*}
\end{table}

In addition to the high resolution imaging, we have utilized Gaia to identify any wide stellar companions that may be bound members of the system.  Typically, these stars are already in the \tess\ Input Catalog and their flux dilution to the transit has already been accounted for in the transit fits and associated derived parameters. We searched for possible widely separated companions based upon similar parallaxes and proper motions \citep{mugrauer2020,mugrauer2021}. Additionally, the Gaia DR3 astrometry provides insight on the possibility of inner companions that may have gone undetected by either Gaia or the high resolution imaging. The Gaia Renormalised Unit Weight Error (RUWE) is a metric, similar to a reduced chi-square, where values that are $\lesssim 1.4$  indicate that the Gaia astrometric solution is consistent with the star being single whereas RUWE values $\gtrsim 1.4$ may indicate an astrometric excess noise, possibily caused the presence of an unseen companion \citep{Zeigler_2020}.

\subsection{Validation of TOI 198 b}\label{subsec:198_val}

\begin{figure}
\centering
 \includegraphics[width=\linewidth]{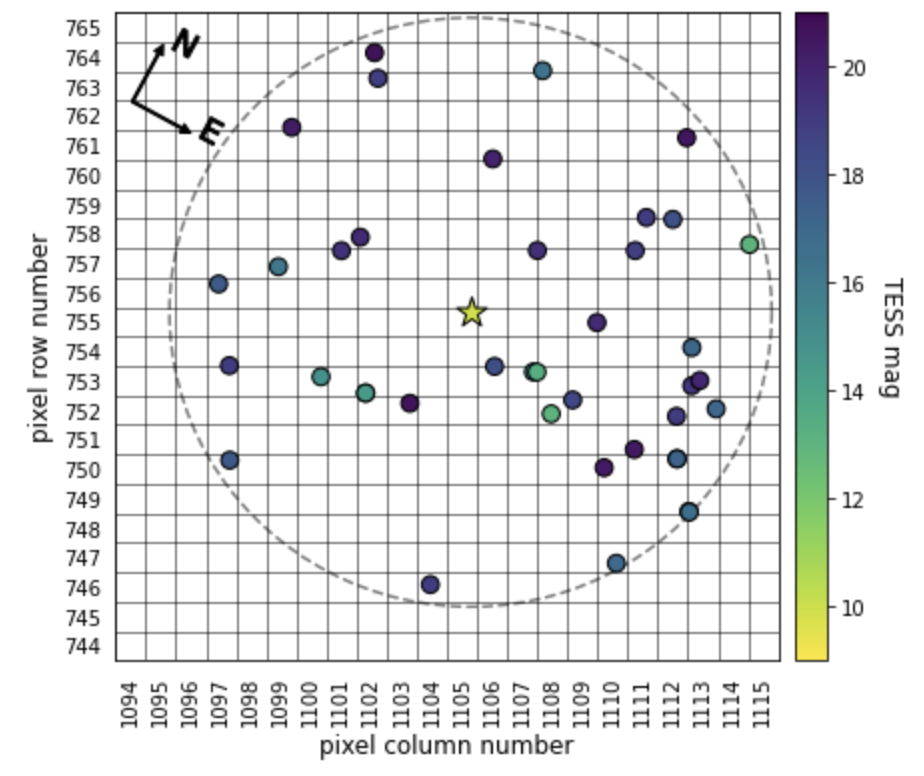}
 \includegraphics[width=\linewidth]{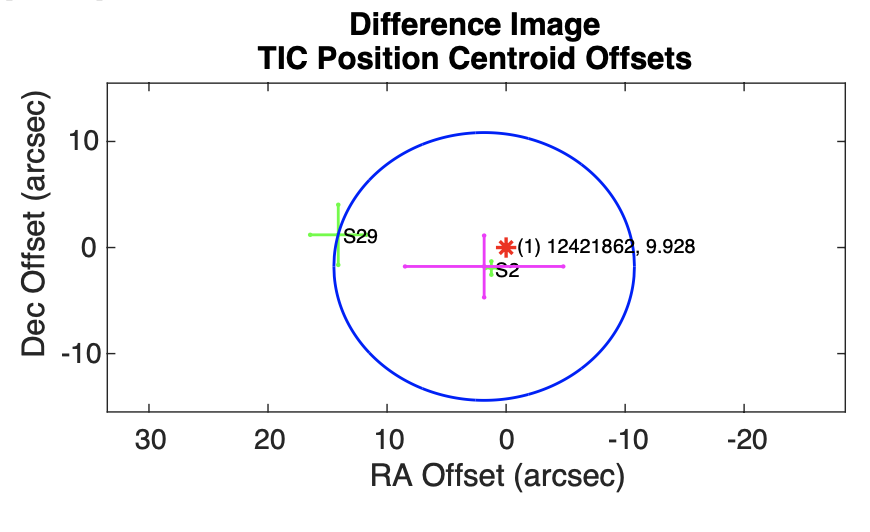}
 \caption{Top: Field stars near TOI 198 (TIC 12421862; indicated as a yellow star) with \tess\ pixels overlaid. TOI 198 resides in a relatively uncrowded portion of the sky, and there are no stars brighter than $\Delta$mag $\approx 3.3$ in the \tess\ band closer than 60 arcseconds, or approximately 3 \tess\ pixels away.  Bottom: Centroid difference image between in-transit flux and out-of-transit flux from \tess\ observations of TOI 198.}
 \label{fig:TOI198_S2}
\end{figure}

We analyzed two \tess\ sectors' worth of data for TOI 198, including one in the PM and one in EM1. Two transit events were initially identified by the SPOC pipeline \citep{Jenkins_2002,Jenkins_2010,2020TPSkdph} at Barycenter-corrected \tess\ Julian Date (BTJD = BJD - 2457000) BTJD = 1356.3754 and 1376.8027 as a potential planet candidate with a 20.427d period in October of 2018. Three transit events were identified by QLP, where the middle transit fell between the two previously-identified signals at a time of BTJD = 1366.574 for a period of 10.218d. The SPOC light curve was later reprocessed, during which the third transit event was identified at the same time as the QLP signal, but the significance of this signal was dubious. During the EM1 sector (S29), only one transit signal was identified, but  due to the timing of the observation, it remained ambiguous whether the period of the signal was truly 10.2d or 20.4d. Further, our \cheops\ observation was scheduled at a time which also did not break the period alias. However, our LCOGT 1\,m observation on UT 2022 September 14 detected the transit-like event on-target relative to known Gaia DR3 stars and confirmed the 10.2\,d alias as the true orbital period. 

We further validated the transit signal by checking the centroid position for each \tess\ sector to assure that the signal was on target. Shown in the bottom of Fig. \ref{fig:TOI198_S2}, the centroid of the difference image agrees very well with the expected position of the target star, meaning the transit signal is indeed on target. We saw no discrepancies between the even and odd-numbered transits, consistent with less than $1\sigma$ depth difference.

We obtained high-resolution imaging observations of TOI 198 with the Gemini-'Alopeke imager on October 4th, 2020, which is shown in the top left panel Fig. \ref{fig:contrast_curves}. The image has a pixel scale of 0.01 arcseconds, with an estimated PSF size of 0.02 arcseconds. We find that this star is single at least out to 1.2 arcseconds, with no companion brighter than 5-8 magnitudes below that of the target star beyond 0.1 arcseconds.

We also obtained three epochs of reconnaissance spectra for TOI 198 with SMARTS-CHIRON. While TOI 198's low temperature (see Table \ref{tab:SED}) renders it unreliable for spectroscopic classification with this telescope, we were able to deduce from these RVs that the lines are narrow, which is indicative of a single, slowly rotating star. Additionally, we see no large RV variation between our epochs, as shown in Table \ref{tab:CHIRON_RVs}.

\begin{table}
\caption{ESPRESSO Radial Velocities for TOI 198}\label{tab:TOI198_RVs}
\begin{tabular}{c|c|c}

\hline
\hline
Time & RV & RV Uncert. \\
(JD) & (m s$^{-1}$) & (m s$^{-1}$) \\
\hline
2458668.84736 & 20278.25 & 0.41  \\
2458669.91176 & 20277.53 & 0.39  \\
2458677.81612 & 20271.54 & 0.36  \\
2458679.87403 & 20275.20 & 0.35  \\
2458687.79316 & 20271.76 & 0.46  \\
2458697.72404 & 20274.94 & 0.56  \\
2458698.80579 & 20275.41 & 0.35  \\
2458699.78268 & 20276.40 & 0.37  \\
2458699.86914 & 20275.54 & 0.35  \\
2458707.85215 & 20272.98 & 0.55  \\
2458708.90561 & 20273.91 & 0.34  \\
2458716.85516 & 20272.36 & 0.30  \\
2458717.74245 & 20273.56 & 0.42  \\
2458717.85335 & 20273.73 & 0.33  \\
2458727.69072 & 20273.65 & 0.58  \\
2458727.89349 & 20274.08 & 0.42  \\
2458728.83222 & 20274.34 & 0.43  \\
2458729.78372 & 20275.43 & 0.56  \\
2458729.90453 & 20274.16 & 0.73  \\
2458738.82735 & 20272.77 & 0.42  \\
2458738.84339 & 20273.81 & 0.39  \\
2458738.86206 & 20272.50 & 0.42  \\
2458738.87890 & 20273.24 & 0.41  \\
\hline
\end{tabular}
\end{table}

\begin{figure*}
    \centering
    \includegraphics[width=.58\linewidth]{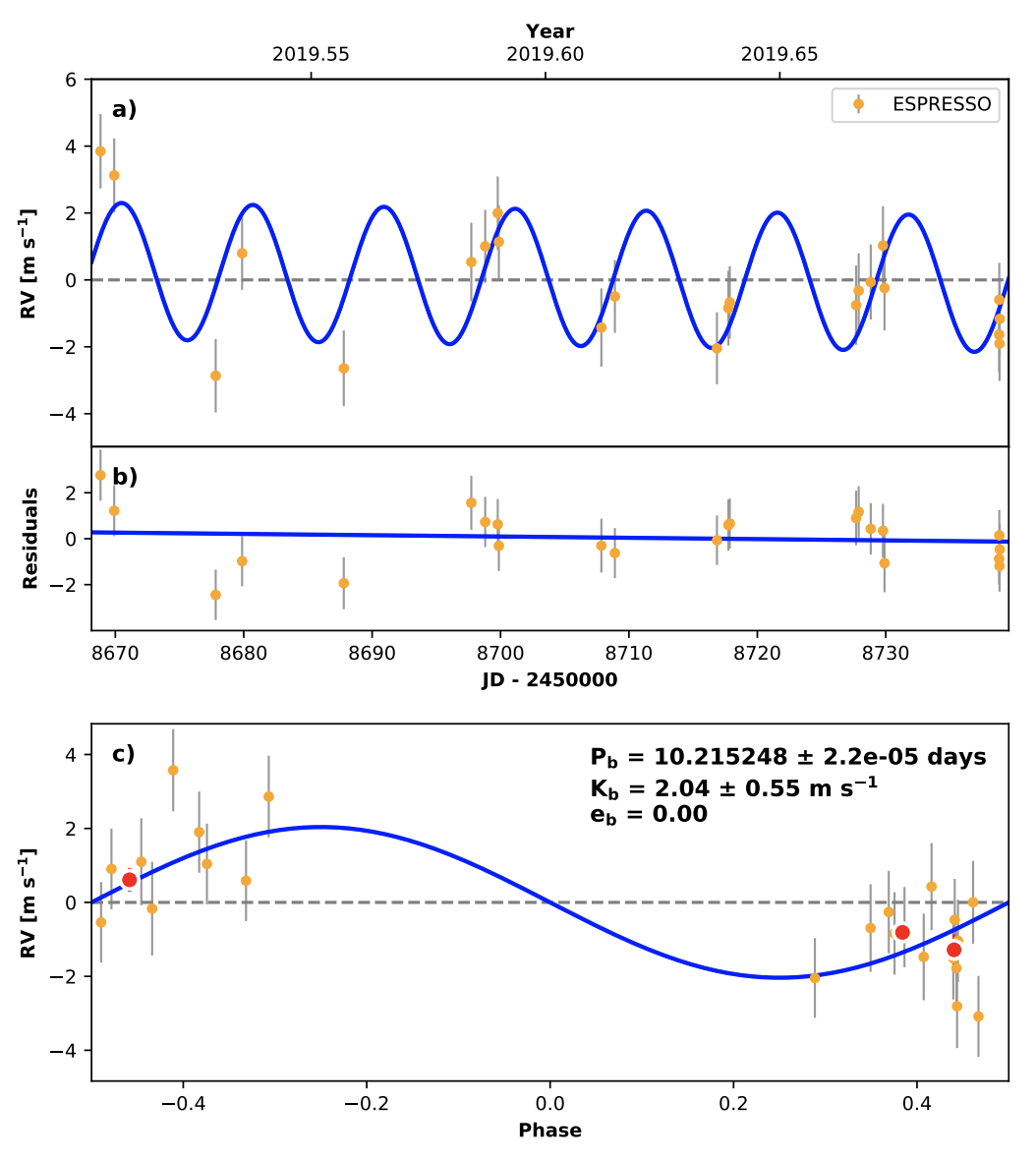}
    \includegraphics[width=.38\linewidth]{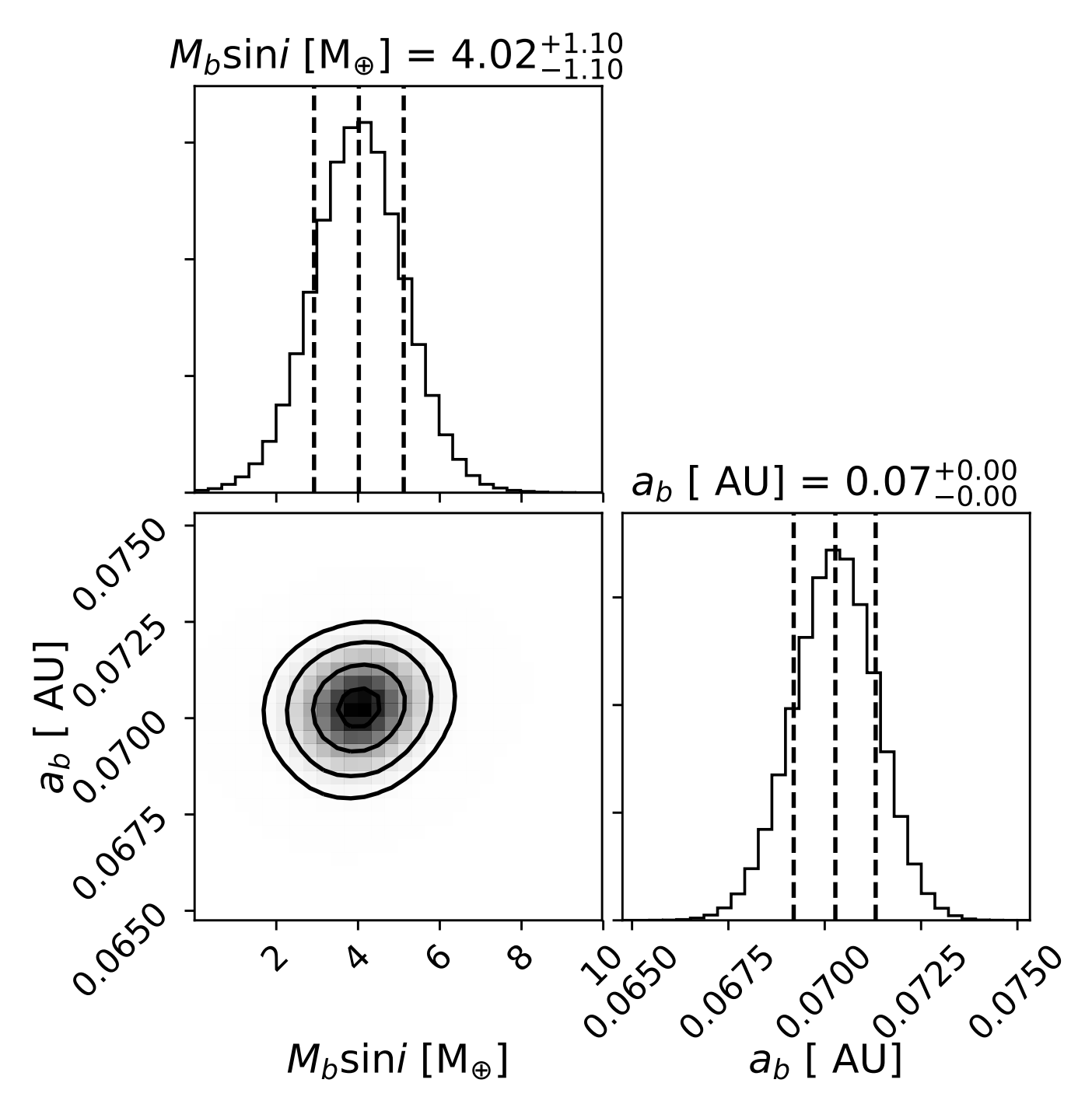}
    \caption{Left: a. Best-fit 1-planet Keplerian orbital model for TOI-198. The maximum likelihood model is plotted. The thin blue line is the best fit 1-planet model. We add in quadrature the RV jitter term(s) listed in Table 2 with the measurement uncertainties for all RVs. b) Residuals to the best fit 1-planet model. c) RVs phase-folded to the ephemeris of planet b. The small point colors and symbols are the same as in panel a.
    Right: Corner plot showing posterior distribution for TOI 198 b mass.}
    \label{fig:TOI198_espresso}
\end{figure*}

Further, we obtained 23 high-precision radial velocity epochs of TOI 198 with the Very Large Telescope (VLT) Echelle SPectrograph for Rocky Exoplanets and Stable Spectroscopic Observations (ESPRESSO; \citealt{Pepe_ESPRESSO2021}) instrument between July 4th, 2019 and September 12th, 2019, shown in Fig. \ref{fig:TOI198_espresso}. With these observations, we were able to not only rule out the presence of a stellar-mass companion to TOI 198, but we are able to constrain the mass of TOI 198 b. The radial velocity time-series were modeled with a Keplerian model using \texttt{Radvel}\footnote{\url{https://radvel.readthedocs.io/en/latest/}} \citep{Fulton2018RadVel}. Our independent analysis of transit data allowed us to use informative priors on the orbital period and time of central transit. We fixed a circular orbit (e = 0) and added a radial velocity jitter term. \texttt{Radvel} performs a Markov Chain Monte Carlo (MCMC) technique to obtain the credible intervals of the parameters. We obtained a semi-amplitude of $K_b=2.04^{+0.55}_{-0.56}$ [m s$^{-1}$], equivalent to a planetary minimum mass of  $4.0\pm1.1$ M$_{\earth}$.

Finally, our assessment of this star with Gaia showed that based upon similar parallaxes and proper motions, there are no additional widely separated companions identified by Gaia. TOI 198 has a Gaia EDR3 RUWE value of 1.09, indicating that the astrometric fits are consistent with the single star model.

Our statistical vetting with \texttt{TRICERATOPS} supports the conclusion that the transit signals are from a transiting planet which orbits the target star, with $FPP = 0.0142\pm0.0016$ and $NFPP = 0.00023\pm0.00005$.

Given the information at hand, we are able to conclude that TOI 198 is a star which has no massive companions, and that the transit signal we see is likely due to a transiting planet. Thus, we consider TOI 198 b to be validated.

\subsection{Validation of TOI 244 b}

\begin{figure}
\centering
 \includegraphics[width=\linewidth]{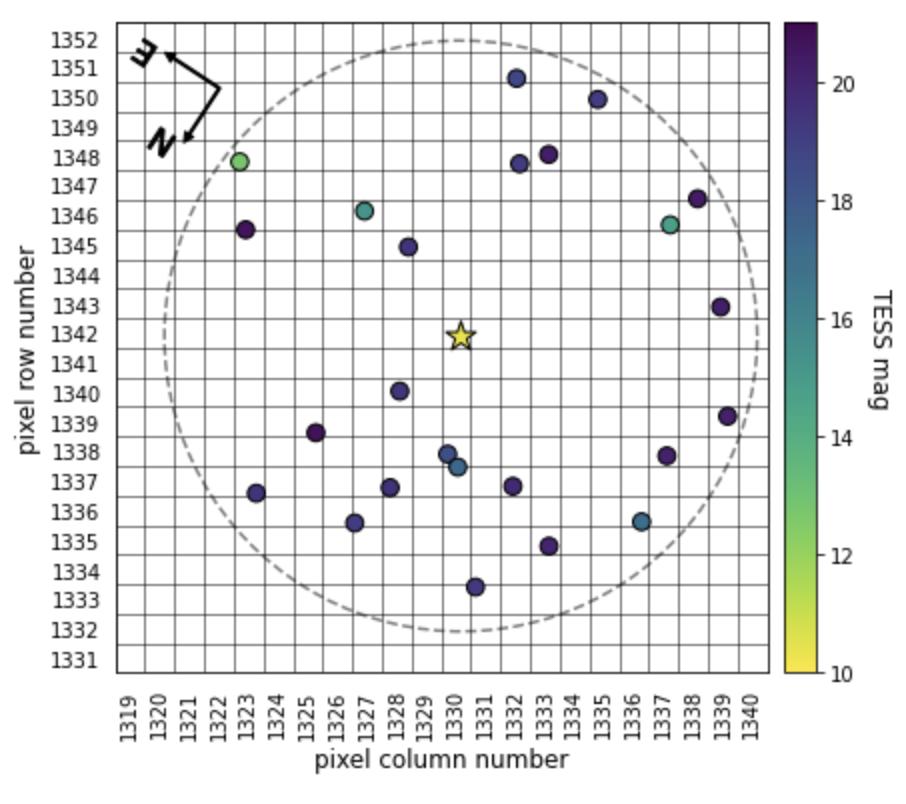}
 \includegraphics[width=\linewidth]{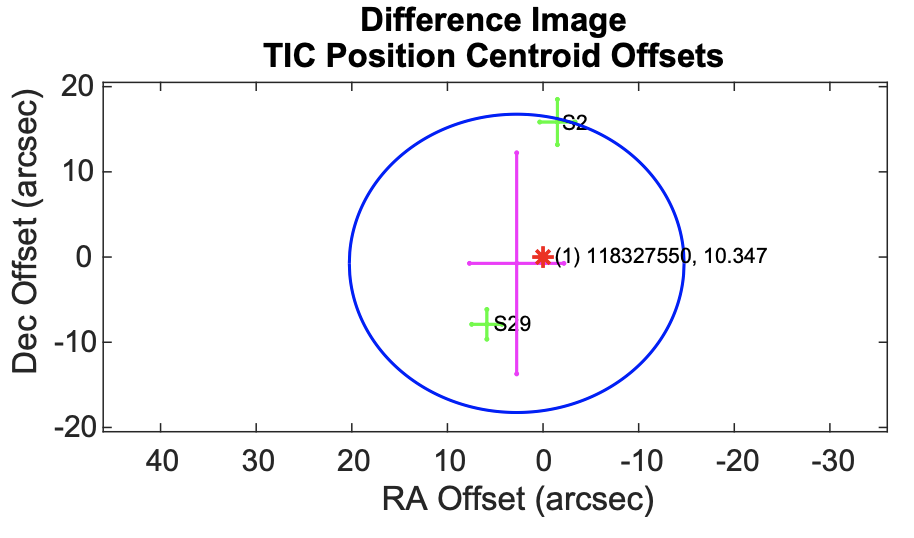}
 \caption{Top: Field stars near TOI 244 (TIC 118327550; indicated as a yellow star) with \tess\ pixels overlaid. TOI 244 resides in a relatively uncrowded portion of the sky, and there are no stars brighter than $\Delta$mag $\approx 5.0$ in the \tess\ band closer than 110 arcseconds, or approximately 5 \tess\ pixels away.  Bottom: Centroid difference image between in-transit flux and out-of-transit flux from \tess\ observations of TOI 244.}
 \label{fig:TOI244_S29}
\end{figure}

We analyzed two \tess\ sectors' worth of data for TOI 244, including one in the PM and one in EM1. Four transit events were identified in each of the sectors by the SPOC pipeline at BTJD = 1357.3647 with a period of 7.39719 d. As shown by the top panel of Fig. \ref{fig:TOI244_S29}, the field around TOI 244 is relatively uncrowded, and there are no stars brighter than $\Delta$mag $\approx 7.5$ in the \tess\ band closer than 110 arcseconds, or approximately 5 \tess\ pixels away. This would indicate that there are no meaningful contributions from nearby stars in the aperture for this target. Our two visits to TOI 244 with \cheops\ also exhibit extremely low levels of flux contributions from nearby stars. The DRP simulated images of nearby stars in both of our \cheops\ visits, which occurred in October of 2021. The light curves of simulated stars near TOI 244 exhibit average contributions of $6.89\times 10^{-4}$\% and $6.74\times10^{-4}$\% for our first and second visits, respectively.

Further, we confirm that the transit signals we see are on the target star. Our centroid difference image of in-transit and out-of-transit flux (bottom panel Fig. \ref{fig:TOI244_S29}) shows that while there is some offset in the position of the centroid position of in-transit flux during sector 2 from its expected position, the centroid positions is overall consistent with being in the expected position. The centroid position of in-transit flux does not appear to move strictly towards any other star. Further, for either of our \cheops\ observations, there is no centroid offset from its central position larger than 2 pix, which means the transit signal is indeed on target. Using \texttt{TRICERATOPS}, we find an NFPP consistent with zero.

TOI 244 was observed by LCOGT on UT 2019 September 30 using a 5.8'' target aperture that excludes flux from the nearest known Gaia DR3 stars and detected a 1 ppt transit-like event on-target. 

We obtained high-resolution speckle images of this star with the 8.0m-Gemini telescope equipped with the 'Alopeke instrument at central wavelengths of 832 nm and 562 nm. Our images had a pixel scale of 0.01 arcsec/pixel, with an estimated PSF of 0.02 arcsec. We obtained an estimated contrast of $\Delta$mag = 5.98 at 0.5" in the 832 nm band, as shown in  Fig. \ref{fig:contrast_curves} (top right). A $\sim1000$ ppm event could be caused by a star as dim as $\Delta$mag $\sim7.5$ Given that the only two stars brighter than $\Delta$mag 7.5 are at a distance from the target so as not to cause significant contamination, we believe they only margninally contaminate the \tess\ aperture. This indicates that TOI 244 is a single star.

Similar to TOI 198, we were unable to reliably use our followup reconnaissance spectra of TOI 244 with SMARTS-CHIRON to classify the star. While our phase coverage was poor, narrow spectroscopic lines indicate that this star is single and slowly rotating. Our two epochs are unable to independently rule out a stellar mass companion at the ephemeris of the transiting candidate, but our high-contrast speckle imaging showed no indication of a luminous companion. Therefore, between these two pieces of evidence, we are able to rule out a stellar companion. Further, given that our RVs span nearly 400 days, we are able to rule out any massive companion at a wider orbit.

Finally, our assessment of this star with Gaia showed that based upon similar parallaxes and proper motions, there are no additional widely separated companions identified by Gaia. TOI 244 has a Gaia EDR3 RUWE value of 1.25, indicating that the astrometric fits are consistent with the single star model.

Our statistical vetting with \texttt{TRICERATOPS} supports the conclusion that the transit signals we see are due to a planet orbiting the target star, as we report $FPP = (1.203\pm0.814)\times 10^{-5}$. For the above reasons, we consider the planetary nature of the transit signals around TOI 244 to be validated.

\subsection{Validation of TOI 262~b}

\begin{table}
 \caption{Radial velocities collected with FLWO-TRES for TOI 262.}
 \label{tab:TOI262_tres}
\begin{tabular}{llll}
\hline
Time           & RV         & RV Uncert. & SNRe  \\
BJD            & m s$^{-1}$ & m s$^{-1}$ &        \\ \hline
2458473.6662 & -75.45     & 12.95      & 44.0      \\
2459502.8796 & -95.30     & 12.95      & 49.3   \\ \hline
\end{tabular}
\end{table}

\begin{figure}
\centering
 \includegraphics[width=\linewidth]{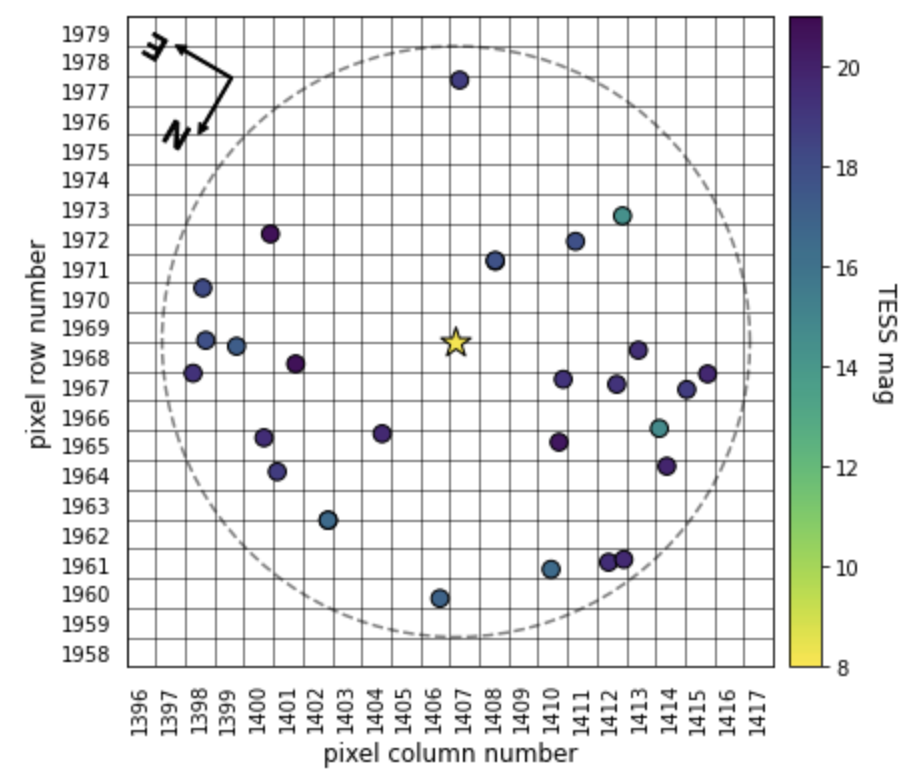}
 \includegraphics[width=\linewidth]{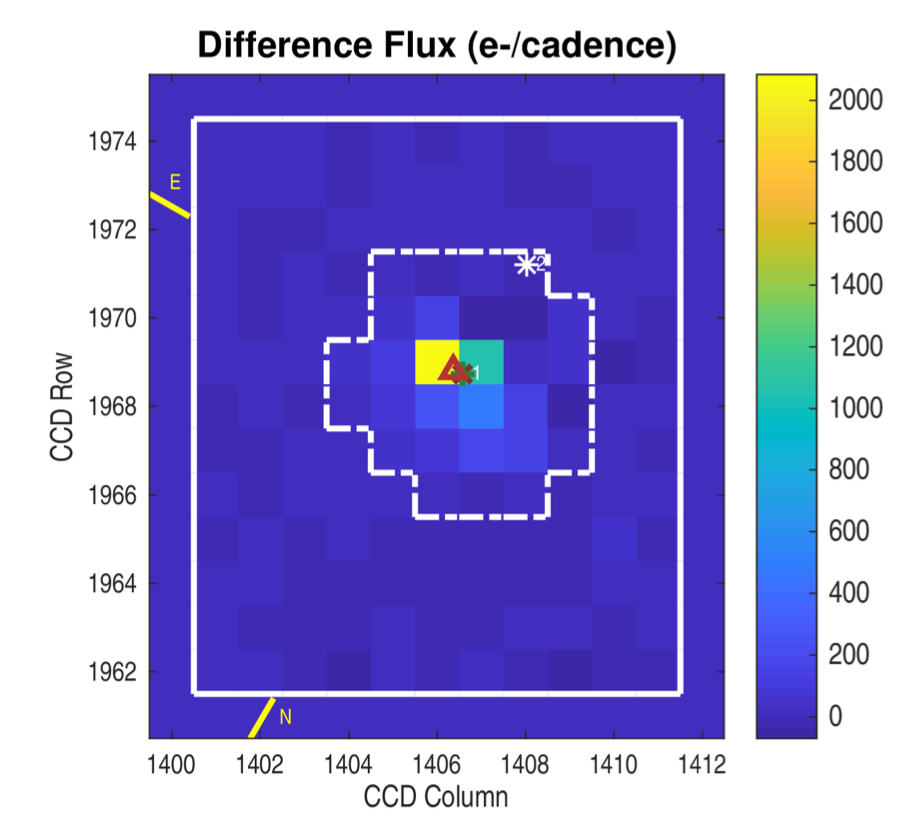}
 \caption{Top: Field stars near TOI 262 (TIC 70513361; indicated as a star) with \tess\ pixels overlaid. The closest star to TOI 262, TIC 70513359, has a \tess\ magnitude of 16.2, and does not contribute significantly in the \tess\ bandpass. Bottom: Difference image between in-transit flux and out-of-transit flux from \tess\ sector 2 observation of TOI 262. The orange triangle shows the centroid in the difference image, which is in good agreement with the expected position of the target star, indicating that transit signal is indeed on target.}
  \label{fig:TOI262_S3}
\end{figure}

TOI 262 was observed in sector 3 of the \tess\ PM and sector 30 of the EM. Fig. \ref{fig:TOI262_S3} shows the field around TOI 262 (top) and the SPOC difference image, which is the difference between in-transit and out-of-transit flux (bottom). The top panel of Fig. \ref{fig:TOI262_S3} shows that the field around TOI 262 is not crowded, and that there are no stars within 63 arcsec, or about 3 \tess\ pixels. The nearest star, TIC 70513359, has a T mag of 16.2, which is about 8 mags dimmer than TOI 262 and represents a negligible flux contribution to the \tess\ light curve. This is further evidenced by our difference image, which shows good agreement between the expected centroid position and the actual photometric centroid position during transit. It is notable that while there is significant flux difference in multiple pixels, this is likely a result of the star's position between two pixels, leading to flux contributions in both. Further, the lack of differences in depth between even and odd transits supports the conclusion that the transit events we see are occultations from the a non-luminous object and not primary and secondary eclipses from a grazing EB.

We verified the lack of stellar or massive companions to TOI 262 with reconnaissance spectroscopy and high-contrast speckle imaging. We obtained spectra from both the FLWO-TRES and SMARTS-CHIRON instruments, which both indicated that this star is a slow rotator and single-lined, both of which are indicators that the star does not have a massive companion. Further, our RVs calculated with these instruments indicate no significant velocity variation. We obtained speckle imaging with the Gemini-'Alopeke instrument, which imaged the star with a contrast to $\Delta$mag $\sim5$ at a separation of 0.02 arcseconds in the 832 nm band, corresponding to a physical distance of 0.88 AU from the star. This contrast would indicate there is no nearby luminous companion. Our statistical vetting with \texttt{TRICERATOPS} in 20 trials returned $FPP = (2.17\pm0.93)\times10^{-4}$ and $NFPP$ consistent with zero, which indicate that there is very little probability that the transit signal is caused by something other than a transiting  planet orbiting the target star.

\begin{figure*}
\centering
 \includegraphics[width=.8\linewidth]{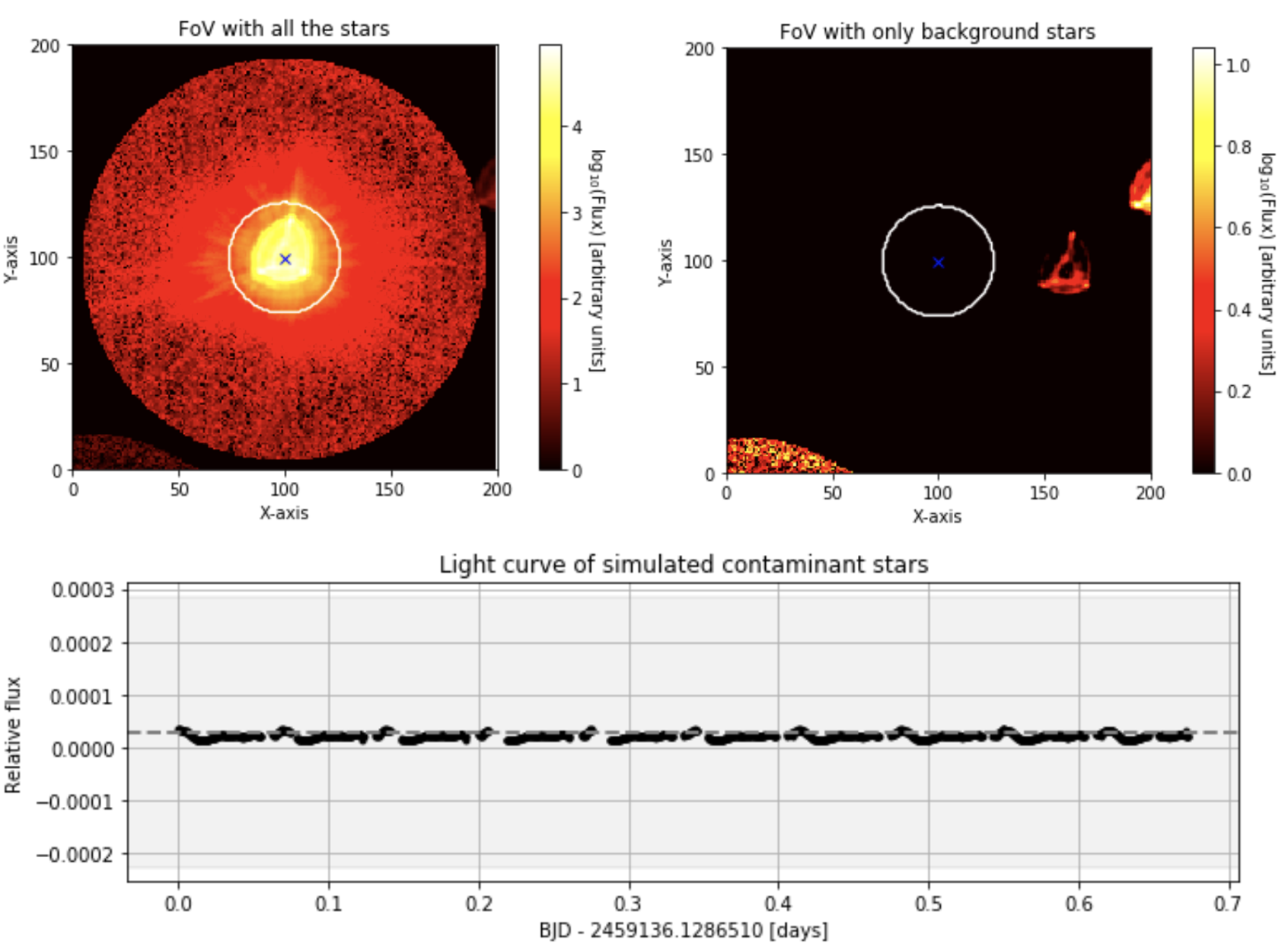}
 \caption{Diagnostics of contamination from nearby stars in the \cheops\ light curve for TOI 262. Top left: \cheops\ field of view, including both our target star (centered on black 'X') and field stars. Top right: DRP simulated image of only field stars without target star, exhibiting relative fluxes of non-target stars. Note difference in color scales between field stars with and without target star. Bottom: Simulated contributions to \cheops\ light curve from field stars, which is subtracted from light curve. Average contribution from off-axis stars is nearly consistent with zero for this light curve.}
 \label{fig:TOI262_contam}
\end{figure*}

Fig. \ref{fig:TOI262_contam} shows the simulated \cheops\ DRP images including background stars based on Gaia DR2 star maps, as well as the estimated contamination from nearby stars as a function of time in the \cheops\ DRP light curve. Although we did not use the DRP light curve, this may still give us a good estimation of contamination from nearby stars as a further check. We see very little evidence of contamination in this light curve, as the contamination from nearby stars in this light curve is consistent with zero. 

Finally, our assessment of this star with Gaia showed that based upon similar parallaxes and proper motions, there are no widely separated companions identified by Gaia. TOI 262 has a Gaia DR3 RUWE of 0.91, indicating that the astrometric fits are consistent with the single star model. 

Given all of the information at hand, we consider TOI 262 b to be validated. 

\subsection{Validation of TOI 444 b}

\begin{table}
 \caption{Radial velocities collected with FLWO-TRES for TOI 444.}
 \label{tab:TOI444_tres}
\begin{tabular}{lllll}
\hline
Time           & RV         & RV Uncert. & SNRe  \\
BJD            & m s$^{-1}$ & m s$^{-1}$ &      \\ \hline
2458516.685280 & -34.50     & 33.86      & 33.8     \\
2458528.614782 & -66.17     & 33.86      & 26.0     \\ \hline
\end{tabular}
\end{table}

\begin{figure}
\centering
 \includegraphics[width=\linewidth]{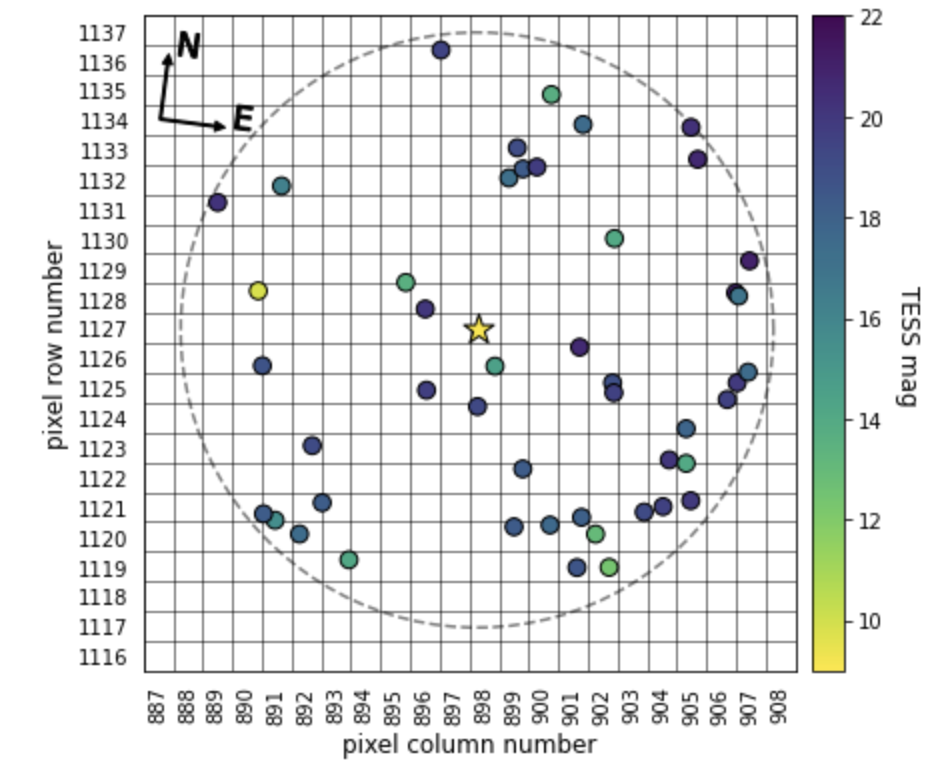}
 \includegraphics[width=\linewidth]{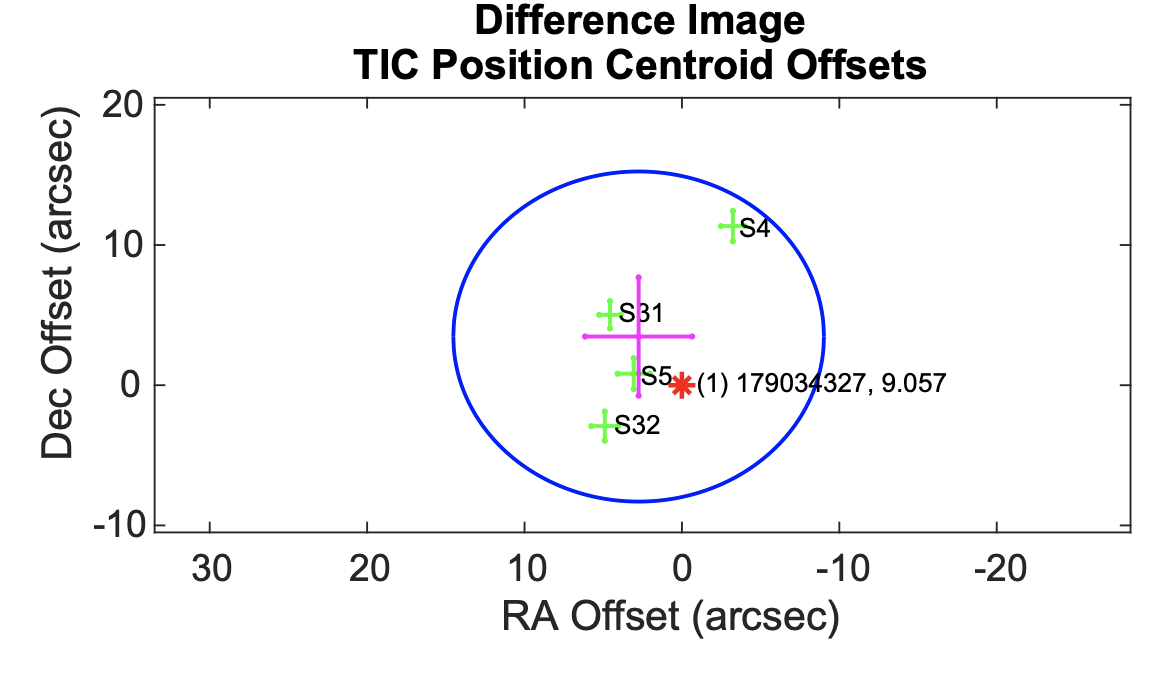}
 \caption{Top: Field stars near TOI 444 (TIC 179034327; indicated as a star) with \tess\ pixels overlaid. The closest star to TOI 444, TIC 179034325, has a \tess\ magnitude of 14.4, and does not contribute significantly in the \tess\ bandpass. Bottom: Centroid difference image between in-transit flux and out-of-transit flux from \tess\ observations of TOI 444.}
 \label{fig:TOI444_S5}
\end{figure}

We analyzed four \tess\ sectors' worth of data for TOI 444, including two each in the PM and EM1. We examined difference images from all four sectors (bottom panel of Fig. \ref{fig:TOI444_S5}), where the centroid position in the difference image is shown by a green cross. Although the sector 4 centroid position approaches the $3\sigma$ circle, we see no significant evidence of centroid offset for this target, indicating that the transit signal is on target. Another potential indicator of a false positive is a difference in the parameters of even-numbered transits in the light curve and odd-numbered transits, which may indicate that they are primary and secondary eclipses of an EB. However, the SPOC report for this TOI showed no statistically significant difference between even and odd transits. 

 The \cheops\ DRP estimated that contamination from background stars accounted for only 0.1\% of the flux in the OPTIMAL aperture, meaning that the signal is very likely uncontaminated. We also checked the centroid location in each \cheops\ image to ensure that the in-transit and out-of-transit flux were on target. The centroid shifted at most 1.0 pixels from the mean in both the X and the Y directions, indicating that the \cheops\ light curve and transit were on target.

 As a further check the transit signal was on target, TOI 444 was observed by LCOGT on UT 2020 October 31 using a follow-up aperture that excludes all flux from the nearest Gaia DR3 neighbors and detected a 1.4 ppt transit-like event on-target.

We performed reconnaissance spectroscopy to determine whether the star had an unresolved massive companion. A radial velocity on the order of km s$^{-1}$ would indicate the presence of such a companion, but observations of this star with both FLWO-TRES and the CHIRON spectrometer at SMARTS showed that it is a single G-dwarf. Further, the star  observed by multiple high-resolution imagers, including the Zorro and 'Alopeke imagers at Gemini-North, HRCam at the Southern Astrophysical Research Telescope in Chile, and NaCo at the Very Large Telescope in Chile. For simplicity, we show only the Gemini contrast curve, but all images of this star indicated that it had no close companion at $\Delta$mag $<5$ at 0.1 arcsec of separation. 
 
 Finally, our assessment of this star with Gaia showed that based upon similar parallaxes and proper motions, there are no additional widely separated companions identified by Gaia. TOI 444 has a Gaia EDR3 RUWE value of 1.13, indicating that the astrometric fits are consistent with the single star model.
 
We applied the \texttt{TRICERATOPS} statistical vetting. \texttt{TRICERATOPS} returned $FPP = (3.26 \pm 6.68) \times10^{-6}$ and $NFPP = (1.44 \pm 0.95) \times 10^{-16}$ for TOI 444 b, and a visual inspection of the stellar field showed no bright stars in or near the \tess\ aperture for this light curve. 

\subsection{Validation of TOI 470 b}

\begin{figure}
\centering
 \includegraphics[width=\linewidth]{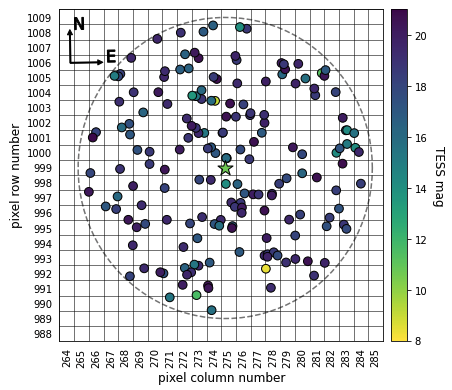}
 \includegraphics[width=\linewidth]{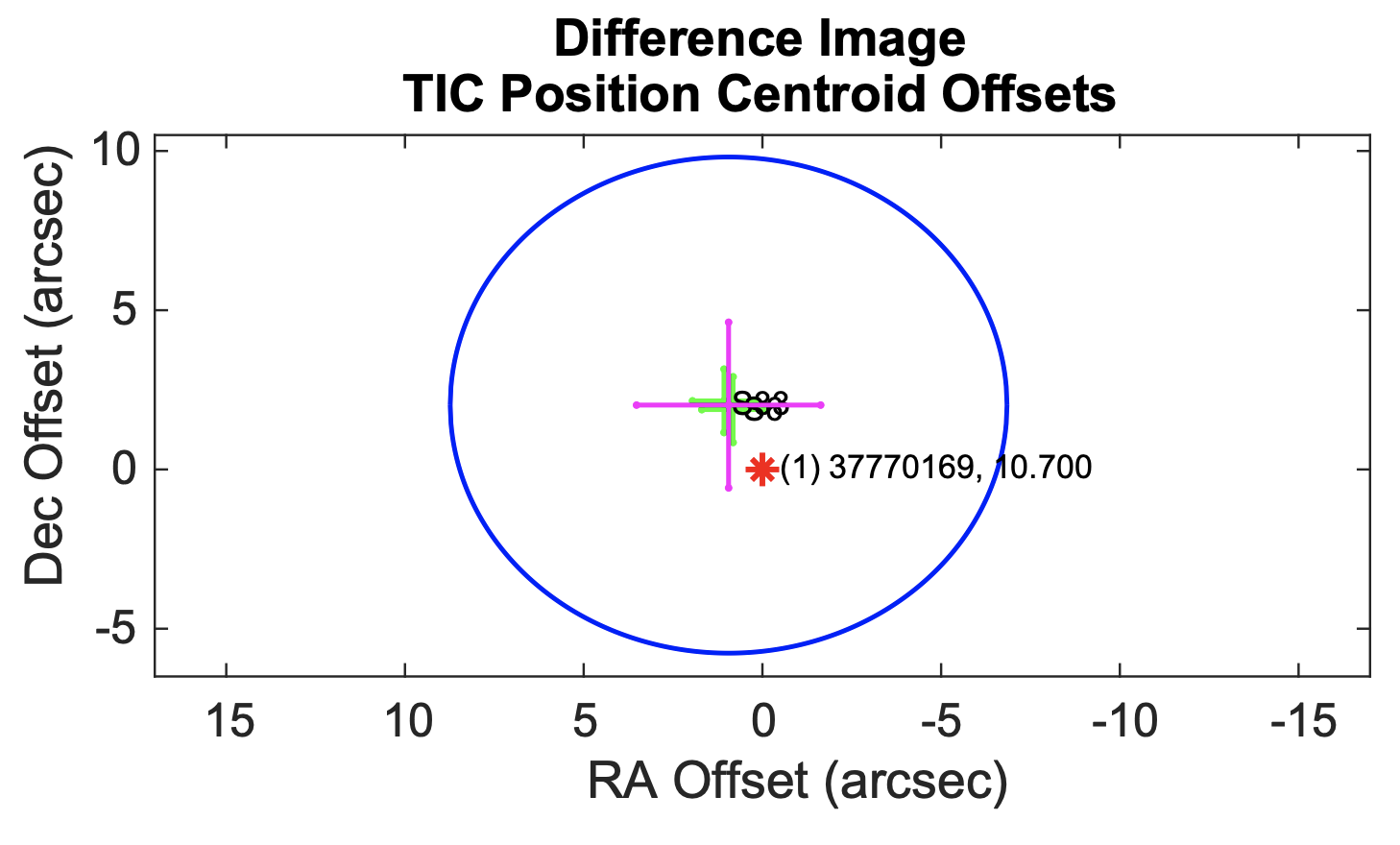}
 \caption{Top: Field stars near TOI 470 (TIC 37770169; indicated as a star) with \tess\ pixels overlaid, and \tess\ magnitude is indicated by the color bar. Although the field appears crowded, there are no stars brighter than 14th magnitude closer than 80 arcseconds, which is several \tess\ pixels away. The field star of note is to the north and west of TOI 470, which is also visible in the difference image. This star, TIC 37770142 (T mag = 9.05) is nearly 100 arcseconds away from TOI 470 and does not contribute significant flux in the aperture for our star. The other bright star in the field, TIC 37794435 (T mag = 8.37), is even farther away at a separation of nearly 150 arcseconds, and does not contribute any flux to the light curve.  Bottom: Centroid difference image between in-transit flux and out-of-transit flux from \tess\ observations of TOI 470.}
 \label{fig:TOI470_S6}
\end{figure}

TOI 470 was observed in sector 6 of the \tess\ PM and sector 33 of the EM1. Fig. \ref{fig:TOI470_S6} shows the field around TOI 470 (top) and the SPOC centroid image (bottom). The top panel is color-coded by \tess\ magnitude, and TOI 470 is shown as a star. Although the field appears to be crowded, there are no stars brighter than 14th magnitude within 80 arcsecsonds of the target star, which contribute little to the light curve. Additionally, there are two bright stars relatively nearby, TIC 37770142 to the north-north-west and TIC 37794435 (T mag = 9.05 and 8.37, respectively), these stars are too far from TOI 470 to contribute significantly to the light curve. This is evident from the centroid image, which is a data product of the SPOC pipeline. The centroid position is consistent in the difference image, and the target is within the $1\sigma$ area, suggesting the transit signal is on target. Further, we saw no statistically significant difference between the depths of even and odd-numbered transits, indicating that these are likely transit events and not primary and secondary eclipses of an EB. 

Fig. \ref{fig:TOI470_contam} shows the simulated \cheops\ DRP images including background stars based on Gaia DR2 star maps, as well as the estimated contamination as a function of time in the \cheops\ light curve. Clearly, there are bright stars near the aperture, and in particular, TIC 37794435 is visible on the right sides of the top two panels in this figure. There is light from this star which bleeds into the aperture for our \cheops\ light curve, despite the fact that this is the smallest aperture we use to make a \cheops\ light curve, and is the main contributor of noise in this light curve. However, this noise contribution appears to be well-characterized throughout the light curve, as the contamination level in the DEFAULT light curve remains relatively consistent around 4.7\%, with an RMS spread of 0.2\%. 

TOI 470 was observed by LCOGT on UT 2021 January 28 in both a blue (Sloan $g'$) and red (zs) filter. Transit-like events with depths consistent with \tess\ were detected on-target in both filters relative to known Gaia DR3 stars.

We checked the singularity of TOI 470 with both reconnaissance spectroscopy and high-resolution imaging. Low-resolution spectroscopy via the TRES instrument at FLWO and the CHIRON instrument at SMARTS, which tests whether the star has a high radial velocity, showed that the target star is indeed a single G-dwarf. Our star was also observed via high-resolution speckle imaging with the Gemini Zorro instrument in both the 562 and 832 nm bands, as well as the SOAR HRCam I-band (centered at 879 nm). Additionally, the star was imaged with adaptive optics at Keck2 on the NIRC2 instrument in K band (centered at 2.196 $\mu$m). The Gemini and Keck2 images showed that TOI 470 is a single star to better than $\Delta$mag $> 6.16$ mag at 0.5 arcseconds, and the SOAR image showed that the star is single to an estimated contrast of $\Delta6.8$ mag at 1.0 arcsecond. For clarity, we show only the Gemini contrast curve in Fig. \ref{fig:contrast_curves} (middle right). 

Finally, our assessment of this star with Gaia showed that based upon similar parallaxes and proper motions, there are no additional widely separated companions identified by Gaia. TOI 470 has a Gaia EDR3 RUWE value of 1.17, indicating that the astrometric fits are consistent with the single star model.

Our statistical vetting with \texttt{TRICERATOPS} supports the conclusion that the transit signals are from a transiting planet which orbits the target star, returning $FPP = 0.0034 \pm 0.0012$ and $NFPP = 0.0009 \pm 0.0003$ for TOI 470 b, just under the threshold for NFPP. 

\subsection{Observations of TOI 518}\label{subsec:val_518}

\begin{figure}
\centering
 \includegraphics[width=\linewidth]{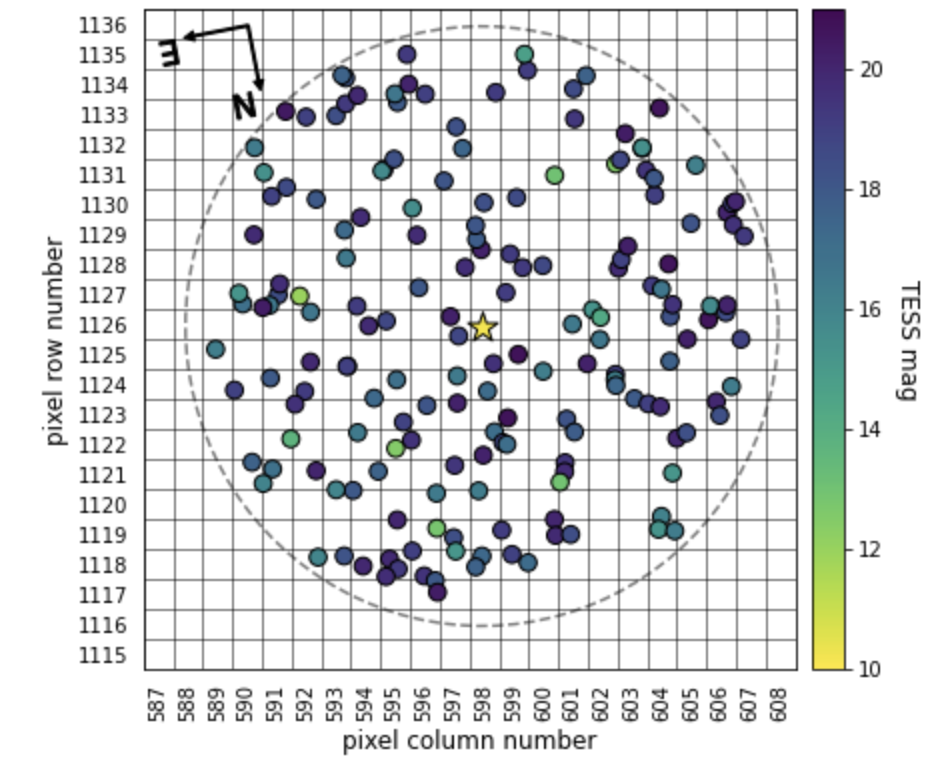}
 \includegraphics[width=\linewidth]{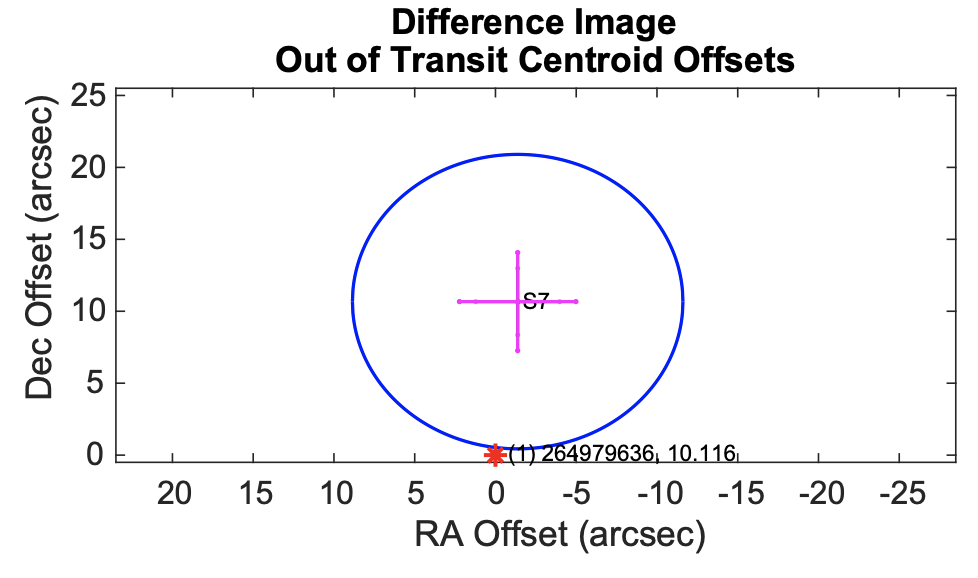}
 \caption{Top: Field stars near TOI 518 (TIC 264979636; indicated as a star) with \tess\ pixels overlaid, and \tess\ magnitude is indicated by the color bar. Bottom: Centroid difference image between in-transit flux and out-of-transit flux from \tess\ observations of TOI 518.}
 \label{fig:TOI518_S7}
\end{figure}

TOI 518 was observed in sector 7 of the \tess\ PM and sector 34 of the EM1. Photometric performance was nominal for sector 7, but there appeared to be stray light which corrupted the sector 34 light curve, rendering it untenable. Therefore, our \tess\ analysis of this target relies on only one sector. Fig. \ref{fig:TOI518_S7} shows the field of stars around TOI 518 (top panel) and the centroid position (bottom panel). Although the field around our target appears to be relatively crowded, there are no stars brighter than 12th magnitude closer than 120 arcseconds, which is several \tess\ pixels away. In the centroid position image, the centroid is south of TOI 518, representing a greater than $3\sigma$ centroid offset. It is thus questionable the transit signal is on target based on our \tess\ observations. 

In both of our \cheops\ visits to this star, the contamination from nearby stars is very low, with a median contribution of 0.047\% for the first visit in December of 2021 and 0.056\% for the second visit in March of 2022. The slight difference between these numbers is due to the relative sky motions of stars near our target as estimated by their proper motions from Gaia DR2. Centroid analysis of the \cheops\ photometry showed that the centroid position during the visit moved no more than 3 arcseconds, indicating that the transit events seen by \cheops\ were indeed on target.

\begin{figure}
\centering
 \includegraphics[width=\linewidth]{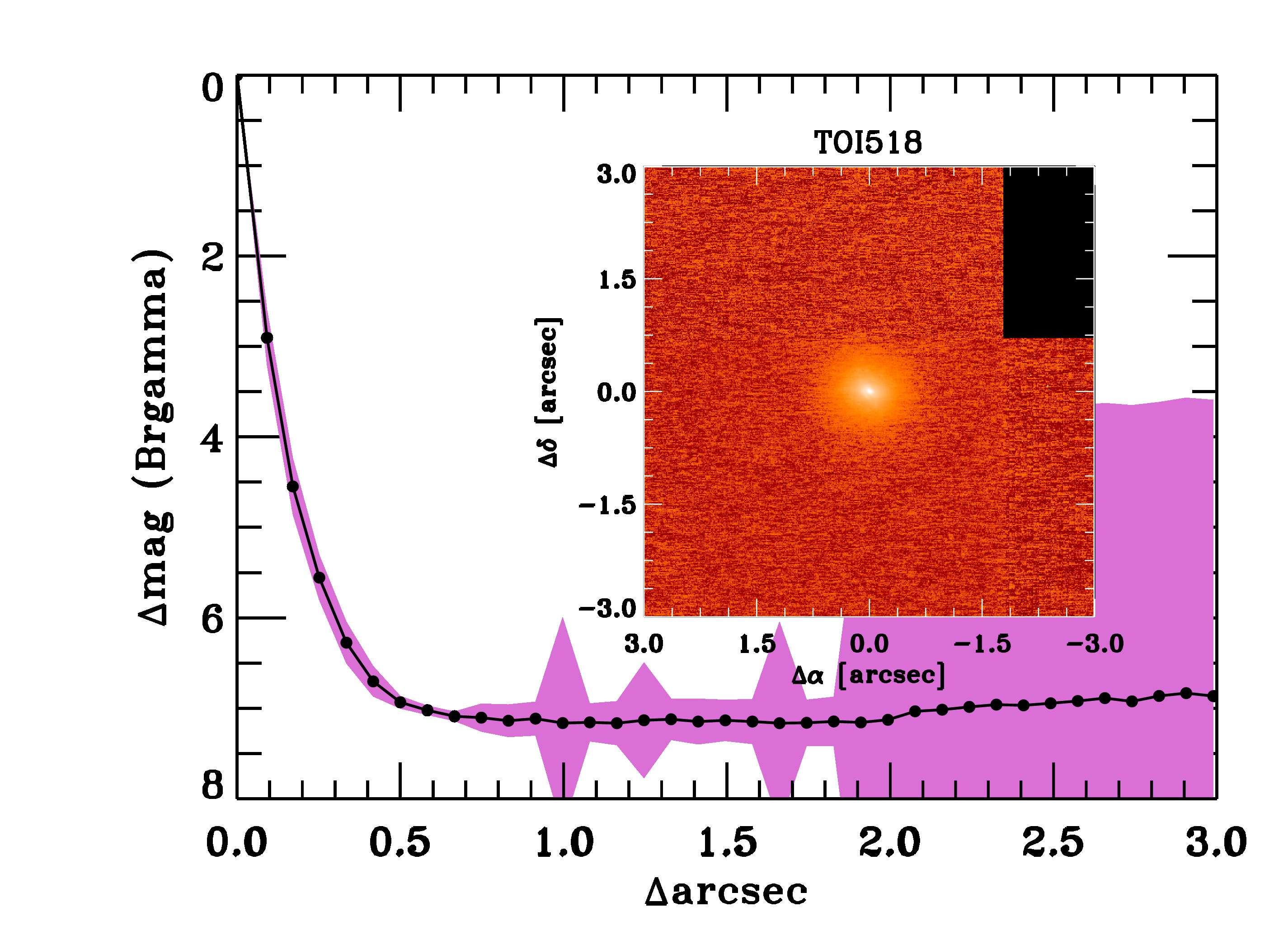}
 \includegraphics[width=\linewidth]{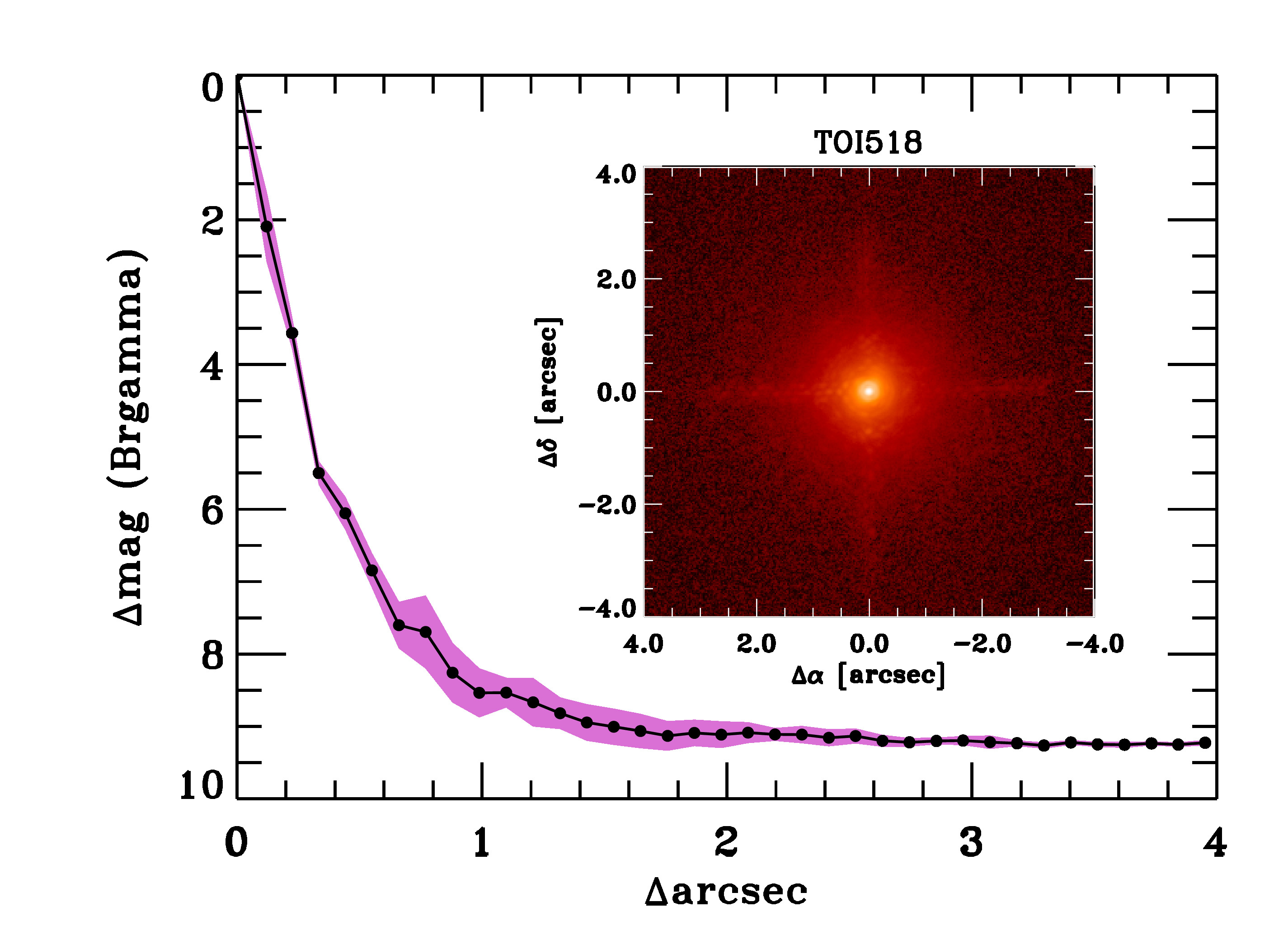}
 \includegraphics[width=\linewidth]{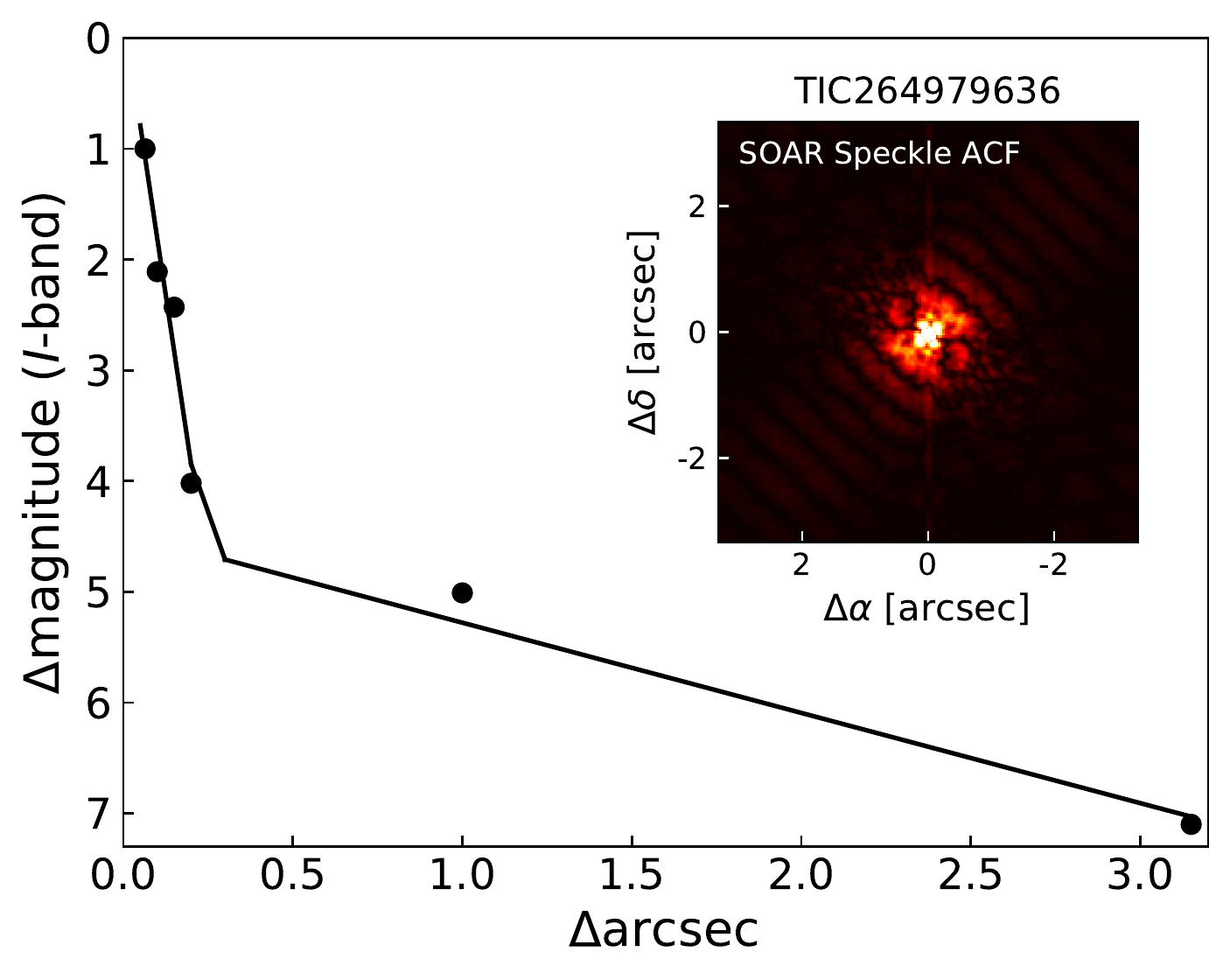}
 \caption{More contrast curves for TOI 518 (TIC 264979636), indicating that this star is single and not a blended EB. Top: Keck2 10m AO sensitivity curve and image inlet. Middle: Palomar 5m AO sensitivity curve and image inlet. Bottom: SOAR speckle sensitivity curve and image inlet.}
 \label{fig:TOI518_contrasts}
\end{figure}

High-resolution imaging observations from multiple large telescopes, including the 8m Gemini North telescope, the 10m Keck2 telescope, the Palomar 5m telescope, and the SOAR 4.1m telescope indicated that the star is single. Keck2 and Palomar employed near-infrared adaptive optics, whereas SOAR and Gemini employed optical speckle imaging, and all of these sensitivity curves are shown in Figs. \ref{fig:contrast_curves} (bottom row) and \ref{fig:TOI518_contrasts} (all panels). %We confirmed this finding with low-resolution spectroscopy with the TRES instrument at FLWO and the CHIRON instrument at SMARTS. Both of these observing runs showed no significant shift in radial velocity, indicating that the star is single and not contaminated by any on-axis star. 
While the optical observations tend to provide higher resolution, the NIR AO tend to provide better sensitivity, especially to lower-mass stars. The combination of the observations in multiple filters enables better characterization for any companions that may be detected. Gaia DR3 is also used to provide additional constraints on the presence of undetected stellar companions as well as wide companions.
    
The Palomar Observatory observations of TOI 518 were made with the PHARO instrument \citep{hayward2001} behind the natural guide star AO system P3K \citep{dekany2013} on 2019~Apr~18 in a standard 5-point quincunx dither pattern with steps of 5\arcsec\ in the narrow-band $Br-\gamma$ filter $(\lambda_o = 2.1686; \Delta\lambda = 0.0326~\mu$m). 
	
The Keck Observatory observations were made with the NIRC2 instrument on Keck-II behind the natural guide star AO system \citep{wizinowich2000} on 2019-Mar-25 UT in the standard 3-point dither pattern that is used with NIRC2 to avoid the left lower quadrant of the detector which is typically noisier than the other three quadrants. 
	
The sensitivities of the final combined AO image were determined by injecting simulated sources azimuthally around the primary target every $20^\circ $ at separations of integer multiples of the central source's FWHM \citep{furlan2017}. The brightness of each injected source was scaled until standard aperture photometry detected it with $5\sigma $ significance. The resulting brightness of the injected sources relative to TOI~518 set the contrast limits at that injection location. The final $5\sigma$ limit at each separation was determined from the average of all of the determined limits at that separation and the uncertainty on the limit was set by the rms dispersion of the azimuthal slices at a given radial distance.  The final sensitivity curves for the Palomar and Keck data are shown in (Figure~\ref{fig:TOI518_contrasts}); no additional stellar companions were detected in agreement with observations from SOAR and Gemini.

We searched for stellar companions to TOI 518 with speckle imaging on the 4.1-m Southern Astrophysical Research (SOAR) telescope \citep{Tokovinin_2018} on 12 December 2019 UT, observing in Cousins I-band, a similar visible bandpass as \tess\. As shown in Fig. \ref{fig:TOI518_contrasts},  this observation was sensitive to a 5.8-magnitude fainter star at an angular distance of 1 arcsec from the target. More details of the observations within the SOAR \tess\ survey are available in \cite{Zeigler_2020}. The $5\sigma$ detection sensitivity and speckle auto-correlation functions from the observations are shown in Figure \ref{fig:TOI518_contrasts}. No nearby stars were detected within 3\arcsec of TOI 518 in the SOAR observations.

Finally, our assessment of this star with Gaia showed that based upon similar parallaxes and proper motions, there are no additional widely separated companions identified by Gaia. TOI 518 has a Gaia EDR3 RUWE value of 1.02 indicating that the astrometric fits are consistent with the single star model. 

Our statistical vetting with \texttt{TRICERATOPS} supported the conclusion that these transit events are not likely to be from a nearby source with $NFPP = 0.0004 \pm 0.00004$ in 20 trials for TOI 518.01. However, initial inspection of the two transits in the \tess\ light curve shows that they are shallow and slightly V-shaped. In some cases, this could indicate an EB with grazing eclipses, but at the very least this makes the radius of the transiting object uncertain given a large uncertainty in the impact parameter, which is the projected distance between the midline of the stellar disc and the planet. As such, \texttt{TRICERATOPS} returned $FPP = 0.1555 \pm 0.2062$, which indicates that there is a significant chance that this signal may be an astrophysical false positive and warrants closer inspection. 

According to our statistical analysis, potential false positive scenarios - from highest to lowest probability - include the following: 1. a planet orbiting the target star at the given period but diluted by an unresolved foreground or background star (known as DTP), 2. an eclipse caused by an unresolved stellar companion with twice the period of the reported period (denoted SEBx2P), or 3. a planet orbiting the primary star of an unresolved stellar binary (PTP). These are scenarios which cannot easily be accounted for with high-contrast imaging alone. 

Reconnaissance RVs may help account for these false positive scenarios by constraining the mass of a potential massive companion. However, our reconnaissance RVs with FLWO-TRES and SMARTS-CHIRON were not taken at quadrature with respect to the estimated orbital period ($\sim17.87$ d), meaning we could not reliably constrain the presence of any massive companion. Further characterization with radial velocities could confirm this system as having no massive companion.

For all of the above reasons, we cannot consider the transit events around TOI 518 validated. Therefore, moving forward we treat this planet candidate cautiously, with the understanding that this transit signal has not yet been statistically validated as a planet, although it is close. Thus, we refer to this signal as TOI 518.01.

\section{Methodology: Photometric Modeling, Fitting, \& Comparisons}\label{sec:meth}

In an effort to reduce the effects of systematics, we applied three independent methods to model and fit physical transiting planet parameters to our photometric data. Prior to fitting transit models to these light curves, we detrended these observations from various sources of both instrumental and astrophysical noise. Detrending, modeling, and fitting are discussed in the following section. 

\subsection{Obtaining and Detrending \tess\ Light Curves}

As previously mentioned, we chose to analyze SPOC PDCSAP light curves from TESS observations of these systems. In order to extract light curves, Simple Aperture Photometry (SAP) is applied to pixel data, which is simply summing the pixel values within a pre-defined aperture as a function of time. The SPOC pipeline applies various calibrations and corrections. Calibrations include standard CCD reduction (bias, dark, and flat field calibrations), smear corrections due to lack of camera shutters, and removal of cosmic ray signals (20 s cadence only). Background flux is estimated and removed per pixel and cadence; scattered light primarily from Earth and Moon is identified and flagged for each light curve and cadence. Systematic errors due to spacecraft pointing and focus changes are encapsulated in Cotrending Basis Vectors (CBVs), which are available for download from MAST. Presearch Data Conditioning (PDCSAP; \citealt{Stumpe_2012,Stumpe_2014,Smith_2012}) light curves are obtained by cotrending SAP light curves against the CBVs. PDCSAP light curves are also corrected for finite photometric aperture and for crowding within the aperture. We downloaded SPOC PDCSAP light curves using \texttt{lightkurve}, a Python package for time-series data analysis. We stitched PDCSAP light curves from different \tess\ sectors together to yield one light curve object, containing time in BTJD, normalized flux, and normalized flux error for each entry. We also removed Not-a-Number entries.

SPOC PDCSAP light curves are already corrected for instrumental noise, but they may contain stellar or other astrophysical sources of noise. To account for this, we removed long-term stellar variability using \texttt{wotan}, which applies a sliding biweight filter to flatten the light curve \citep{hippke2019wotan}. In many instances, detrending with \texttt{wotan} did not significantly change the shape of the light curve. However, there were some obvious trends of stellar variability in our \tess\ light curves for TOIs 444, 455, and 560 that we eliminated with this detrending. Since our results are sensitive to transit depth, we did not want to over-correct the light curves, and thus applied only minimal detrending. We did not apply regressions, splines, or Gaussian Processes (GPs) to our light curves. Our \tess\ light curves are shown in Figs. \ref{fig:TESS_LCs}. 

\subsection{Detrending \cheops\ Light Curves}\label{cheops_detrend}

As described in section \ref{subsec:CHEOPS}, we extracted PSF photometry with \texttt{PIPE} to generate our \cheops\ light curves from imagettes. Then, we detrended our \cheops\ light curves using \texttt{pycheops}, which is a Python library developed for easy and efficient use with \cheops\ data products \citep{Maxted2021pycheops}. We used \texttt{pycheops} routines to trim outliers, decorrelate the light curves from systematic effects, and fit transit models. Photometric model fitting is described in the following section. 

We began by trimming outliers from the light curves, which were those points which were $5\sigma$ away from the median value. Given that \cheops\ rolls about its optical axis as it observes, the shape of the PSF changes throughout an observation. Telescope roll angle with respect to a reference CCD position is reported. As such, we detrended against multiple parameters including $(X,Y)$ pixel position and telescope roll angle as first, second, and third order sine and cosine functions. We also detrended \cheops\ observations against background flux dominated by zodiacal light or scattered light from Earth, observation time and stellar contamination in the aperture. Finally, we checked nearby solar system objects in order to account for glint from these objects. In order to avoid overfitting, we introduced each of these detrending parameters independently of one another and calculated the Bayes Factor with and without the parameter, as in \citet{Trotta2007Bayesian}. This allowed us to determine whether each parameter was necessary for the model. We detrended for all of the above sources while simultaneously fitting a transit model with \texttt{pycheops} to avoid removing transit features. Both our undetrended and detrended \cheops\ light curves are shown in Figs. \ref{fig:CHEOPS_LCs}.

\subsubsection{Noise comparison to other \cheops\ targets}

\begin{figure*}
    \centering
    \includegraphics[width=\linewidth]{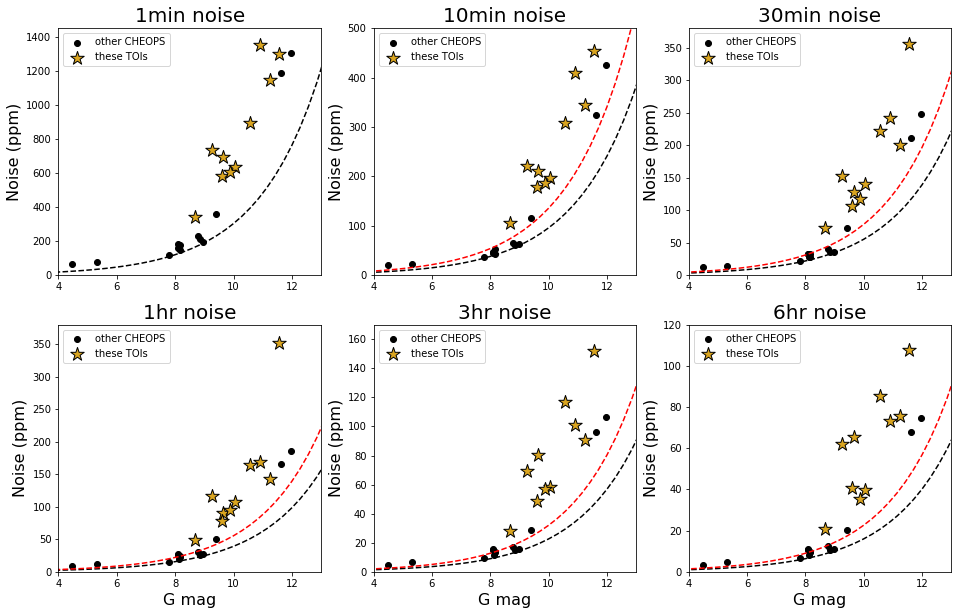}
    \caption{Measured photometric precision for these TOIs (yellow stars) and other \cheops\ targets (black circles, courtesy of Thomas G. Wilson) as a function of Gaia G-band magnitude. Black dashed lines represent photon noise limits assuming 100\% efficiency, whereas red dashed lines represent photon noise limits assuming 50\% observing efficiency. Different panels represent different integration times, where photon noise is averaged over the referenced timescale.}
    \label{fig:CHEOPS_noise}
\end{figure*}

We report measured photometric noise as a function of Gaia G-band magnitude for our \texttt{PIPE}-extracted \cheops\ and detrended light curves and compare against other stars targeted by \cheops. Further, by comparing to photon-limited noise, this serves as a check of in-flight performance. Using the "minimum errors" method described in \citet{Maxted_2022}, we calculated light curve noise levels after subtraction of the best-fit transit model by finding the transit depth which can be detected at S/N of 1 at timescales including 1 minute, 10 minutes, 30 minutes, 1 hour, 3 hours, and 6 hours, as specified in Fig. \ref{fig:CHEOPS_noise}, assuming that our flux errors are minimum bounds on true errors in flux values. We compare to other \cheops\ targets (black points; courtesy of Thomas G. Wilson, priv. comm.). We also compare to photon-limited noise at 100\% and 50\% observing efficiency (black and red dashed lines, respectively), which were calculated using the \cheops\ exposure time calculator (ETC\footnote{\url{https://cheops.unige.ch/pht2/exposure-time-calculator/}}). 

In general, noise in our light curves seems to follow trends from other \cheops\ targets. While we do not add any targets to the \cheops\ sample brighter than 8th magnitude, we are able to fill a gap between magnitudes about 9.5 to 11.5, where most of our targets reside. As such, it is evident that our targets exhibit slightly more noise than the photon-limited predictions shown by the dashed line in each panel of Fig. \ref{fig:CHEOPS_noise}. This may be due to in-flight noise sources which were not well-constrained prior to launch, including atmospheric airglow, \cheops's large PSF, hot pixels, and cosmic rays. Despite these sources of noise, in-flight performance appears largely to match what is expected, which is supported by our noise estimates (Fortier et al., in prep.).

We note that while there appears to be low noise in many of our light curves, there is one notable exception. In particular, our TOI 244 \cheops\ light curve exhibits noise on mid-range timescales (30 minutes to 3 hours) at much higher levels than other targets. TOI 244, our dimmest and therefore rightmost star on each panel, pulls farther away from the dashed line than any other target. This may have been impactful for our model fitting, as we will discuss later. This is especially true given that noise on this timescale is approximately equivalent to the duration of a transit for a typical short-period planet ($\sim30$ min to 3 hrs).

\subsection{Treatment of Time-correlated Red Noise}

\begin{table}
    \centering
    \begin{tabular}{c|c|c}
        \hline\hline
        TOI & \cheops\ & \tess\ \\
        \hline
        TOI 118 & 1.87 & 1.29 \\
        TOI 198 & 1.91 & 1.35 \\
        TOI 244 & 1.23 & 1.19 \\
        TOI 262 & 1.45 & 1.25 \\
        TOI 444 & 1.58 & 2.20 \\
        TOI 455 & 2.08 & 2.35 \\
        TOI 470 & 1.39 & 1.26 \\
        TOI 518 & 1.62 & 1.27 \\
        TOI 560 & 2.25 & 1.90 \\
        TOI 562 & 1.23 & 1.29 \\
        \hline
    \end{tabular}
    \caption{$\beta$ factors for each of our light curves, which were the factors by which we inflated our flux errors in order to account for red noise.}
    \label{tab:betas}
\end{table}

Time-correlated red noise in light curves can significantly bias fit results, and in some cases can lead to a non-detection of a transiting planet \citep{Pont_2006}. Therefore, it must be accounted for in model fitting. We do so by following methods similar to those described in \citet{Winn_2008} and \citet{Wong_2021}. Without including corrections for red noise in an initial run, we fit transit models to our \tess\ and \cheops\ light curves as described below. Then, we binned the residuals from the joint fit of both light curves for each system using our HOMEBREW method (described below) with $n$ points in $m$ bins (for a total of $\sim n\times m$ points in a whole light curve) and calculated the RMS. This quantity represents the RMS deviation from the model, which should follow a Gaussian $\sqrt{n}$ in the case of uncorrelated errors. We did this for bin sizes $n$ from 1 to 1000, or the maximum number of points in the light curve if the number of points was less than 1000. 

Although the RMS deviation should follow a $\sqrt{n}$ trend with bin size, this quantity deviated from this trend in every case, presumably due to the presence of time-correlated noise. In order to account for red noise on the same timescale as a transiting planet, we found the mean deviation $\beta$ from this trend when binning at 14 and 60 minutes, representing between 7 points and 30 points in a bin for our 2 min cadence \tess\ light curves and between 14 and 60 points per bin for our 60 s cadence \cheops\ light curves. This factor $\beta$ differed for each light curve and was typically between 1 and 2 (as shown in Table \ref{tab:betas}). We then inflated our flux errors for each point in each light curve by this factor $\beta$. We did so because our previous flux error values were calculated using purely Poisson statistics and assumed to have no correlated noise, and inflating them in this way is one way to account for red noise to first order \citep{Pont_2006}. We then used these new flux error values to rerun our fits and calculate new model parameters, which are reported as our final model fits.

\subsection{Modeling}

Accurate modeling of a transiting planet across the face of a star is the cornerstone of transcribing photometric data to a set of physical parameters which describe the system. An important piece of accurately modeling transits is the function by which stellar limb darkening is parameterized, which is the subject of much discussion in recent literature \citep{Morello_2017,Neilson_2017,Espinoza2016limb,Muller2013limbdark}. The power-2 limb darkening law \citep{Hestroffer_1997} is a  two-parameter form of limb darkening which is both fast and accurate to compute \citep{Maxted2019qpower2}. It is the limb darkening law which is implemented in \texttt{pycheops} as qpower2, so we use this parameterization throughout our fits in order to maintain consistency in modeling. The 'power2' limb-darkening law follows the functional form

\begin{equation}
    I(\mu) = I_o\left[1-c_1\left(1-\mu^{c_2}\right)\right]
\end{equation}
where $\mu = \sqrt{1-x^2}$ is the projected radial coordinate and $I_o$ is a normalization constant. 

In all our models and fits, we use the Python Limb Darkening Toolkit (PyLDTK; paper: \citealt{Parviainen2015LDTK}; code: \citealt{2015ascl.soft10003P}) to calculate stellar limb darkening coefficients. Given a set of stellar parameters as input, PyLDTK uses the library of PHOENIX-generated specific intensity spectra by \citet{Husser2013PHOENIX} to calculate stellar limb darkening coefficients for a given model. We provide estimations of stellar effective temperature, surface gravity (log$g$), and metallicity, which were obtained from previous publications for known planets or obtained from our SED analysis or spetroscopic characterization for newly-validated planets. In addition, we provide the bandpass for which the coefficients are calculated, by providing preexisting \tess\ and \cheops\ throughput curves. This meant that these coefficients were slightly different for \tess\ and \cheops\ photometry. We specified the model input as power2, which gave us limb darkening coefficients for this law, as well as uncertainties on these values. We chose to keep limb darkening coefficients for each star in each bandpass constant through all fits and models for two reasons. First, we did so in order to reduce uncertainties on other values and second, because neither the \tess\ nor the \cheops\ photometry is sufficiently precise to allow a meaningful constrain on the limb darkening coefficients.

From fitting a model of a transiting planet to a light curve, we can recover many system parameters. These may include transit depth $D$ (related to the radius ratio of the planet and star as $D = (R_p/R_*)^2$), the time of mid-transit $T_o$ (reported as a time in BJD), the orbital period $P$ (reported in days), and the impact parameter $b$, which is the sky-projected distance between the stellar midline and the chord traced by the planet across the face of the star, and is thus a scaled value from 0 to 1. $b$ in turn is related to the semi-major axis $a$ of the orbit for the planet, scaled to the radius of the star, and the planet's orbital inclination with respect to Earth $i$ (reported in degrees). However, different models report these orbital and physical parameters in different ways. For example, the \texttt{batman} \citep{Kreidberg2015batman} transit model directly fits for orbital inclination $i$ in degrees, meaning impact parameter is a derived quantity, whereas \texttt{juliet} directly fits for a parameterization of impact parameter $b$, meaning orbital inclination is a derived quantity. Therefore, we delineate between fitted and derived quantities for different models in Table \ref{tab:fit_derive}. The ways in which our model parameterizations differ from one another may account for some differences in results, despite the fact that parameterizations of many parameters frequently depend on one another. 

\begin{table*}
\centering
\caption{Distinction between fitted and derived properties for our various models.}
\label{tab:fit_derive}
\begin{tabular*}{.53\linewidth}{|c|c|c|c|}
\hline
Parameter                   & \texttt{pycheops} & HOMEBREW                 & \texttt{juliet} \\ \hline
Time of mid-transit         & \multirow{2}{*}{fitted}            & \multirow{2}{*}{fitted}  & \multirow{2}{*}{fitted}          \\
T$_o$                       &                                    &                          &                                  \\ \hline
Orbital period              & \multirow{2}{*}{fitted}            & \multirow{2}{*}{fitted}  & \multirow{2}{*}{fitted}          \\
$P$                         &                                    &                          &                                  \\ \hline
Planet-to-star radius ratio & \multirow{2}{*}{\textbf{derived}}           & \multirow{2}{*}{fitted}  & \multirow{2}{*}{fitted}          \\
$k = R_p/R_s$                         &                                    &                          &                                  \\ \hline
Transit depth               & \multirow{2}{*}{fitted}            & \multirow{2}{*}{\textbf{derived}} & \multirow{2}{*}{\textbf{derived}}         \\
$D = (R_p/R_s)^2$           &                                    &                          &                                  \\ \hline
Impact parameter            & \multirow{2}{*}{fitted}            & \multirow{2}{*}{\textbf{derived}} & \multirow{2}{*}{fitted}          \\
$b = (a/R_s)\cos(i)$        &                                    &                          &                                  \\ \hline
Scaled semi-major axis      & \multirow{2}{*}{fitted}           & \multirow{2}{*}{fitted}  & \multirow{2}{*}{\textbf{derived}}         \\
$(a/R_s)$                   &                                    &                          &                                  \\ \hline
Orbital inclination         & \multirow{2}{*}{\textbf{derived}}           & \multirow{2}{*}{fitted}  & \multirow{2}{*}{\textbf{derived}}         \\
$i$                         &                                    &                          &                                  \\ \hline
Stellar density             & \multirow{2}{*}{\textbf{derived}} & \multirow{2}{*}{\textbf{derived}} & \multirow{2}{*}{fitted} \\
$\rho_s$ & & & \\ \hline
Planet radius               & \multirow{2}{*}{\textbf{derived}}           & \multirow{2}{*}{\textbf{derived}} & \multirow{2}{*}{\textbf{derived}}         \\
R$_{\earth}$        &                                    &                          &                                  \\\hline
\end{tabular*}
\end{table*}

One of the primary goals of this work is to compare photometric performance between \tess\ and \cheops. One way in which we do this is by comparing model values and uncertainties in transit depth. Transit depth represents a solid metric of comparison due to that fact that in all cases it is either computed directly by our models or singularly calculated from one other model parameter which is itself fitted directly. Therefore, uncertainties in transit depth are propagated directly from our fits (in the case of \texttt{pycheops}) or from only planet-to-star radius ratio.

\subsection{Fitting}

We used a variety of fitting methods including MCMC sampling with \texttt{emcee} \citep{Foreman_Mackey_2013} and nested sampling with \texttt{dynesty} \citep{Speagle2020dynesty}. We compare these methods of fitting in order to gauge their effects on the model output uncertainties.

We used different fitting codes for different datasets. \texttt{pycheops} is primarily useful for detrending and fitting transit models to \cheops\ light curves, so we used it exclusively on our \cheops\ light curves. Then, we developed our HOMEBREW code for use with both \cheops\ and \tess\ light curves. Finally, we used \texttt{juliet} for both \cheops\ and \tess\ light curves. We describe each below.

\subsubsection{\texttt{pycheops}: \texttt{lmfit} + \texttt{emcee}}

For our runs with \texttt{pycheops} \citep{Maxted_2022}, we used the built-in nonlinear least-squares minimization \texttt{lmfit} to initially constrain model parameters, and then passed these results to the MCMC sampler \texttt{emcee}, which is also built into the functionality of \texttt{pycheops}. We employ \texttt{lmfit.Parameter} objects, which are model parameters that can be either varied or kept constant, depending on the part of the model one is constraining. We supplied a loose set of priors for the initial model generation, which were retrieved by the authors from ExoFOP \tess. As previously stated, detrending parameters and limb darkening coefficients are held constant, and we fit for only physical and orbital parameters of the planet. We apply this fitting framework only to \cheops\ photometry. For our MCMC sampler runs, we initialize 80 walkers around the best-fit values provided by the nonlinear least-squares minimization, run 300 steps of burn-in, and then run for 1000 steps, at which point we verified the sampler converged for each system by checking the convergence time. 

\subsubsection{HOMEBREW: least squares minimization + \texttt{emcee}}

We built a similar framework as in the previous subsubsection to fit both \tess\ and \cheops\ photometry, which we call our HOMEBREW method. This code is distinct because we used it to fit models to both datasets, including individual fits and joint fits to both datasets, whereas we used \texttt{pycheops} exclusively for \cheops\ photometry. Therefore, we generate three results for each system with this framework, including one each for \tess\ and \cheops\ and one which was jointly fitting both datasets. We begin by initializing a model in \texttt{batman}, which is informed by the same initial values drawn from ExoFOP \tess\ as for \texttt{pycheops}. For our fits to \cheops\ and \tess\ separately, our model fits orbital parameters jointly, but computes a different depth for each dataset. For our joint fits to both datasets, all model parameters converge to the same value, including transit depth. We designed our fitting code in this way to strictly examine transit depths and uncertainties on this value, thus maintaining consistency between orbital parameters.

We find an initial model with least-squares minimization by employing the \texttt{scipy.optimize} function. These model parameters are then passed to the MCMC sampler \texttt{emcee} with uniform priors. We initialize 48 walkers for fits to \cheops\ and \tess\ data, and 40 walkers for fits to both datasets jointly. This is due to the fact that we fit for one fewer parameter for joint fits, as there is only one transit depth to compute across both datasets. For both types of fits, we run the MCMC sampler for a burn-in phase of 500 steps, and then run for 1,500 more steps for a total of 2,000 steps. The results of these fits are displayed as corner plots using \texttt{corner} \citep{corner}.

\subsubsection{\texttt{juliet}: Nested sampling with \texttt{dynesty}}

As a final fitting framework, we used nested sampling as implemented in \texttt{juliet} \citep{Espinoza2019juliet}. Nested sampling is a method of estimating Bayesian evidence for a set of model parameters and allows the ability to robustly sample from a high-dimensional, multi-modal parameter space. Such an algorithm relies on integrating the prior in nested shells of constant likelihood \citep{Speagle2020dynesty}. We modified the base code of \texttt{juliet} slightly to include the power2 limb darkening law. We initialize a transit model in \texttt{juliet}, which is built from the \texttt{batman} transit model, and then apply nested sampling with \texttt{dynesty} to find convergence. We apply this framework to \tess\ and \cheops\ photometry separately and also use it to find a global fit across both datasets, generating three more results for each system. Given \texttt{juliet}'s model parameterization, we were not able to completely maintain consistency between fits by fitting for certain parameters separately. Therefore, we generated one complete set of model parameters each for \tess\ and \cheops\ light curves separately, and one more complete set of model parameters for our global fits. In section \ref{sec:results}, we verify consistency between parameters generated with HOMEBREW and \texttt{juliet}.

\subsection{Comparing \tess\ and \cheops\ precision}

An overall goal of this work is to compare the relative photometric performances of \tess\ and \cheops. As previously stated, \cheops's larger primary aperture size and smaller pixel scale makes it a higher-precision instrument relative to \tess. However, as a survey mission, \tess\ has the advantage of capturing more transits on average per target relative to \cheops. Further, the lower observing efficiency $\epsilon_{CHEOPS}$ of \cheops\ will increase model uncertainties. As such, we expect that uncertainties in transit depth should theoretically scale according to photon-limited precision as
\begin{equation}\label{eq:N_transits_te}
\sigma_{1,TESS} = \sigma_{TESS}\sqrt{N_{tr,TESS}\epsilon_{TESS}}
\end{equation}
and
\begin{equation}\label{eq:N_transits_ch}
\sigma_{1,CHEOPS} = \sigma_{CHEOPS}\sqrt{N_{tr,CHEOPS}\epsilon_{CHEOPS}}.
\end{equation}
Here, we define $\epsilon_{CHEOPS}$ to be the \cheops\ observing efficiency, which we calculate as the number of data points in transit divided by the number of points which should constitute the whole transit given the sampling cadence of 60 s. $\epsilon_{TESS}$ is similarly defined for \tess\ observations, calculated according to the \tess\ observing cadence. Further, we define $N_{tr,\ldots}$ as the number of transits in our light curves for either \tess\ or \cheops. Finally, $\sigma_{CHEOPS}$ and $\sigma_{TESS}$ represent depth uncertainties as reported by our models, and $\sigma_{1,CHEOPS}$ and $\sigma_{1,TESS}$ represent depth uncertainties on 1 ideal \cheops\ or 1 ideal \tess\ transit, respectively. As shown by eqns. \ref{eq:N_transits_te} and \ref{eq:N_transits_ch}, the reported model uncertainty would be modified by the number of transits captured by a given telescope, as well as the observing efficiency of that telescope for a given target. Uncertainty on 1 ideal transit would be larger than reported model uncertainty if more than 1 transit is observed, whereas uncertainty on 1 ideal transit would be smaller when accounting for observing efficiency. Importantly, we assume that transit coverage with \tess\ is 100\%, meaning we set $\epsilon_{TESS} = 1$ in all cases. This is a safe assumption for our targets as shown in Figs. \ref{fig:phase_folded_plots}, which demonstrate that our transit coverage with \tess\ for each target is full. However, this is not always the case for \tess\ observations, as there are instances in which data downlink interrupts transit coverage or photometry is corrupted by stray light. Therefore, \tess\ observing efficiency ought to be included in general applications of this analysis.

Theoretical photon-limited performance depends on the photon flux of a given star in the relevant bandpass, as well as background and read noise. However, we neglect the contributions from background and read noise, as these are assumed to be sufficiently accounted for during image subtraction and light curve generated. To compute the ideal theoretical performance for a system, we ought to evaluate the number of captured photons per time. This will be an SED-dependent ratio between \cheops\ and \tess. The effective wavelengths of the two bandpasses are about 581\,nm and 746\,nm for \cheops\ and \tess, respectively, meaning there are about 28\% more photons per energy in the \tess\ band. However, \cheops\ has a primary aperture size which is three times larger than that of \tess, meaning both of these factors should be accounted for.  

This implies the following:
\begin{equation}\label{eq:q}
    \frac{\sigma_{1,TESS}}{\sigma_{1,CHEOPS}} = \sqrt{\frac{N_{photons,CHEOPS}}{N_{photons,TESS}}} = \sqrt{q}
\end{equation}
where 
\begin{equation}\label{eq:photons}
    N_{photons} = A\int \textup{SED}(\lambda)\textup{F}(\lambda)d\lambda
\end{equation}
and
\begin{equation}
    \textup{SED}(\lambda) \propto \frac{\textup{Energy}}{\textup{area}\cdot\lambda\cdot\textup{time}}.
\end{equation}
We have defined $q$ to be the photon ratio between \cheops\ and \tess, $\textup{SED}(\lambda)$ to be the wavelength-dependent energy flux per unit time, and $\textup{F}(\lambda)$ to be the filter function of a given telescope. To find $q$, we computed the expected energy fluxes for CHEOPS and TESS for a range of model spectra, assuming 100\% efficiency for both telescopes. Combining the above, we expect the following ratio of uncertainty in depth:

\begin{equation}\label{eq:final_comp}
    \frac{\sigma_{CHEOPS}}{\sigma_{TESS}} = \sqrt{\frac{N_{tr,TESS}}{N_{tr,CHEOPS}}\frac{\epsilon_{TESS}}{q\epsilon_{CHEOPS}}}
\end{equation}

Physically, this means that given the $\sim3:1$ primary size ratio of \cheops\ to \tess, we would expect that the precision on the depth of one \cheops\ transit should be equivalent to the precision in depth of nine phase-folded \tess\ transits ($q \approx 9$ in most cases) in the case that the star is equally as bright in both bandpasses and assuming perfect observing efficiency. Equation \ref{eq:final_comp} allows us an effective tool for comparison between these two telescopes even given different observing efficiencies and numbers of transits for each target.

\subsubsection{How many \tess\ transits \it{would} we need to reach the same precision as a single \cheops\ transit?}

Using the above formulation, we can approach the question of how many \tess\ transits it would take to reach the precision in one \cheops\ transit. Similar to the way that equations \ref{eq:N_transits_te} and \ref{eq:N_transits_ch} show that noise scales as $1/\sqrt{N_{tr}}$, we can solve for the number of transits needed to equate the precision on one idealized \cheops\ transit and the precision on one idealized \tess\ transit, as

\begin{equation}\label{eq:N_T_equiv}
    \sigma_{1,CHEOPS} = \sigma_{1,TESS}\sqrt{N_{tr,equiv}}
\end{equation}

According to photon-limited noise, the number of \tess\ transits we expect ought to satisfy the above equation is equal to the photon ratio $q$ for any given system. However, by solving equation \ref{eq:N_T_equiv} for the number of equivalent transits, $N_{tr,equiv}$, and applying equations \ref{eq:N_transits_te} and \ref{eq:N_transits_ch}, we arrive at our actual value for number of equivalent transits as 

\begin{equation}\label{eq:N_tess_transits}
    N_{tr,equiv} = \left(\frac{\sigma_{TESS}}{\sigma_{CHEOPS}}\right)^2 \frac{N_{tr,TESS}}{N_{tr,CHEOPS}}\frac{\epsilon_{TESS}}{\epsilon_{CHEOPS}}.
\end{equation}

The above equation is generalizable when comparing depth precisions, number of transits obtained, and in-transit observing efficiencies for any two telescopes. Then, equation \ref{eq:N_tess_transits} can then be interpreted as the total number of transits required with a given telescope to match the precision in depth as measured with the other telescope. 

\section{Results}\label{sec:results}

\begin{sidewaystable}
\caption{Calculated transit depths for our 10 systems, reported in ppm.}
\begin{tabular*}{\linewidth}{|l|l|l|l|l|l|l|l|l|l|l|l|}\label{tab:depths}
%\hline
              &           & \textbf{TOI 118} & \textbf{TOI 198} & \textbf{TOI 244} & \textbf{TOI 262} & \textbf{TOI 444} & \textbf{TOI 455} & \textbf{TOI 470} & \textbf{TOI 518} & \textbf{TOI 560} & \textbf{TOI 562} \\ \hline
CHEOPS only   & pycheops  & $1477 \pm 79$                     & $1030 \pm 102$                    & $630 \pm 85$                      & $538 \pm 97$                      & $1476 \pm 151$                    & $1180 \pm 85$                     & $2563 \pm 92$                     & $750 \pm 150$                     & $1500 \pm 80$                     & $1102 \pm 55$                     \\
              & HOMEBREW  & $1936 \pm 238$                    & $1018 \pm 89$                     & $650 \pm 112$                     & $454 \pm 66$                      & $1129 \pm 81$                     & $1024 \pm 90$                     & $2450 \pm 135$                    & $640 \pm 91$                      & $1459 \pm 74$                     & $973 \pm 45$                      \\
              & juliet    & $1697 \pm 247$                    & $1009 \pm 140$                    & $831 \pm 100$                     & $600 \pm 100$                     & $1130 \pm 81$                     & $1171 \pm 99$                     & $2177 \pm 168$                    & $646 \pm 61$                      & $1541 \pm 196$                    & $1148 \pm 78$                     \\ \hline
TESS only     & HOMEBREW  & $1624 \pm 210$                    & $807 \pm 102$                     & $841 \pm 93$                      & $471 \pm 65$                      & $930 \pm 79$                      & $2172 \pm 168$                    & $2061 \pm 136$                    & $471 \pm 100$                     & $1122 \pm 60$                     & $949 \pm 37$                      \\
              & juliet    & $1702 \pm 124$                    & $871 \pm 130$                     & $1100 \pm 113$                    & $566 \pm 65$                      & $994 \pm 126$                     & $2335 \pm 126$                    & $2119 \pm 157$                    & $622 \pm 100$                     & $1218 \pm 140$                    & $1046 \pm 39$                     \\ \hline
TESS + CHEOPS & HOMEWBREW & $1414 \pm 68$                     & $894 \pm 64$                      & $773 \pm 95$                      & $484 \pm 66$                      & $1056 \pm 46$                     & $1665 \pm 82$                     & $2275 \pm 76$                     & $605 \pm 64$                      & $930 \pm 56$                      & $955 \pm 62$                      \\
              & juliet    & $1549 \pm 202$                    & $844 \pm 64$                      & $967 \pm 93$                      & $586 \pm 61$                      & $1068 \pm 49$                     & $1652 \pm 105$                    & $2071 \pm 200$                    & $660 \pm 62$                      & $1235 \pm 63$                     & $1056 \pm 34$                     \\ \hline
\end{tabular*}
\end{sidewaystable}

In this section, we report physical and orbital parameters for each of our systems. We will highlight some specific examples of interest, but ultimately discuss the results of our fitting in the aggregate. Discussion of particular systems is left to section \ref{sec:disc}. Our goal is to compare the performance of \cheops\ relative to the foreseen pre-launch performance estimates via comparison of their depth uncertainties using different fitting methods. Aggregate depth results for each of our fits are shown in Table \ref{tab:depths}.

For each target, we have calculated the orbital properties with both our HOMEBREW method and \texttt{juliet}. However,  for simplicity, we report final orbital and physical properties from only our HOMBREW method.

\subsection{Fractional Depth Uncertainty Comparison Between TESS and CHEOPS}

\begin{figure*}
\centering
 \includegraphics[width=.8\linewidth]{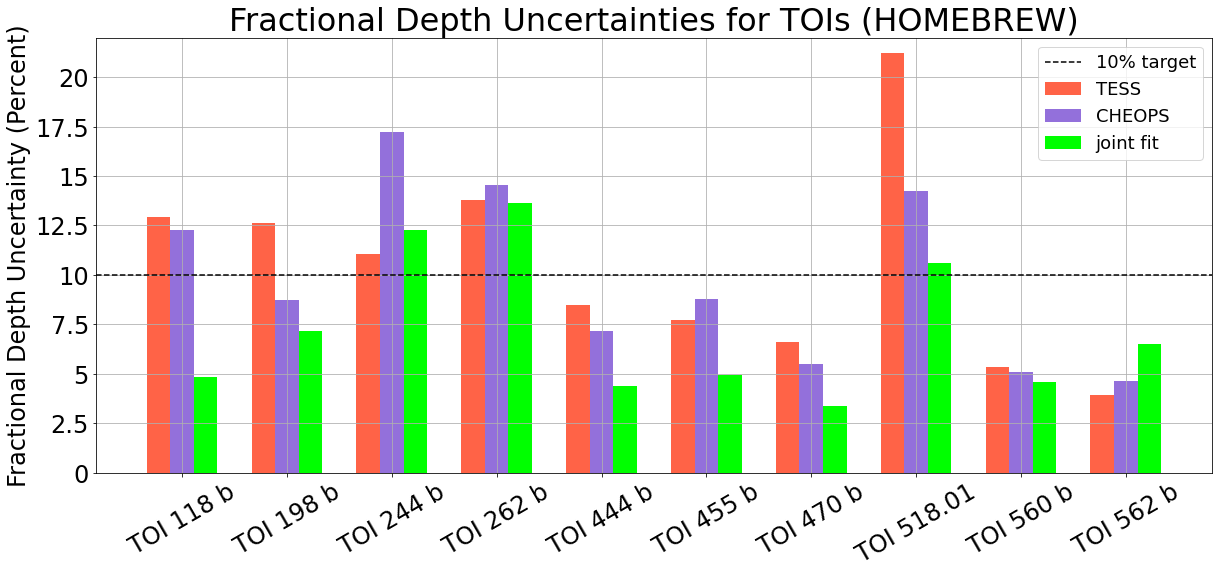}
 \includegraphics[width=.8\linewidth]{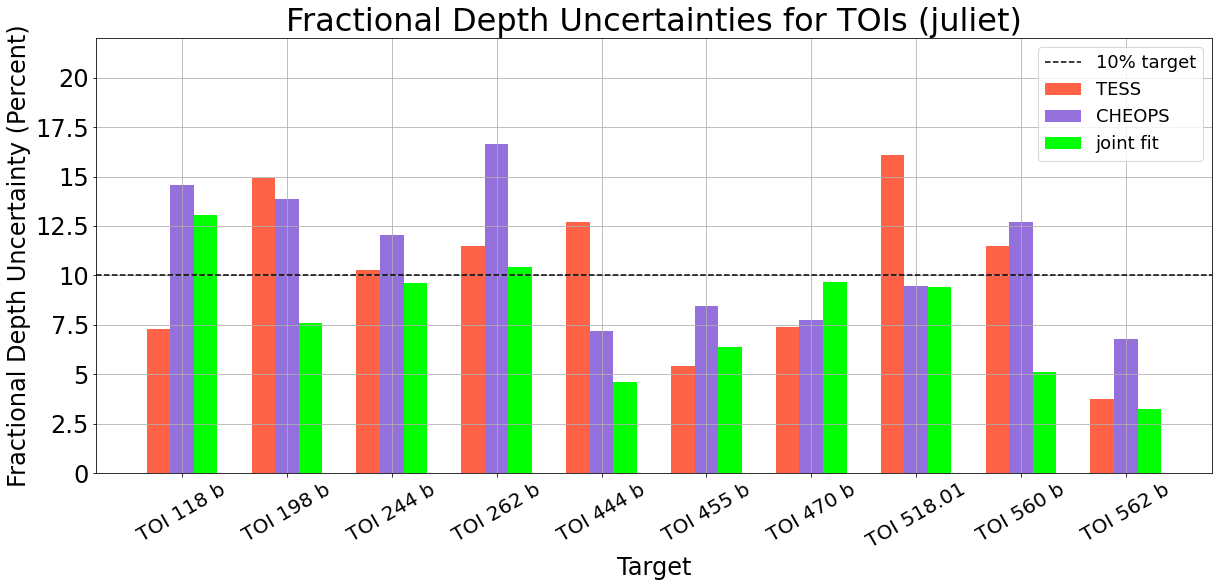}
 \includegraphics[width=.8\linewidth]{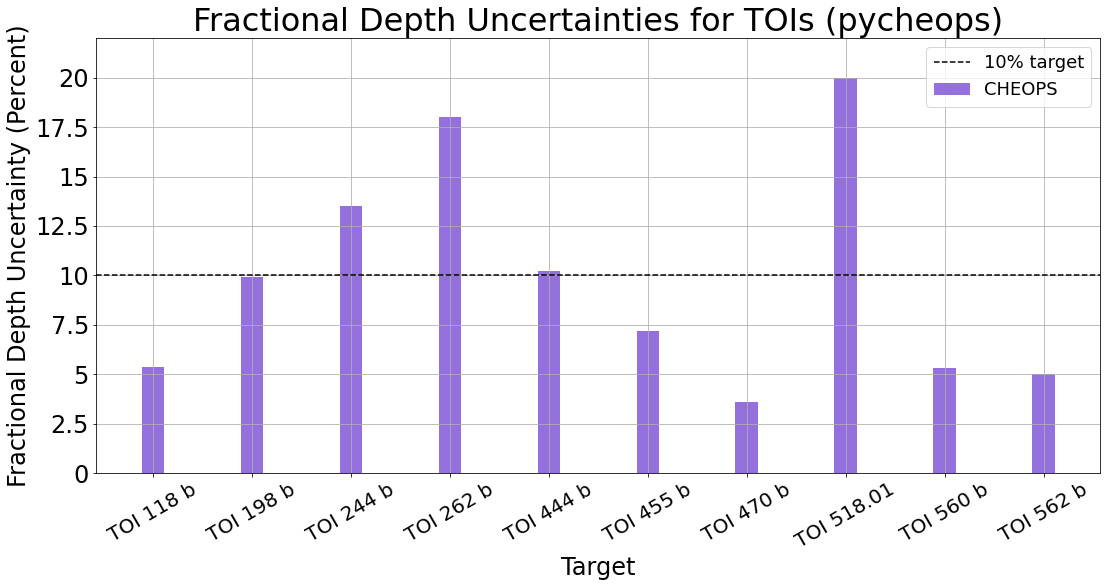}
 \caption{Fractional depth uncertainties for our 9 planets and 1 planet candidate, which is uncertainty in transit as a fraction over calculated transit depth. Top: Fractional depth uncertainty for our HOMEBREW method, which consists of fitting a \texttt{batman} transit model with \texttt{emcee}. Middle: Fractional depth uncertainty using \texttt{juliet} nested sampling. Bottom: Fractional depth uncertainty using \texttt{pycheops} transit model with \texttt{emcee}. For all panels, models fitted to \tess\ data only are represented by red bars, models fitting to \cheops\ data only are represented by purple bars, and models fitting to both datasets jointly are represented by light green bars. A 10\% fractional uncertainty line is shown for both panels as a guide.}
 \label{fig:frac_depths}
\end{figure*}

Here we seek to compare the fractional depth uncertainties yielded by different fitting methods for our photometric datasets. We define fractional depth uncertainty as the reported model uncertainty in transit depth divided by transit depth normalized to a percentage. Comparing fractional uncertainty in depth allows us to compare uncertainties in our fitted model parameters without propagating uncertainties from stellar radius, giving us a straightforward comparison of model performance. Fig. \ref{fig:frac_depths} shows our calculated fractional depth uncertainties for our different modeling and fitting methods, where each bar plot represents a different method for each of our datasets. The top panel shows fractional depth uncertainties for our HOMEBREW method, which consists of fitting a \texttt{batman} transit model to our photometric data with \texttt{emcee}, the MCMC sampler. The middle panel shows these quantities for our fits using \texttt{juliet}, which fits a \texttt{batman} transit model using \texttt{dynesty}, the dynamic nested sampler. The bottom panel shows the fractional depth uncertainty for fits using \texttt{pycheops}, which fits a transit model using \texttt{emcee}, the MCMC sampler. The first two panels show fractional depth uncertainties represented as percentages for models computed with \tess\ data alone (red), \cheops\ data alone (purple), and both datasets jointly (light green), whereas the bottom panel is only for \cheops\ data alone (purple). We have included a horizontal 10\% fractional uncertainty target line as a guide for comparison. Comparing fractional uncertainties between datasets and fit methods allows us to compare the photometric properties of these light curves. All panels are normalized to the same vertical axis for uniform comparison.

Inspection of the top panel of Fig. \ref{fig:frac_depths} shows that many of our fits computed with our HOMEBREW method are below the 10\% fractional uncertainty threshold, although there are some notable exceptions. In comparing the fits to \tess\ photometry alone and \cheops\ photometry alone, we see that for five out of ten systems, the fractional uncertainties when fitting \cheops\ data alone are lower than those fitting \tess\ data alone. For eight of our ten planets, excluding only TOI 244 b and TOI 562 b, we report the lowest fractional uncertainties in depths when using jointly-computed models as opposed to either \tess\ or \cheops\ alone. In the case of both TOI 244 b and TOI 562 b, we report the lowest fractional depth uncertainty using the HOMEBREW method for \tess\ data alone. Given these results, we can see that in more cases our HOMEBREW model fits yield lower relative uncertainties to \tess\ data alone as compared to fits to \cheops\ data alone. Additionally, joint fits to both datasets yield lower uncertainties in more cases and also yield a lower average fractional uncertainty in depth across the ensemble relative to fits to individual datasets.

%moving the analysis of particular targets to subsubsections, saving this for overall results reporting
We examine the cases of larger-than-10\% fractional depth uncertainty calculated with the HOMEBREW method in section \ref{sec:disc}, including individual fits to \tess\ data and \cheops\ data for TOI 118 b, our fit to \tess\ data alone for TOI 198 b, and all our fits to TOI 244 b, TOI 262 b, and TOI 518.01. 

We may also compare our fractional depth uncertainties as computed using \texttt{juliet}, as shown in the middle panel of Fig. \ref{fig:frac_depths}. There are more fits which do not meet the 10\% fractional depth uncertainty threshold using \texttt{juliet} compared to our HOMEBREW method. This may have been due to the fact that orbital parameters were fitted separately for individual fits to \tess\ and \cheops\ photometry. For seven out of our ten systems, including TOI 198 b, TOI 244 b, TOI 262 b, TOI 444 b, TOI 518.01, TOI 560 b, and TOI 562 b, we report the lowest fractional uncertainties in depths when using jointly-computed models as opposed to either \tess\ or \cheops\ alone. When comparing fits to \tess\ data alone and to \cheops\ data alone for \texttt{juliet}, we see that fits to \cheops\ data alone yield smaller fractional depth uncertainties for only three out of ten planets compared to fits to \tess\ data alone. As an ensemble, these results indicate that while fits to \cheops\ data yield lower uncertainties in fewer cases compared to \tess\ data, our fits to both datasets jointly yield lower uncertainties more often than either dataset alone. 

We examine cases of fractional depth uncertainty larger than 10\% calculated with \texttt{juliet} in section \ref{sec:disc}, including fits to \tess\ data alone for TOI 198 b, TOI 244 b, TOI 262 b, TOI 444 b, TOI 518.01, and TOI 560 b, as well as our fits to \cheops\ data alone for TOI 118 b, TOI 198 b, TOI 244 b, TOI 262 b, and TOI 560 b.

Finally, we compare fractional depth uncertainties for our \texttt{pycheops} fits to \cheops\ visits to each target. There are four out of ten instances in which the fractional depth uncertainty is larger than 10\%. We report the largest fractional depth uncertainty with this fit method was for TOI 518 b, which is also the case for the other fit methods as above.  We discuss potential causes for these high uncertainties below. 

\subsection{Radius estimates}

\begin{figure*}
\centering
 \includegraphics[width=.75\linewidth]{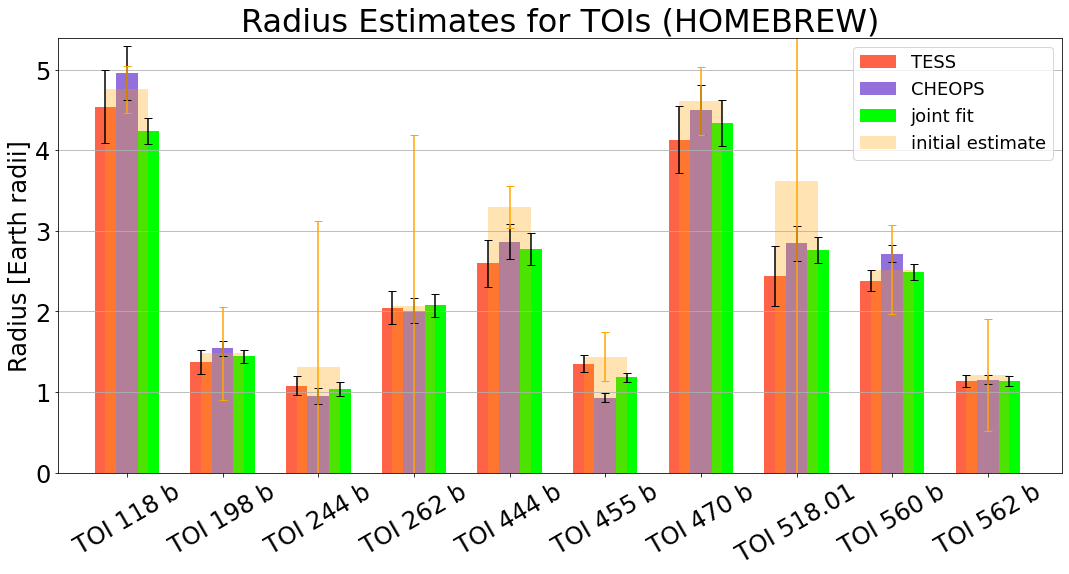}
 \includegraphics[width=.75\linewidth]{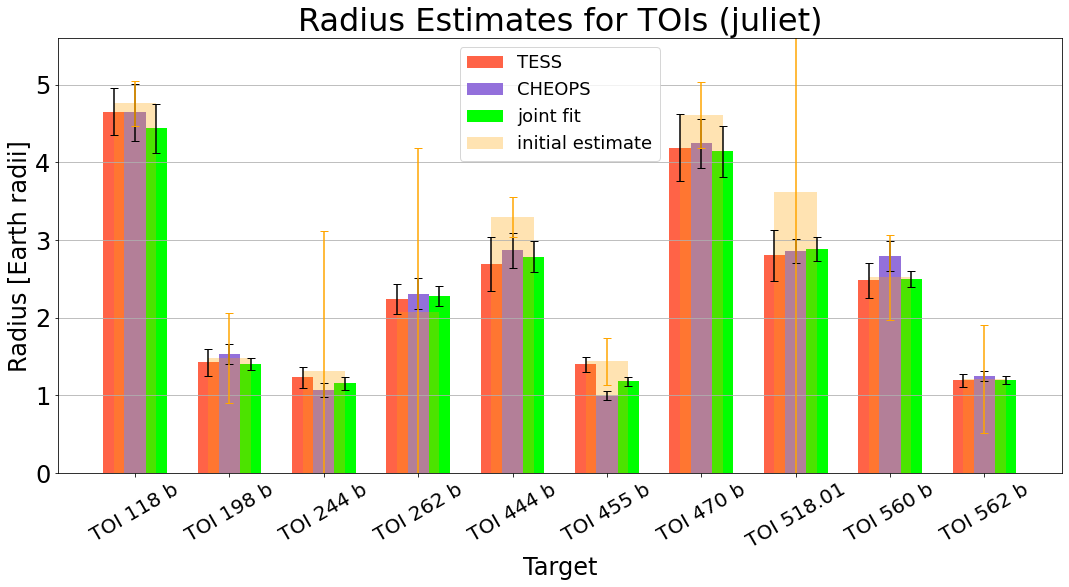}
 \includegraphics[width=.75\linewidth]{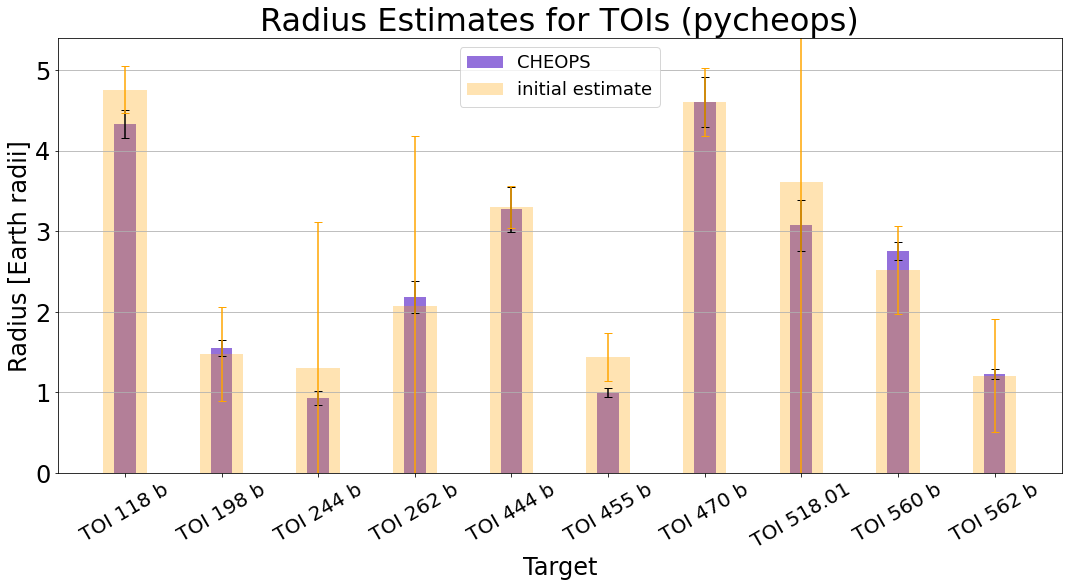}
 \caption{Radius estimates for our ten planets. Top: Radius estimates for our HOMEBREW method, which consists of fitting a \texttt{batman} transit model with \texttt{emcee}. Middle: Radius estimates using \texttt{juliet} nested sampling. Bottom: Fractional depth uncertainty using \texttt{pycheops} transit model with \texttt{emcee}. For all panels, models fitted to \tess\ data only are represented by red bars, models fitting to \cheops\ data only are represented by purple bars, and models fitting to both datasets jointly are represented by light green bars. The lighter-shaded orange bar represents the initial radius estimates from \tess\ PM sectors.}
 \label{fig:radii}
\end{figure*}

Here we compare our radius estimates for different fit methods to those from SPOC DV reports to their PM and EM sector light curves. We calculate planet radius by multiplying the radius ratio of the planet to the star as returned by our models by the radius of the star. For consistency, we use stellar radius values from published results for those systems which have been previously validated, and stellar radius values from our SED analysis of new systems, as in section \ref{sec:stell_pars}. SPOC DV report planet radii are calculated by multiplying the radius ratio by the stellar radius as listed in the TIC. Fig. \ref{fig:radii} shows our radius estimates and their uncertainties, propagating uncertainties from both stellar radius and uncertainties in model parameters. The top panel shows our radius estimates from fits using our HOMEBREW method, the middle panel shows our radius estimates from fits using \texttt{juliet}, and the bottom panel shows our radius estimates using \texttt{pycheops}. In all panels, red bars represent fits to \tess\ data alone, purple bars represent fits to \cheops\ data alone, and light green bars represent joint fits to both datasets. In each panel, we also include SPOC (a.k.a. TESS project) values and their uncertainties in lightly-shaded orange, which allows us a side-by-side comparison of our fits and their uncertainties to these values. 

In nearly every case, our fits and models yield a decrease in radius uncertainty relative to their initial uncertainties as calculated by the SPOC pipeline. The only instances in which the radius uncertainties we report with our models are larger than the SPOC radius estimates are our HOMEBREW fits to \tess\ data alone and \cheops\ data alone for TOI 118, as well as our \texttt{juliet} fits to \cheops\ data alone and our joint fit for TOI 118, as well as our \texttt{pycheops} fit to \cheops\ data for TOI 444. Notably, all of our radius estimates are reported with smaller uncertainties for our HOMEBREW joint fits. 

We can check consistency between different fits and methods for each system. To do so, we compare radius estimates within $1\sigma$. For our fits using the HOMEBREW method, our radius estimates are self-consistent for all systems, except for TOI 455 b. Whereas radius estimates generated with \cheops\ data, \tess\ data, or both for all other systems were consistent with one another, our radius estimates computed with \tess\ and \cheops\ separately are discrepant to nearly $4\sigma$ in the case of TOI 455 b as calculated with our HOMEBREW results. Further, discrepancies between the joint fit value and the value calculated with either dataset alone are discrepant to $\sim2\sigma$, where the joint fit estimate sits between the estimate calculated with either dataset alone. We discuss potential reasons for this in section \ref{subsec455}.

Similar to our HOMEBREW method, the only system which exhibits a radius estimate discrepancy as calculated with \texttt{juliet} is TOI 455 b, where radius estimates calculated from \tess\ photometry and calculated jointly are discrepant to $\sim2\sigma$. The radius estimate for this planet calculated with \cheops\ photometry is not consistent with those calculated with either \tess\ or jointly. We discuss potential reasons for this in the following section. 

The fits which most frequently exhibit the largest differences between radius estimates are those calculated with \texttt{pycheops} for \cheops\ photometry only and with our HOMEBREW method for \tess\ photmetry. In six out of ten cases, our fits with \texttt{pycheops} to \cheops\ photometry yield larger radius estimates relative to those fits calculated with our HOMEBREW method to \tess\ photometry. However, many of these fits are still consistent, with the exception of TOI 455 b. 

We report radius measurements to better than 10\% precision for all 10 of our systems even after incorporating uncertainties in stellar radius. This shows that high-precision results can be obtained from analysis of photometry alone, although our results would be improved by further high-precision characterization of the host stars. Future work could include combining our photometric analysis methods with more precise characterization of the host stars, such as with an ultra-high precision instrument like \emph{Gaia} or closer analysis of spectroscopic observations.

\subsection{Comparing \tess\ and \cheops\ photometry}\label{sec:comparison}

\begin{figure*}
    \centering
    \includegraphics[width=.85\linewidth]{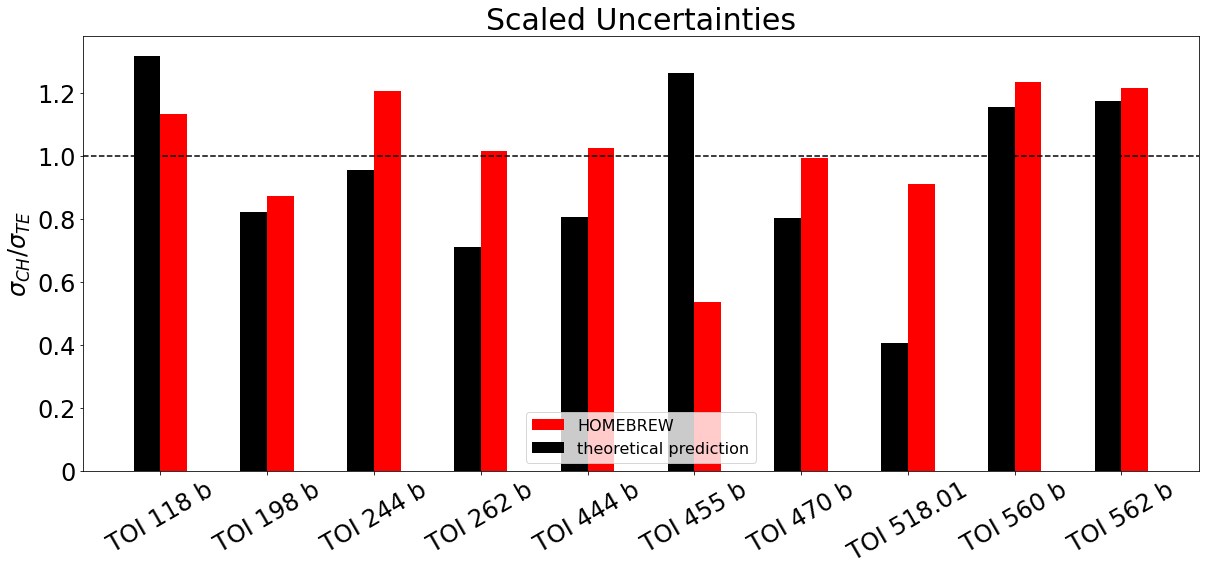}
    \caption{Ratio of depth uncertainties for models using our HOMEBREW method. The black bar for each planet represents our theoretical prediction for the uncertainty ratio, whereas the red bar represents our actual calculated depth uncertainty ratio from \cheops\ models vs \tess\ models.}
    \label{fig:uncert_ratio}
\end{figure*}

\begin{table*}
    \centering
    \caption{Values for calculation of theoretical depth uncertainty comparison, including \cheops\ observing efficiency $\epsilon$, number of transits captured by \cheops\ and \tess\, respectively, and ratio of flux of \tess\ to \cheops\ band. We also give number of equivalent transits, calculated according to \ref{eq:N_T_equiv}, and compare to transit depth and uncertainty as obtained with our HOMEBREW joint fits.}
    \begin{tabular}{c|c|c|c|c|c|c}
        System & $\epsilon_{CHEOPS}$ & N$_{tr,CHEOPS}$ & N$_{tr,TEESS}$ & $q$ & N$_{tr,equiv}$ & Depth [ppm]\\
        \hline
         TOI 118 b & 0.488 & 1 & 8 & 9.47 & 12.76 & $1414\pm68$ \\
         TOI 198 b & 0.955 & 1 & 4 & 6.23 & 5.50 & $894\pm64$ \\
         TOI 244 b & 0.744 & 2 & 8 & 5.89 & 3.70 & $773\pm95$ \\
         TOI 262 b & 0.870 & 1 & 4 & 9.10 & 4.46 & $484\pm66$ \\
         TOI 444 b & 0.864 & 1 & 5 & 8.95 & 5.50 & $1056\pm46$ \\
         TOI 455 b & 0.882 & 1 & 8 & 5.70 & 31.60 & $1665\pm82$ \\
         TOI 470 b & 0.700 & 1 & 4 & 8.89 & 5.80 & $2275\pm76$ \\
         TOI 518.01 & 0.610 & 2 & 2 & 10.02 & 1.98 & $605\pm64$ \\
         TOI 560 b & 0.680 & 1 & 7 & 7.72 & 6.77 & $1225\pm56$ \\
         TOI 562 b & 1.000 & 1 & 10 & 6.15 & 6.76 & $955\pm62$ \\
         \hline
    \end{tabular}
    \label{tab:uncert_params}
\end{table*}

Table \ref{tab:uncert_params} shows our values used for calculation of theoretical predictions for the uncertainty ratio as in equation \ref{eq:final_comp}. The table includes \cheops\ observing efficiency, number of transits captured \cheops\ and \tess\ for each target, and the SED-dependent flux ratio, which wraps contributions from both aperture size and stellar spectral type. Although the \cheops\ DRP gives us a value for observing efficiency which is the total number of points in the light curve over the number of points that should be in the light curve given the observing cadence of 60 s, we recalculate this value specifically for points in transit. Thus, our observing efficiency is calculated as the number of points in transit over the number of points that should be in transit assuming full coverage with the given the observing cadence. This is an important distinction in the case of TOI 470 b, for example, where we only captured a partial transit and had no pre-transit baseline.

The comparison between our theoretical uncertainty ratios given by eqn. \ref{eq:final_comp} and our actual uncertainty ratios are shown in Fig. \ref{fig:uncert_ratio}, where the black bar represents our theoretical value for the ratio of uncertainty in depth from \cheops\ photometry vs \tess\ photometry, which incorporates \cheops\ observing efficiency and number of transits captured by both telescopes. Our actual depth uncertainty ratios for our HOMEBREW fits are given by the red bar for each target. A horizontal dotted line has been placed at 1.0 to guide the eye, which represents a theoretical system with 1 \cheops\ transit with perfect observing efficiency and 9 \tess\ transits. In many cases, the theoretical prediction is below 1.0, meaning we would expect for model uncertainties as calculated with \cheops\ to be lower than those calculated with \tess\, even accounting for the number of transits and \cheops\ observing efficiency. However, given the relative number of photons collected by these telescopes in their respective bandpasses and the number of observed transits, there are four systems in which we may have expected slightly higher uncertainties as calculated with our \cheops\ light curves relative to \tess\ light curves. 

In order to judge relative performance and compare model uncertainties for each target, we compare the heights of the bars for each of our planets in Fig. \ref{fig:uncert_ratio}. Our model uncertainties should approach the theoretical values, and thus be represented by the black bars, but are represented in reality by the red bars. The metric for comparison is uncertainty in depth calculated with \cheops\ photometry vs that calculated with \tess\, so a red bar which is higher than the black indicates that the uncertainty in depth as calculated with \cheops\ is higher than theoretically expected. In eight of ten cases, we see that our calculated uncertainty ratio is higher than our theoretical value, indicating that our model uncertainties as calculated with \cheops\ are slightly higher than predicted, even when incorporating the number of transits and the observing efficiency of \cheops. In two cases, including TOI 118 and TOI 455, we see that the uncertainty as calculated with our \cheops\ light curves is lower than expected. This may be due to one of two causes. First, this may indicate that our model uncertainties as calculated with CHEOPS photometry were smaller than predicted. Conversely, this may indicate that our model uncertainties as calculated with TESS photometry were larger than expected given the number of transits we observed. In the case of TOI 118 b, we expect that this uncertainty ratio should be $\frac{\sigma_{CH}}{\sigma_{TE}}\sim1.3$, which may have been informed by a low measure for \cheops\ in-transit visit efficiency for this system at $\epsilon_{CHEOPS}=0.488$. However, we found an uncertainty ratio $\frac{\sigma_{CH}}{\sigma_{TE}}\sim1.1$, meaning that our measure for uncertainty in the \cheops\ depth was lower than predicted relative to \tess. We believe this may have been due to extremely low noise in our light curve, which exhibited a MAD = 220 ppm in the detrended light curve, as well as solid in-transit and out-of-transit baselines. Therefore, despite the poor transit coverage, which would otherwise inflate the expected uncertainty, the reported uncertainty was low.

We also examine the case of TOI 455 b more closely. We believe evidence supports the second case, where model uncertainties calculated with \tess\ photometry were inflated relative to those calculated with \cheops\ photometry. The fact that this is a hierarchical triple-star system complicates the light curve immensely and would dilute the transit signals, which may have led to actual differences in transit depth between different transits in our \tess\ light curve. This in turn would increase model uncertainties as convergence would be more difficult to reach. On the other hand, given that we only had one \cheops\ transit to analyze, convergence would more easily be reached in this case. In general, this showcases the impact that contamination from nearby stars can have on results and their uncertainties, and thus we recommend careful treatment of contamination in future studies.

Further, in three out of six cases when we expected that model uncertainties calculated with \cheops\ photometry would be lower than those calculated with \tess\ photometry, we actually find that our uncertainty ratio is higher than 1.0, indicating that model uncertainties as calculated with \tess\ were lower in these instances. These findings would indicate that our models fitted to \cheops\ photometry yield higher uncertainties than we might have predicted, which has implications for fitting models to \cheops\ photometry in the future. We believe that this finding is due to two primary reasons, including 1. important data gaps in \cheops\ visits, and 2. the way in which both \cheops\ and \tess\ light curves are generated and detrended. In a few instances, we saw inflated model uncertainties in fits to \cheops\ light curves which may have been the result of gaps in the transit in important places. These gaps are difficult to account for and are a consequence of the \cheops\ observing strategy and its low-earth orbit. On the other hand, we may have seen larger-than-expected errors on models fitted to \cheops\ photometry due to the way in which these light curves are detrended. Each of our \cheops\ light curves were detrended individually by introducing detrending parameters one-by-one, meaning there is no standard method with which to detrend \cheops\ photometry. On the other hand, \tess\ light curves are all processed by the same pipeline which treats the data uniformly. We believe that while the methods we used to detrend \cheops\ data from both instrumental and astrophysical systematics (outlined in section \ref{sec:meth}) were effective, they may not have treated the data uniformly, which would introduce errors relative to a uniform treatment of \tess\ data.

\subsubsection{How many \tess\ transits \it{do} we need to reach the same precision as our \cheops\ transits?}

\begin{figure*}
    \centering
    \includegraphics[width=.8\linewidth]{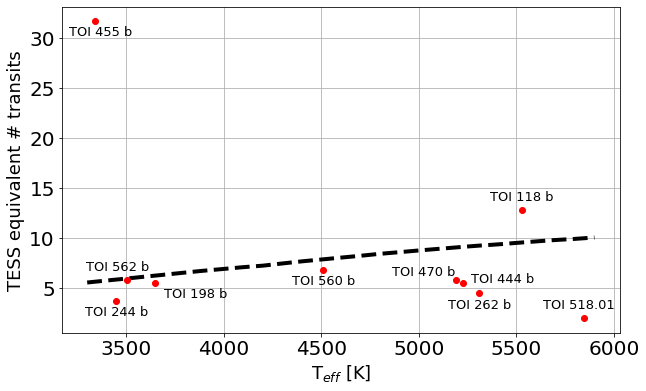}
    \caption{Scatter plot showing equivalent transits vs stellar effective temperature as calculated with our HOMEBREW joint fits, from Table \ref{tab:uncert_params}. The dashed line shows our theoretical expectation for the photon ratio $q$ from eqn. \ref{eq:q}, which is equivalent.}
    \label{fig:equiv_transits}
\end{figure*}
 
Fig. \ref{fig:equiv_transits} shows the number of \tess\ transits necessary to match the precision we obtained on our \cheops\ transits, where our sample of planets is shown as a series of red dots. These are directly calculated from equation \ref{eq:N_T_equiv}. We compare to the theoretical expectation, $q$, which depends on stellar spectral type, shown as a black dashed line. For points which sit below the line, we obtained at least enough \tess\ transits to match the precision of our \cheops\ transits, but for the two systems above the line, we would require $N_{tr,equiv} - N_{tr,TESS}$ more \tess\ transits to match the precision of our \cheops\ transits. Clearly, according to our calculations we would require many more transits with \tess\ to match the reported precision with our \cheops\ transit for TOI 455 b, where $N_{tr,equiv} = 31.6$. However, given that there is clear evidence of blended light in our light curves for this target, we do not treat this estimate as physical. Rather, this represents further evidence that more careful treatment of contamination is necessary in future studies of this system.

\citet{Bonfanti2021CHEOPS} also report equivalent number of \tess\ transits by comparing uncertainty in depth for their model fits to \tess\ data and \cheops\ data. For a V mag $\sim9$ star, they reported that their precision on one \cheops\ transit with a depth of 500 ppm was equivalent to 8 \tess\ transits, whereas the precision for a \cheops\ transit of depth 250 ppm was equivalent to seven \tess\ transits, and the precision on a \cheops\ transit of depth 1000 pm was equivalent to two \tess\ transits. However, given that our analysis differed from theirs in the inclusion of theoretical photon noise limits (i.e. calculation of $q$), we cannot make a straight-forward comparison between our results and theirs. 

Finally, Table \ref{tab:uncert_params} shows the depth reported by our HOMEBREW joint fits in comparison. We see no trend with transit depth, supporting our prior claim that greater uncertainty in modeling \cheops\ transits may stem from data gaps and non-homogeneous detrending of \cheops\ light curves, rather than system parameters.

\section{Discussion}\label{sec:disc}

\begin{sidewaystable*}
\vspace{270pt}
\caption{Final fitted physical and orbital parameters for all systems, from HOMEBREW joint fits.}
\begin{tabular*}{\linewidth}{cl|ccccc}\label{tab:all_params}
%\hline
Parameter & Unit & \textbf{TOI 118 b} & \textbf{TOI 198 b} & \textbf{TOI 244 b} & \textbf{TOI 262 b} & \textbf{TOI 444 b} \\ \hline
\multicolumn{1}{l}{\textbf{Stellar Parameters:}} &  & \multicolumn{1}{l}{} & \multicolumn{1}{l}{} & \multicolumn{1}{l}{} & \multicolumn{1}{l}{} & \multicolumn{1}{l}{} \\
CHEOPS limb darkening & $c_1$ & 0.695 & 0.657 & 0.762 & 0.706 & 0.716 \\
CHEOPS limb darkening & $c_2$ & 0.826 & 0.591 & 0.539 & 0.852 & 0.848 \\
TESS limb darkening & $c_1$ & 0.627 & 0.700 & 0.712 & 0.637 & 0.651  \\
TESS limb darkening & $c_2$ & 0.722 & 0.685 & 0.468 & 0.746 & 0.746  \\
Stellar Radius & R$_s$ (R$_\sun$) & $1.03\pm0.03$ & $0.441\pm0.019$ & $0.399\pm0.019$ & $0.853\pm0.021$ & $0.779\pm0.053$ \\
Stellar Mass & M$_s$ (M$_\sun$) & $0.92\pm0.03$ & $0.467\pm0.023$ & $0.424\pm0.021$ & $0.913\pm0.029$ & $0.96\pm0.13$ \\
Effective Temperature & T$_{eff}$ (K) & $5527\pm65$ & $3650\pm75$ & $3450\pm75$ & $5310\pm124$ & $5225\pm70$  \\ \hline
\multicolumn{1}{l}{\textbf{Orbital \& Transit Parameters}} &  &  &  &  &  &  \\
Orbital Period & days & $6.04345\pm$1e-5 & $10.2152\pm$6e-5 & $7.39726\pm$2e-5 & $11.14529\pm$3e-5 & $17.96360\pm$4e-5 \\
Time of mid-transit & BTJD & $2083.5109\pm0.0011$ & $2480.048\pm0.0004$ & $2489.1482\pm0.0025$ & $2136.5766\pm0.0010$ & $2190.0391\pm0.0008$ \\
Scaled radius & $\frac{Rp}{Rs}$ & $0.0376\pm0.0009$ & $0.0299\pm0.0022$ & $0.0278\pm0.0017$ & $0.0220\pm0.0015$ & $0.0325\pm0.0007$  \\
Scaled semi-major axis & $\frac{a}{Rs}$ & $21.747\pm2.023$ & $49.086\pm3.003$ & $39.663\pm7.539$ & $41.211\pm8.024$ & $36.768\pm2.283$ \\
Inclination angle & i (deg) & $88.938\pm0.505$ & $89.890\pm0.102$ & $89.206\pm0.657$ & $89.011\pm0.413$ & $89.647\pm0.314$ \\
Transit duration & T$_{dur}$ (hrs) & $2.028\pm0.194$ & $1.631\pm0.155$ & $1.233\pm0.118$ & $1.501\pm0.143$ & $3.761\pm0.356$ \\ \hline
\multicolumn{1}{l}{\textbf{Physical Planet Parameters:}} &  &  &  &  &  &   \\
Impact parameter &  & $0.402\pm0.193$ & $0.094\pm0.086$ & $0.554\pm0.494$ & $0.719\pm0.328$ & $0.225\pm0.193$ \\
Semi-major axis & $a$ (AU) & $0.104\pm0.005$ & $0.100\pm0.006$ & $0.073\pm0.003$ & $0.163\pm0.002$ & $0.133\pm0.006$ \\
Radius & R$_p$ (R$_{\earth}$) & $4.24\pm0.16$ & $1.44\pm0.08$ & $1.03\pm0.08$ & $2.07\pm0.15$ & $2.77\pm0.20$ \\
Predicted Mass & M$_p$ (M$_{\earth}$) & $8.1^{+12.5}_{-4.8}$ & $4.0\pm1.1$\footnote{Derived from the Keplerian fit to our ESPRESSO RVs.} & $0.8^{+2.0}_{-0.6}$ & $5.5^{+7.9}_{-3.2}$ & $6.6^{+9.3}_{-3.8}$ \\
Equilibrium Temperature & T$_{eq}$ (K) & $838\pm40$ & $368\pm26$ & $387\pm30$ & $584\pm45$ & $609\pm61$ 
\end{tabular*}

\begin{tabular*}{\linewidth}{cl|ccccc}%\label{tab:470_518_params}
%\hline
Parameter & Unit & \textbf{TOI 455 b} &\textbf{TOI 470 b} & \textbf{TOI 518.01} & \textbf{TOI 560 b} & \textbf{TOI 562 b}   \\ \hline
\multicolumn{1}{l}{\textbf{Stellar Parameters:}} &  & \multicolumn{1}{l}{} & \multicolumn{1}{l}{} & \multicolumn{1}{l}{} & \multicolumn{1}{l}{} & \multicolumn{1}{l}{}  \\
CHEOPS limb darkening & $c_1$ & 0.824 & 0.709 & 0.705 & 0.733 & 0.751   \\
CHEOPS limb darkening & $c_2$ &  0.484 & 0.778 & 0.745 & 0.881 & 0.578   \\
TESS limb darkening & $c_1$ & 0.778 & 0.635 & 0.626 & 0.675 & 0.707 \\
TESS limb darkening & $c_2$ & 0.415 & 0.689 & 0.658 & 0.771 & 0.499  \\
Stellar Radius & R$_s$ (R$_\sun$) & $0.265\pm0.011$ & $0.831\pm0.021$ & $1.027\pm0.025$ & $0.65\pm0.02$ & $0.337\pm0.015$ \\
Stellar Mass & M$_s$ (M$_\sun$) & $0.257\pm0.014$ & $0.87\pm0.09$ & $1.07\pm0.06$ & $0.73\pm0.02$ & $0.342\pm0.011$ \\
Effective Temperature & T$_{eff}$ (K) & $3340\pm150$ & $5190\pm90$ & $5845\pm70$ & $4511\pm110$ & $3505\pm51$ \\ \hline
\multicolumn{1}{l}{\textbf{Orbital \& Transit Parameters}} & &  &  &  &  &  \\
Orbital Period & days & $5.35876\pm$1e-5 & $12.19148\pm$3e-5 & $17.87712\pm$7e-5 & $6.39805\pm$1e-5 & $3.93060\pm$2e-6 \\
Time of mid-transit & BTJD & $2152.2189\pm0.0005$ & $2205.9825\pm0.0011$ & $2568.4107\pm0.0014$ & $2240.6702\pm0.0007$ & $2272.6757\pm0.0004$ \\
Scaled radius & $\frac{Rp}{Rs}$ & $0.0408\pm0.0010$ & $0.0477\pm0.0008$ & $0.0246\pm0.0013$ & $0.0350\pm0.0007$ & $0.0309\pm0.0010$ \\
Scaled semi-major axis & $\frac{a}{Rs}$ & $29.963\pm3.128$ & $30.917\pm2.369$ & $47.733\pm4.451$ & $23.743\pm3.338$ & $22.890\pm1.211$ \\
Inclination angle & i (deg) & $89.203\pm0.658$ & $89.527\pm0.432$ & $89.723\pm0.226$  & $89.428\pm0.441$ & $89.228\pm0.483$ \\
Transit duration & T$_{dur}$ (hrs) & $1.392\pm0.133$ & $3.063\pm0.292$ & $2.844\pm0.228$ & $2.069\pm0.197$ & $1.291\pm0.123$ \\ \hline
\multicolumn{1}{l}{\textbf{Physical Planet Parameters:}} &  &  &  &  &  &  \\
Impact parameter &   & $0.840\pm0.222$ & $0.253\pm0.216$ & $0.249\pm0.169$ & $0.248\pm0.171$ & $0.307\pm0.160$  \\
Semi-major axis & $a$ (AU) & $0.022\pm0.003$ & $0.119\pm0.008$ & $0.227\pm0.002$ & $0.072\pm0.002$ & $0.036\pm0.0004$ \\
Radius & R$_p$ (R$_{\earth}$) & $1.18\pm0.06$ & $4.34\pm0.29$ & $2.77\pm0.16$ & $2.49\pm0.10$ & $1.20\pm0.06$ \\
Predicted Mass & M$_p$ (M$_{\earth}$) & $0.9^{+1.7}_{-0.5}$ & $8.4^{+7.5}_{-5.6}$ & $6.7^{+7.3}_{-4.3}$ & $5.7^{+8.3}_{-3.3}$ & $0.8^{+2.5}_{-0.5}$ \\
Equilibrium Temperature & T$_{eq}$ (K) & $555\pm48$ & $660\pm70$ & $598\pm55$ & $655\pm41$ & $518\pm44$
\end{tabular*}
\end{sidewaystable*}

\subsection{Contextualizing these systems}

Here we report physical and orbital properties for each of our planets/planet candidate individually. Our final reported properties are shown in Table \ref{tab:all_params}. We highlight instances of fractional depth uncertainties which are larger than 10\%, and discuss these results in context with the \tess\ and \cheops\ light curves and our fitting methods, where appropriate. Additionally, with the exception of TOI 198, we report predicted masses of these planets using the non-parametric formulation from \citet{Ning_2018}, who developed R code\footnote{\url{https://github.com/Bo-Ning/Predicting-exoplanet-mass-and-radius-relationship}} using the relations therein, which was also translated to the Python package \texttt{MRExo}\footnote{\url{https://github.com/shbhuk/mrexo}}. For planets which are orbiting M dwarfs, we incorporate mass predictions from \citet{Kanodia_2019}, which is also wrapped into \texttt{MRExo}. For classification of planets, we use the framework of \citet{Chen_2016}, who categorized planets as 'Terran', 'Neptunian', 'Jovian', or 'Stellar' defined by mass cutoffs at 2 M$_{\oplus}$, 0.41 M$_{J}$, 0.08 M$_{\sun}$, respectively. We report these masses and uncertainties to one decimal place, with the understanding that these are not well-constrained values. In the case of TOI 198 b, our ESPRESSO RVs allowed us an estimation of the planet's mass, so we use this mass estimate when calculating density. Mass and bulk density are important parameters to constrain for all of these planets, as these values would help to contextualize formation and evolution for these systems. However, we leave characterization of planet mass to future work, as it is beyond our present scope.

In our reporting of our final parameters for each of these planets/planet candidate we use the HOMEBREW joint fit. This represents a reasonable choice since a joint fit uses all available data, while our HOMEBREW method of fitting was constructed to yield one global set of model parameters. Further, this fit method yielded lowest overall uncertainties.

\subsubsection{TOI 118 b}

We find that TOI 118 b is a Neptunian world orbiting a sun-like star on a 6.034 d orbit. We find it has a radius of $4.24 \pm 0.16$ R$_{\earth}$ and a predicted mass of $8.1^{+12.5}_{-4.8}$ M$_{\earth}$. TOI 118 b's size places it well above the radius valley, but given the planet's short $\sim6$ day orbit, it may be experiencing photo-evaporation of its outer atmospheric layers. SPOC characterization of this planet with \tess\ placed it in a similar part of parameter space on the period-radius diagram, but we improved the radius estimate of this planet by a factor of two relative to its initial uncertainty. \citet{Esposito2019HD219666} reported the radius of this planet as $4.71\pm0.17$ R$_{\earth}$, which is discrepant with our radius estimate to $\sim1.7\sigma$. Interestingly, our joint characterization with HOMEBREW was the only fit method which exhibited a $\geq1\sigma$ discrepancy; our other results for this planet were either larger radius measurements or exhibited larger uncertainties. Further precise photometric characterization is needed to reconcile these differences for this planet.

We report the lowest fractional depth uncertainty for this target for our joint fit using the HOMEBREW method, and our \texttt{pycheops} fit to \cheops\ data and \texttt{juliet} to \tess\ data exhibit similarly low fractional uncertainties in depth. However, we report fractional depth uncertainties at greater than 10\% for both the fits to \cheops\ data alone for this target. This may be due to a relatively low in-transit observing efficiency for the \cheops\ visit to this system of 48.8\%, where significant gaps in the \cheops\ light curve as a result of Earth occultations of the star may be the cause of the large fractional uncertainty relative. 

Errors may have been introduced in joint fitting relative to \tess\ data alone due to a perceived difference in depth in our \texttt{juliet} fit. A perceived difference in transit depth between \tess\ and \cheops\ may be a result of the slight differences in bandpass coverage between these two telescopes. The \tess\ bandpass is optimized for nearby M dwarf stars, whereas \cheops\ has a  bluer filter optimized for sun-like stars. We computed two pairs of limb darkening coefficients for each star for these two bandpasses, but this may not fully account for a discrepancy. Given that a star may appear brighter or dimmer in the \tess\ bandpass relative to the \cheops\ bandpass, computed transit depths may appear to be different in these bands relative to one another. This may, in part, account for the discrepancy between fits to these light curves. However, as the fractional uncertainty in transit depth for joint fits to both datasets are not statistically different for this target, we cannot meaningfully make the claim that our errors were influenced by this effect. Further, a relatively large fractional depth uncertainty when jointly fitting both datasets is only seen in our \texttt{juliet} fit. We report the lowest overall fractional depth uncertainty for this target for our HOMEBREW joint fit. 

\subsubsection{TOI 198 b}

We find that TOI 198 b is a Terran world orbiting an early M dwarf star on a 10.215 d orbit. We find it has a radius of $1.44 \pm 0.08$ R$_{\earth}$ and a mass from our ESPRESSO RVs of $4.0\pm1.1$ M$_{\earth}$, making its likely mean density 7.3 g cm$^{-3}$. This size and density indicate that this planet is likely a dense super-Earth. Using the non-parametric mass from above, we predict its mass as $1.3^{+6.0}_{-1.0}$ M$_{\earth}$, which encompasses the measurement from our ESPRESSO RVs.

Given TOI 198 b's radius and orbital period, it sits just below the radius valley, as shown in Fig. \ref{fig:period-radius}. This density and position in period-radius space would indicate that this planet is likely a bare rock which either never formed with a gaseous envelope or lost such an envelope quickly after formation. Interestingly, initial characterization of this planet with \tess\ PM photometry upon discovery could not resolve whether it was in the gap or not, but our revised radius estimate places it more precisely.

The fractional depth uncertainty for TOI 198 b fitted with \tess\ data alone is above 10\% for fits with our HOMEBREW method and with \texttt{juliet}, whereas the fractional uncertainty with \cheops\ data alone is below this threshold for our HOMEBREW fit but not our \texttt{juliet} fit, with the fractional uncertainty in the joint dataset being lower still in both cases. This relatively large uncertainty in the fit to \tess\ data may have been a result of noise in the light curve, which had a Mean Absolute Difference (MAD) after detrending of 1649 ppm across both sectors, but a higher MAD of 1901 ppm in the EM1 sector for this star. We calculated a transit depth of $\sim 1000$ ppm for this planet, meaning that the scatter in the light curve was approximately twice as large in magnitude as the transit signal itself. Therefore, it is sensible that fits to \tess\ data alone would yield larger uncertainties for this target, given that the noise in \cheops\ photometry was 522 ppm after detrending, representing a significant reduction in noise. Further, only four transits were captured in \tess\ data, potential leading to an increase in uncertainty given the low number of transits. Finally, this relatively large uncertainty in our model fit may have been a result of a degeneracy between transit depth and impact parameter, particularly for the \texttt{juliet} fit to the \tess\ photometry, where orbital parameters were computed separately from those computed for fits to the \cheops\ light curve. A high impact parameter would indicate that the planet is transiting away from the stellar midline, which would would inflate the planet-to-star radius ratio for a constant transit depth. In particular, our \texttt{juliet} fit yields a higher impact parameter for \tess\ photometry relative to the fit to \cheops\ photometry, potentially accounting for this higher uncertainty in transit depth.

\subsubsection{TOI 244 b}

Similar to TOI 198 b, we find that TOI 244 b is a Terran world orbiting an M dwarf star, with an orbital period of 7.397 d. We find it has a radius of $1.03 \pm 0.08$ R$_{\earth}$ and a predicted mass of $0.8^{+2.0}_{-0.6}$ M$_{\earth}$. We report the predicted mass of TOI 244 b using the nonparametric function for M dwarf exoplanets from \citet{Kanodia_2019}, which predicts smaller masses for small planets compared to the \emph{Kepler} sample, which is comprised of primarily sun-like stars. Previous studies indicate that even when accounting for observational biases of the \emph{Kepler} mission, M dwarfs typically yield more small planets compared to FGK stars and fewer giant planets \citep{Mulders_2015}. The planet's short period may indicate that photoevaporation may have played a significant role in stripping the planet of a gaseous envelope, if it ever accreted one. Further precise characterization of this planet, particularly with extreme-precision RVs (EPRVs) may illuminate its properties, and thus its formation and evolutionary history, in greater detail. Initial estimation of this planet's radius with \tess\ PM photometry was imprecise, and determination of its position relative to the radius valley was not originally possible. Our period and radius estimates place this planet squarely below the radius valley in the Earth-sized regime. This represents an interesting system which includes an Earth-sized planet in a close orbit around an M dwarf star.

We note that the fractional uncertainty in two \cheops\ visits to TOI 244 b was larger than 10\% for all three fitting methods. This may be due to the relatively high level of noise in our \cheops\ light curves for this target, as shown in Fig. \ref{fig:CHEOPS_noise}. When fitting \cheops\ photometry with our HOMEBREW method, the fractional depth uncertainty is larger than the fractional uncertainty in the fit to \tess\ data alone. In the case of fitting our \tess\ and \cheops\ datasets independently with \texttt{juliet}, the fractional depth estimates are greater than 10\%. In the case of \texttt{juliet}, jointly fitting both photometric datasets results in lower than 10\% fractional uncertainty, representing an improvement, but this threshold is not met by our joint fit for the HOMEBREW method. This may have been a result of the gaps in both of our \cheops\ visits, which are due to Earth occultations. A significant gap in \cheops\ coverage during our first visit occurred near the mid-point of the transit, which may have made the bottom of the transit difficult to identify. Further, a significant gap in \cheops\ coverage occurred during our second visit during the transit egress, which could have obscured the general shape of the transit and increased the uncertainty. These gaps may have contributed to an overall degeneracy between planet size and impact parameter, increasing model uncertainties for both quantities. It is also notable that our \cheops\ light curves exhibited high noise on timescales which are relevant to characterization of transits, as shown in Fig. \ref{fig:CHEOPS_noise}. Further, noise in the \tess\ light curve, which had a MAD of 2146 ppm across both sectors, may have contributed to a relatively large fractional uncertainty in this case, given that the calculated transit depth was smaller by a factor of two, which is similar to the case of TOI 198 b. However, it seems this uncertainty was slightly more constrained during model fitting to both datasets jointly for both fitting methods. 

\subsubsection{TOI 262 b}

We find that TOI 262 b is a Neptunian world orbiting a sun-like star with an orbital period of 11.145 d. We find it has a physical size of $2.07 \pm 0.15$ R$_{\earth}$ and a predicted mass of $5.5^{+7.9}_{-3.2}$ M$_{\earth}$. Similar to TOI 244 b, the position of this planet was not well constrained with \tess\ PM photometry, as it could have been either a sub-Neptune, a super-Earth, or a planet in the radius valley. TOI 262 b represents an interesting case which closely borders - or perhaps falls into - the radius valley. While our models do not conclusively place this planet outside of the radius valley, we are able to more tightly constrain its radius compared to prior estimates.

We report a fractional depth uncertainty of 10\% or larger for all fits to TOI 262 photometry, regardless of dataset or fit method (where our joint fit to both datasets with \texttt{juliet} yields a fractional depth of uncertainty of precisely 10\%). When calculating fractional uncertainties, a shallower transit with a similar uncertainty to a deeper one will yield a relatively larger fractional uncertainty. As shown by our calculated transit depths in Table \ref{tab:depths}, our models vary around $\sim500-600$ ppm for this target. As such, our model uncertainties, which are otherwise comparable in magnitude to those for other systems, mean that the fractional uncertainty is slightly inflated for this target. 

Further, the noise in the \tess\ light curve was relatively low at MAD $\sim 400$ ppm and the noise in the detrended \cheops\ light curve was lower than this by a factor of two. These are indicative of light curves with low amounts of scatter, and we might expect for our models to be very well-constrained. Additionally, our light curves in and out of transit are well-sampled and systematics appear to be well-constrained given the apparent flatness of the light curves, further giving credence to the notion that we may have expected our models to be well-behaved. However, for all of our models, our comparatively large uncertainties arose from degeneracy between transit depth and impact parameter. For all of our models, we report an impact parameter $b \geq 0.7$, with large uncertainties in both $\frac{a}{R_*}$ and inclination angle. A high impact parameter would indicate the planet is transiting towards the edge of the stellar disk. Even though we hold limb darkening coefficients constant, the prospect of transiting away from the stellar midline introduces more error in our models, thus contributing to the larger-than-expected uncertainties in depth.

\subsubsection{TOI 444 b}

We find that TOI 444 b is a Neptunian world orbiting a K dwarf star on a 17.964 d orbit. We find it has a radius of $2.77 \pm 0.20$ R$_{\earth}$ and a predicted mass of $6.6^{+9.3}_{-3.8}$ M$_{\earth}$. Our characterization of the planet's period and radius place it firmly in the sub-Neptune regime. Our findings match previous findings from SPOC PM sector models, which initially constrained this planet to the sub-Neptune regime. 

While all of our fits with the HOMEBREW method yield relatively small fractional depth uncertainties, our fit to the \tess\ light curves with \texttt{juliet} yield a fractional depth uncertainty greater than 10\%. Our joint fit and \cheops\ fit with \texttt{juliet} yield similarly small uncertainties as our HOMEBREW method, making the aforementioned \texttt{juliet} fit to \tess\ data somewhat anomalously large. This may be a result of the way in which \texttt{juliet} handles orbital parameters separately for our individual fits. Specifically, since we compute one set of orbital parameters but different transit depths for our HOMEBREW fits, our errors on these calculated depths are propagated in the same way for both datasets. This is not true for our individual fits with \texttt{juliet}, which may have been the cause of such a large uncertainty in depth for our \tess\ fit for this target. As shown in Table \ref{tab:all_params}, the impact parameter for our joint HOMEBREW fit is relatively low. However, for our \texttt{juliet} fit to \tess\ data for this target, the calculated impact parameter was significantly larger and carried a larger uncertainty. Given the aforementioned degeneracy between transit depth and impact parameter, this may have contributed to a larger uncertainty on transit depth for this particular fit.
%I am almost certain there's a better way to say all this, but succinct-ness is not on the agenda right now clearly

\subsubsection{TOI 455 b}\label{subsec455}

\begin{figure}
    \centering
    \includegraphics[width=\linewidth]{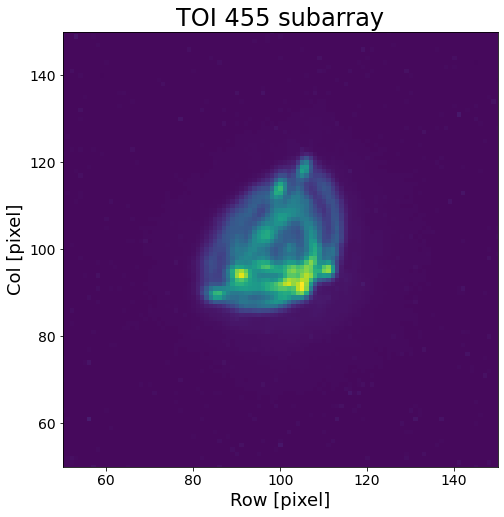}
    \caption{One zoomed-in frame from the \cheops\ subarray for TOI 455 (TIC 98796344), which clearly shows two sources in the center of the image.}
    \label{fig:TOI455_subarray}
\end{figure}

We find that TOI 455 b is a Terran world orbiting an M dwarf star on a 5.359 d orbit. We find it has a radius of $1.18 \pm 0.06$ R$_{\earth}$ and a predicted mass of $0.9^{+1.7}_{-0.5}$ M$_{\earth}$.

Our results yield no large fractional depth uncertainties above the 10\% threshold, indicating our results were well-constrained. However, as previously mentioned, our radius measurements are discrepant from one another in many cases for this planet. Interestingly, this planet has been found to be orbiting an M dwarf which is part of a hierarchical triple system \citep{Winters2019LTT,Winters_2022}. Therefore, it is likely that the transits we detect have been severely contaminated by light from its partner stars, as shown by Fig. \ref{fig:TOI455_subarray}. Since we were not able to use \texttt{PIPE} to satisfactorily disentangle the strongly confused nearby companion from TOI-455, we instead analyzed the aperture integrated signal of both components as provided by the DRP. We detrended our \cheops\ and \tess\ light curves as specified in section \ref{sec:meth}, but our discrepant radius measurements may mean that our detrending contributed to these differences. Whereas we subtracted and corrected for background flux in the aperture in our \cheops\ detrending, we merely flattened the \tess\ light curve, which accounted for short-term variations in flux due to stellar eclipses. This could have led the transits in the \tess\ light curve to appear systematically deeper as compared to \cheops\. This differing treatment may have led to an over-estimation of the planet radius when examining \tess\ photometry and an under-estimation of the planet radius when examining \cheops\ photometry. 

Indeed, \citet{Winters2019LTT} reported a radius of this planet of $1.38_{-0.12}^{+0.13}$ R$_{\earth}$ and \citet{Winters_2022} later constrained the radius to $1.30 \pm 0.06$ R$_{\earth}$, both of which are consistent with our final reported value, which uses both datasets. However, these estimates are not consistent with our estimates using \tess\ or \cheops\ alone, which are systematically too high or too low, respectively. %This can most clearly been seen in Fig. \ref{fig:TOI455_pfs}, which shows our models overlaid on \cheops\ phase-folded photometry (top panel), \tess\ phase-folded photometry (middle panel), and both datasets (bottom panel). 
It is apparent that while our models that fit \tess\ data or \cheops\ data separately match each individual dataset well, our joint models fall between these datasets, suggesting the final value is influenced by both datasets nearly equally. 

We believe this represents a cautionary tale. We believe it is beyond the scope of this work to systematically account for contamination from nearby sources in both \tess\ and \cheops\ bandpasses, as it was our goal to demonstrate the effects of base-level detrending. However, we believe that fully accounting for contamination from nearby stars is warranted, as it may have helped assuage differences in transit depth in this case.

\subsubsection{TOI 470 b}

We find TOI 470 b is a Neptunian world orbiting a late G dwarf on a 12.191 d orbit. We report that it has a radius of $4.34\pm0.29$ R$_{\earth}$ and a predicted mass of $8.4^{+7.5}_{-5.6}$ M$_{\earth}$. This size supports the claim that this planet is an ice giant world which has held onto its gaseous envelope and may be comprised of as much as 2\% H$_2$ by mass \citep{doi:10.1073/pnas.1812905116}. However, further precise characterization with EPRVs of this planet will illuminate its properties.

We report no models which have a fractional depth uncertainty greater than 10\%. Our single \cheops\ visit for the target missed the transit ingress, and as such as we report an in-transit observing efficiency of only 70\%. This means we did not have a baseline for the flux prior to the beginning of the transit, which may have contributed to some uncertainty regarding the true depth of the planet. However, it seems our models did not significantly suffer from this lack of pre-transit baseline. This may have been aided by the fact that there was a very low amount of noise in the \cheops\ light curve, which had a MAD of 557 ppm after detrending.

\subsubsection{TOI 518.01}

TOI 518.01 is the only system which we were not able to validate as a planet. As such, we treat conclusions drawn for this system with more caution, as these results will clearly depend on whether the system is later validated. Should these signals be validated, we find that TOI 518.01 may be a Neptunian world orbiting an early G dwarf star on a 17.877 d orbit. We report a potential radius of $2.77 \pm 0.16$ R$_{\earth}$ and a predicted mass of $6.7^{+7.3}_{-4.3}$ M$_{\earth}$. Given the size estimate for this candidate planet, this would be a world which has retained its gaseous envelope, and it sits well above the radius valley.

The model fit to \tess\ data alone for TOI 518.01 yielded the largest fractional depth uncertainty for our HOMEBREW method. The fractional depth uncertainties for HOMEBREW fits to \cheops\ and both datasets jointly on this target were 14.2\% and 10.5\%, respectively, but the fractional uncertainty in the depth for \tess\ data alone was 21.2\%, which also represented the largest overall fractional depth uncertainty for any fit method or dataset combination. This may be due to the fact that this was the only TOI for which we were only able to use one \tess\ sector, given that our attempts to salvage the EM sector (contaminated with stray light) for TOI 518 were unsuccessful. Therefore, we were only fitting photometry which included two transits in a light curve which had a noise level of MAD $\approx 1000$ ppm, which is larger than the transit depth.

Additionally, relatively large uncertainties in fits to our two \cheops\ visits may be due to important data gaps while the planet was in transit, which may obscure the true shape of the transit and thus contribute to a larger uncertainty. These data gaps obscured crucial parts of our two transits, including transit ingress and parts of the bottom of the transit. 

\subsubsection{TOI 560 b}

\begin{figure}
    \centering
    \includegraphics[width=\linewidth]{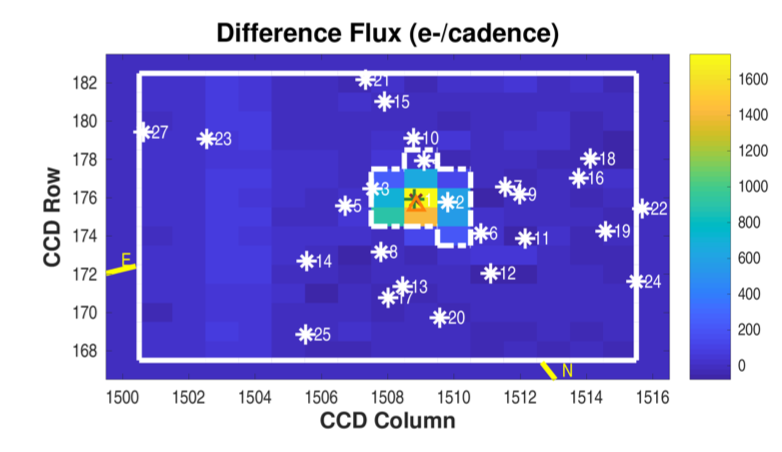}
    \caption{Star field around TOI 560 (TIC 101011575) from the SPOC report for PM sector 8 for this star. There are 7 TIC entries within 60. arcseconds of TOI 560, which may be evidence of possible contamination from nearby field stars, which would dilute the transits in the \tess\ light curves.}
    \label{fig:TOI560_field}
\end{figure}

We find that TOI 560 b is a Neptunian planet orbiting a K dwarf star on a 6.398 d orbit. It has a radius of $2.49\pm0.10$ R$_{\earth}$ and a predicted mass of $5.7^{+8.3}_{-3.3}$ M$_{\earth}$. This planet's radius places it above the radius valley, meaning it has held onto its gaseous envelope and likely a rocky core enveloped by a thick H/He atmosphere, despite a relatively short orbital period. 

Our final reported radius measurement is not consistent with the radius measurement of $2.79\pm0.10$ R$_{\earth}$ from \citet{Barragan2022TOI560}. Interestingly, our models fitted to the \cheops\ light curve are indicative of deeper transits relative to models fitted to \tess\ light curves. We compute a planet radius of $2.72\pm0.11$ R$_{\earth}$ with our HOMEBREW method from \cheops\ photometry, but we compute a planet radius of $2.38\pm0.10$ R$_{\earth}$ with our HOMEBREW method from \tess\ photometry. Consequentially, these radius estimates are not consistent with one another, and only the radius estimate from our fit to \cheops\ data is consistent with that from \citet{Barragan2022TOI560}.

In a similar vein to our discussion for TOI 455 b, this possible discrepancy in radius may be a result of contamination in the \tess\ aperture for this star. Fig. \ref{fig:TOI560_field}, which was produced as part of a SPOC report for this planet from sector 8 PM photometry, shows the star field around TOI 560. There are seven TIC entries within 60.0 arcseconds, one of which is TIC 101011568, which has a Tmag of 11.79. This indicates that there may be flux from other stars present in the light curve for this target, which would dillute any transit signals therein. This in turn would make a planet appear smaller than if this contamination were not present, which may be the case for this system. Indeed, as previously stated, we merely flattened our \tess\ light curves to account for astrophysical noise rather than fully correcting for possible sources of contamination. 
%I would bet that a reviewer will say "if you realized this, then why didn't you account for it towards the beginning?

\subsubsection{TOI 562 b}

We find that TOI 562 b is a Terran world orbiting an M dwarf star on a 3.931 d period. We report that it has a radius of $1.20\pm0.06$ R$_{\earth}$ and a predicted mass of $0.8^{+2.5}_{-0.5}$ M$_{\earth}$. Given these physical parameters, we find that this planet is likely a super-Earth with a high silicon-to-iron ratio, although with no mass estimate, we cannot say for sure. Our radius measurement is consistent with the reported radius measurement from \citet{Luque2019GJ357} of $1.217^{+0.084}_{-0.083}$ R$_{\earth}$. Based on the fact that none of our models for this target had fractional depth uncertainties above the 10\% threshold, and that the largest fractional depth uncertainty for any fit was our \texttt{juliet} fit to \cheops\ data at 6.8\%, our model parameters are well-constrained for this planet.

\subsection{Position Relative to the Radius Valley}

\begin{figure*}
    \centering
    \includegraphics[width=.8\linewidth]{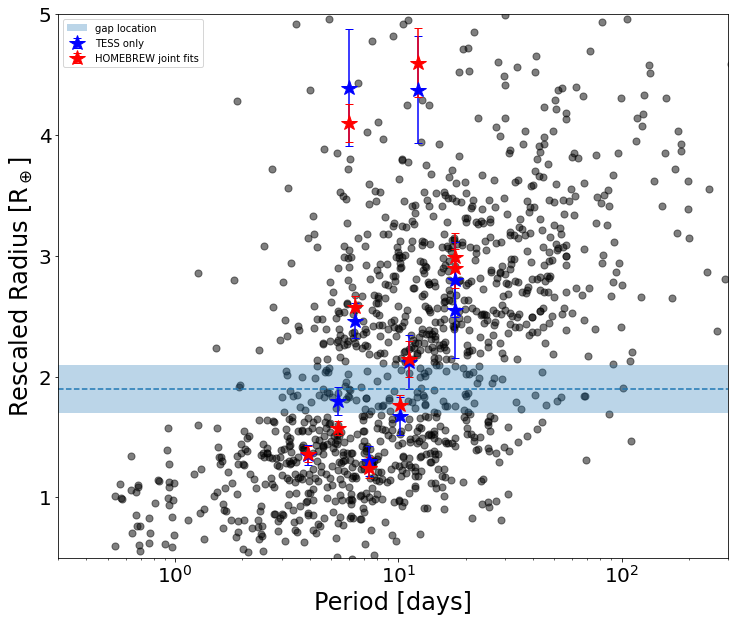}
    \caption{Period-scaled radius diagram for the California-Kepler survey (CKS) sample of planets (black circles;\citet{petigura2022california}), with initial estimates from TESS-SPOC (blue stars) and our HOMEBREW joint fit results overlaid (red stars). All radii have been rescaled according to eqn. \ref{eq:scaled_radius}, clearly exhibiting a deficiency of planets around 2 R$_\oplus$.}
    \label{fig:period-radius}
\end{figure*}

We have improved radius estimates for these TOIs by jointly fitting \tess\ and \cheops\ photometry, allowing us to more effectively place them in period-radius space. As noted in \citet{Zhu_Dong_ARAA_2021}, the location of the valley decreases with orbital period and increases with host mass. These authors report the location of the valley as a function of scaled radius, which is a double power law of the following form:

\begin{equation}\label{eq:scaled_radius}
    \tilde{R}_p(P,M_*) = R_{p}\left(\frac{P}{10 \textup{days}}\right)^{-g} \left(\frac{M_*}{M_\sun}\right)^{-h}
\end{equation}
where $\tilde{R}_p$ is the rescaled radius of the planet, and $g$ and $h$ represent power law coefficients for the gap location according to orbital period and host stellar mass, respectively. Similar to \citet{Zhu_Dong_ARAA_2021}, we choose $g=-0.09$ \citep{vaneylen2018astroseismic} and $h=0.26$ \citep{berger2020gaia}.

Fig. \ref{fig:period-radius} shows our period-scaled radius diagram, which includes confirmed planets from the California-Kepler survey (CKS) sample. In this figure, our systems are overplotted as red stars, with initial values are reported by SPOC analysis plotted as blue stars. We also include the gap as a light blue shaded region from $R_{p,0} = 1.9 \pm 0.2$ R$_\earth$. We note that the location of the radius valley with respect to the variables in eqn. \ref{eq:scaled_radius} has important implications for distinguishing between different planet formation and evolution models, and do not claim that the values we chose are definitive. Rather, we chose these values to illuminate the positions of the planets in our sample with respect to the valley.

We can also examine our final reported radius measurements in relation to their initial reported values from TESS-SPOC. Our radius measurements appear to be systematically smaller than those from TESS-SPOC. Planet radii are typically calculated by multiplying the model planet-to-star radius ratio by the radius of the host star. TESS-SPOC values may therefore have been greater than our reported values due to the fact that stellar radii as given in the TIC are systematically slightly larger than our reported radii.

Two of the planets we present in this work may fall into the radius gap according to the functional form we present here. Upon comparison of our uncertainties in planet radii relative to TESS-SPOC values, we see that our uncertainties are smaller, which corresponds to fewer planets potentially falling into the gap. With a relatively small sample of 10 planets/planet candidate, we do not claim to resolve the valley, but rather seek to add to the sample of small planets which are characterized with high precision. Improving the precision with which we place these planets and other small planets in period-radius space will inform theories of small planet formation and evolution (see section \ref{sec:intro}), and this work had added a valuable contribution to the overall small planet sample. 

\section{Summary and Conclusion}\label{sec:conc}

In this work, we have presented our characterization of 10 small planets via observations with both \tess\ and \cheops. We vetted and validated transit signals as being planetary in nature for planets which had not yet been published, including TOI 198 b, TOI 244 b, TOI 444 b, and TOI 470 b, although we could not conclusively validate transit signals for  TOI 518.01. To this end, we introduced and analyzed followup observations of these systems to verify these stars do not have previously-unseen companions and the transit signals we analyze are due to transiting planets on the target star. We detrended our \tess\ and \cheops\ light curves from sources of instrumental and astrophysical noise. We fitted transit models to these light curves three different ways to check consistency between our fits.

We summarize our results and findings below:

\begin{itemize}
    \item We report updated physical and orbital properties for 10 planets, including 4 Terran worlds and 6 Neptunian worlds. 
    \begin{itemize}
        \item We find that TOI 118 b is a Neptunian world orbiting a sun-like star on a 6.034 d orbit, and that it has a radius of $4.24 \pm 0.16$ R$_{\earth}$. Our fits for this system may have yielded slightly higher fractional depth uncertainties due to differences in perceived transit depth between different filters.
        \item We find that TOI 198 b is a Terran world orbiting an M dwarf star on a 10.215 d orbit, and that it has a radius of $1.44 \pm 0.08$ R$_{\earth}$. While most fits for this target were well-constrained, fits to the \tess\ light curve for this target might have suffered from relatively high photon noise.
        \item We find that TOI 244 b is a Terran world orbiting an M dwarf star on a 7.397 d orbit, and that it has a radius of $1.03 \pm 0.08$ R$_{\earth}$. Our fits to \cheops\ photometry for this target may have been affected by data gaps which obscured crucial parts of our transits.
        \item We find that TOI 262 b is a Neptunian world orbiting a sun-like star on a 11.145 d orbit, and that it has a radius of $2.07 \pm 0.15$ R$_{\earth}$. We report a fractional depth uncertainty of 10\% or larger for almost all fits to TOI 262 photometry. This may have been due to either a comparatively shallow computed transit depth or a degeneracy between transit depth and impact parameter.
        \item We find that TOI 444 b is a Neptunian world orbiting a sun-like star on a 17.964 d orbit, and that it has a radius of $2.77 \pm 0.20$ R$_{\earth}$. The fractional depth uncertainty for the \texttt{juliet} fit to \tess\ photometry was anomalously large, perhaps due to the fact that orbital parameters are computed separately for \texttt{juliet} fit, leading to a degeneracy between transit depth and impact parameter in this case.
        \item We find that TOI 455 b is a Terran world orbiting an M dwarf star on a 5.359 d orbit, and that it has a radius of $1.18 \pm 0.06$ R$_{\earth}$. Different detrending methods for our \cheops\ and \tess\ light curves may have led to different transit depths as seen by these telescopes, and may have contributed to an inflation of model uncertainties when fitting \tess\ data relative to model uncertainties when fitting \cheops\ data.
        \item We find that TOI 470 b is a Neptunian world orbiting a sun-like star on a 12.191 d orbit, and that it has a radius of $4.34 \pm 0.29$ R$_{\earth}$. Despite not having an out-of-transit baseline prior to the transit in the \cheops\ visit for this target, our models for this planet were well-constrained.
        \item We find that TOI 518.01 may a Neptunian world orbiting a sun-like star on a 17.877 d orbit, with a potential radius of $2.77 \pm 0.16$ R$_{\earth}$. Our models for this system yielded the largest fractional depth uncertainties relative to any other system, perhaps due to data gaps in our \cheops\ transits and photon noise in our \tess\ light curves.
        \item We find that TOI 560 b is a Neptunian world orbiting a sun-like star on a 6.398 d orbit, and that it has a radius of $2.49 \pm 0.10$ R$_{\earth}$. A possible discrepancy between our fits to \tess\ photometry and the radius valley reported by \citet{Barragan2022TOI560} may have been the result of contamination from nearby field stars in the \tess\ aperture for this star.
        \item We find that TOI 562 b is a Neptunian world orbiting an M dwarf star on a 3.931 d orbit, and that it has a radius of $1.20 \pm 0.06$ R$_{\earth}$. Our models were well-constrained for this target, and matched well with previously-published values.
    \end{itemize}
    \item We improved radius estimates with all fitting methods relative to initial characterization with \tess\ PM estimates. We report radius measurements to better than 10\% precision for all 10 of our planets/candidate planet, even when wrapping in uncertainties in stellar parameters. This shows that high-precision results can be obtained from analysis of photometry alone, although our results would be improved by further high-precision characterization of the host stars. Further, interesting information regarding these systems' formation and evolutionary histories could be gained from mass measurements via spectroscopic observations.
    \item We compared relative photometric performances of \tess\ and \cheops\, finding that our models fitted to \cheops\ photometry under-performed relative to our predictions in most cases, as indicated by the fact that we needed fewer \tess\ transits than we obtained to match depth precision on \cheops\ transits. We believe that finding is due to two primary reasons, including 1. important data gaps in CHEOPS visits, and 2. the way in which both CHEOPS and TESS light curves are generated and detrended. There is no standard method with which to detrend CHEOPS photometry, whereas TESS light curves are all processed by the same pipeline which treats the data uniformly. Regardless, when we compare precision from one \cheops\ transit, our results indicate that precision of our \cheops\ observations are equivalent to between 2 and 12 \tess\ transits, excluding TOI 455 b as anomalous.
    \item Finally, we were able to place these planets precisely in period-radius space. Two of the planets from our sample may fall into the gap, or immediately border it. With a relatively small sample of 10 planets, we do not claim to resolve the valley, but rather seek to add to the sample of small planets which are characterized with high precision. Improving the precision with which we place these planets and other small planets in period-radius space will inform theories of small planet formation and evolution (see section 1), and this work had added a valuable contribution to the overall small planet sample.
\end{itemize}

\section{Acknowledgments}
We graciously thank the anonymous referee and data editor who both provided valuable feedback on this publication.

D. D. acknowledges support from the TESS Guest Investigator Program grants 80NSSC21K0108 and 80NSSC22K0185.

We acknowledge contributions from Thomas G. Wilson, Andrea Fortier, and other members of the \cheops\ GTO team, regarding estimation of noise in \cheops\ light curves and comparisons to other systems observed by \cheops.

This paper includes data collected by the \tess\ mission. Funding for the \tess\ mission is provided by NASA's Science Mission Directorate.

We acknowledge the use of public TESS data from pipelines at the TESS Science Office and at the TESS Science Processing Operations Center. Resources supporting this work were provided by the NASA High-End Computing (HEC) Program through the NASA Advanced Supercomputing (NAS) Division at Ames Research Center for the production of the SPOC data products.

Some/all of the data presented in this paper were obtained from the Mikulski Archive for Space Telescopes (MAST) at the Space Telescope Science Institute. The specific observations analyzed can be accessed via \dataset[10.17909/dshz-jz09]{https://doi.org/DOI}

This work makes use of observations from the LCOGT network. Part of the LCOGT telescope time was granted by NOIRLab through the Mid-Scale Innovations Program (MSIP). MSIP is funded by NSF.

This research has made use of the NASA Exoplanet Archive, which is operated by the California Institute of Technology, under contract with the National Aeronautics and Space Administration under the Exoplanet Exploration Program.

This research made use of Lightkurve, a Python package for Kepler and TESS data analysis (Lightkurve Collaboration, 2018).

Some of the observations in the paper made use of the High-Resolution Imaging instruments ‘Alopeke and Zorro obtained under Gemini LLP Proposal Number: GN/S-2021A-LP-105. ‘Alopeke and Zorro were funded by the NASA Exoplanet Exploration Program and built at the NASA Ames Research Center by Steve B. Howell, Nic Scott, Elliott P. Horch, and Emmett Quigley. Alopeke (Zorro) was mounted on the Gemini North (South) telescope of the international Gemini Observatory, a program of NSF’s OIR Lab, which is managed by the Association of Universities for Research in Astronomy (AURA) under a cooperative agreement with the National Science Foundation. on behalf of the Gemini partnership: the National Science Foundation (United States), National Research Council (Canada), Agencia Nacional de Investigación y Desarrollo (Chile), Ministerio de Ciencia, Tecnología e Innovación (Argentina), Ministério da Ciência, Tecnologia, Inovações e Comunicações (Brazil), and Korea Astronomy and Space Science Institute (Republic of Korea).

\vspace{5mm}
\facilities{TESS, CHEOPS, LCOGT, VLT-ESPRESSO, CTIO/SMARTS-CHIRON, FLWO-TRES, Gemini-'Alopeke/Zorro, Keck2-NIRC2, Palomar-PHARO, SOAR-HRCam}

\software{AstroImageJ \citep{Collins_2017},
          astropy \citep{astropy:2013,astropy:2018,astropy:2022},
          numpy \citep{harris2020array},
          matplotlib \citep{Hunter:2007}, 
          lightkurve \citep{lightkurve},
          pycheops \citep{Maxted2021pycheops},
          PyLDTK \citep{Parviainen2015LDTK},
          juliet \citep{Espinoza2019juliet},
          emcee \citep{Foreman_Mackey_2013},
          batman \citep{Kreidberg2015batman},
          corner \citep{corner},
          MRExo \citep{Kanodia_2019},
          TAPIR \citep{Jensen_2013}
          }

\appendix

\section{\tess\ light curves}\label{app:te_LCs}

\begin{figure*}
\centering
 \includegraphics[width=.8\linewidth]{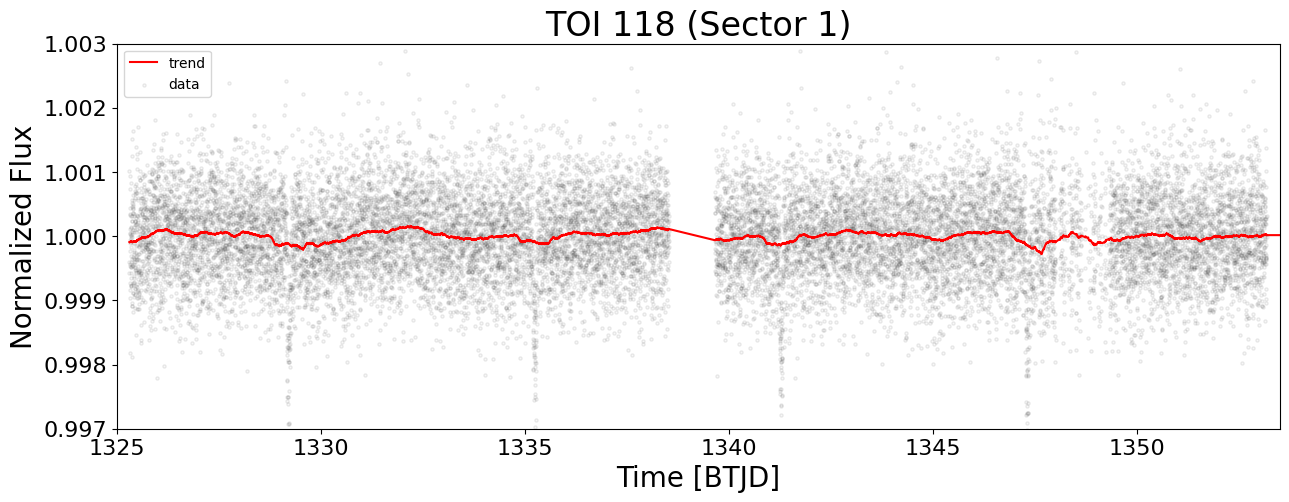}
 \includegraphics[width=.8\linewidth]{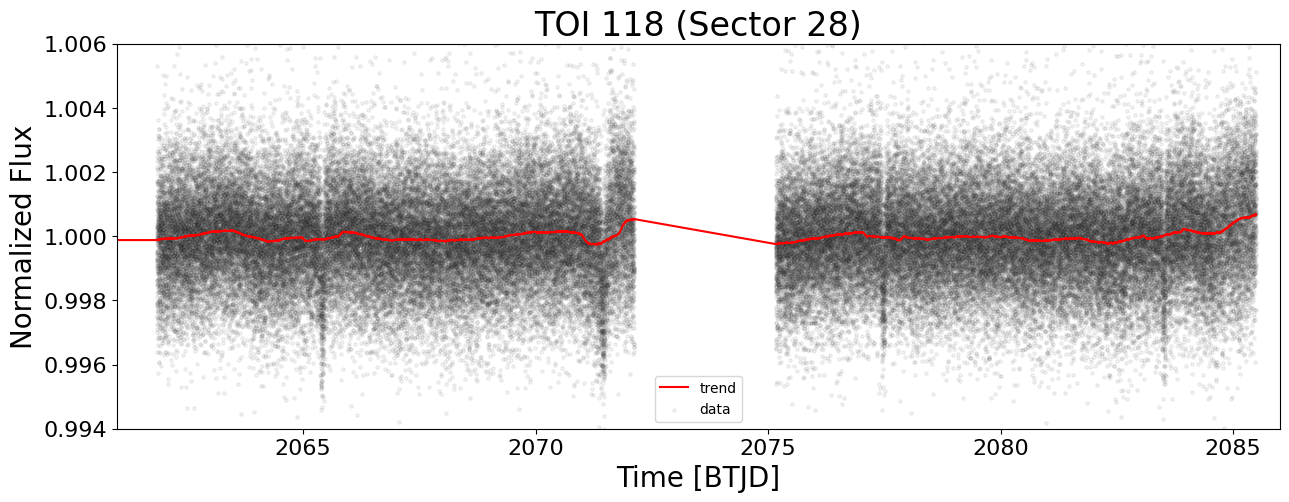}
 \caption{TESS sectors for TOI 118. Red line represents the function by which we smoothed the light curve to eliminate stellar noise using \texttt{wotan}.}
 \label{fig:TOI118_TESS}
\end{figure*}

\begin{figure*}
\centering
 \includegraphics[width=.8\linewidth]{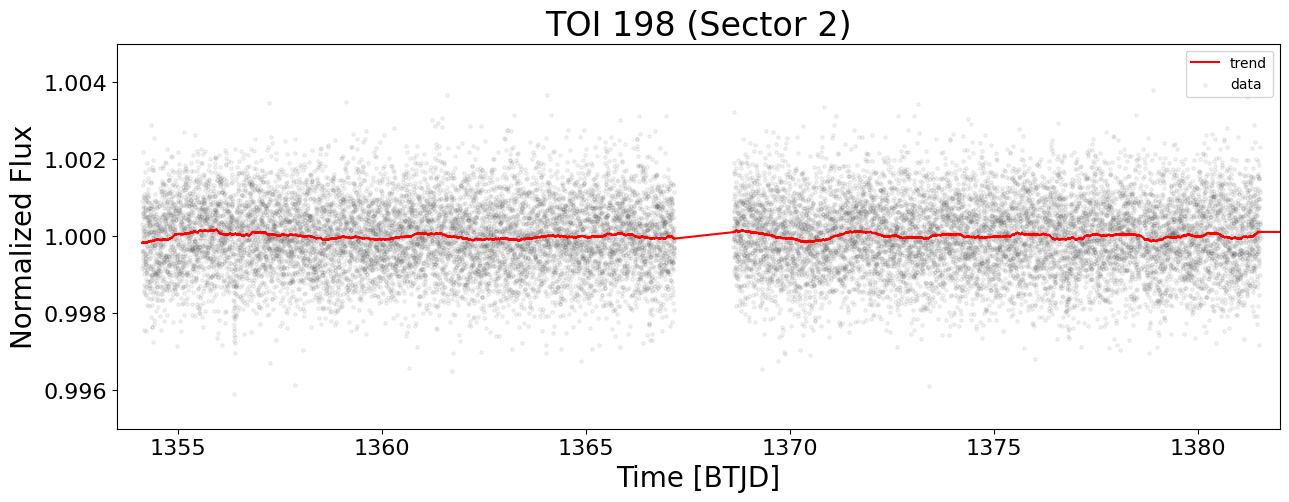}
 \includegraphics[width=.8\linewidth]{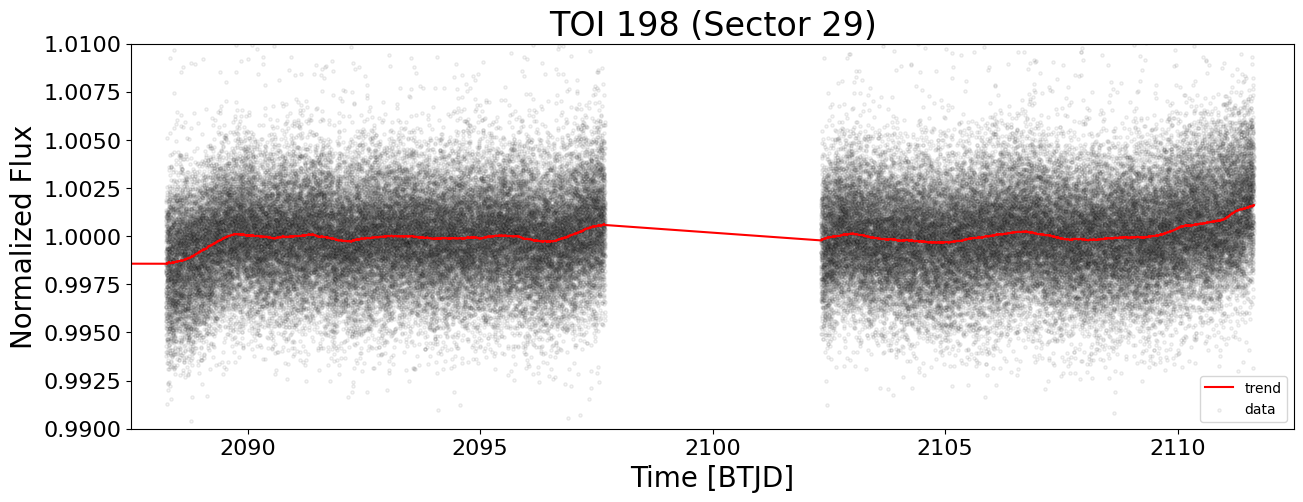}
 \caption{TESS sectors for TOI 198. Red line is the same as Fig. \ref{fig:TOI118_TESS}.}
 \label{fig:TOI198_TESS}
\end{figure*}

\begin{figure*}
\centering
 \includegraphics[width=.8\linewidth]{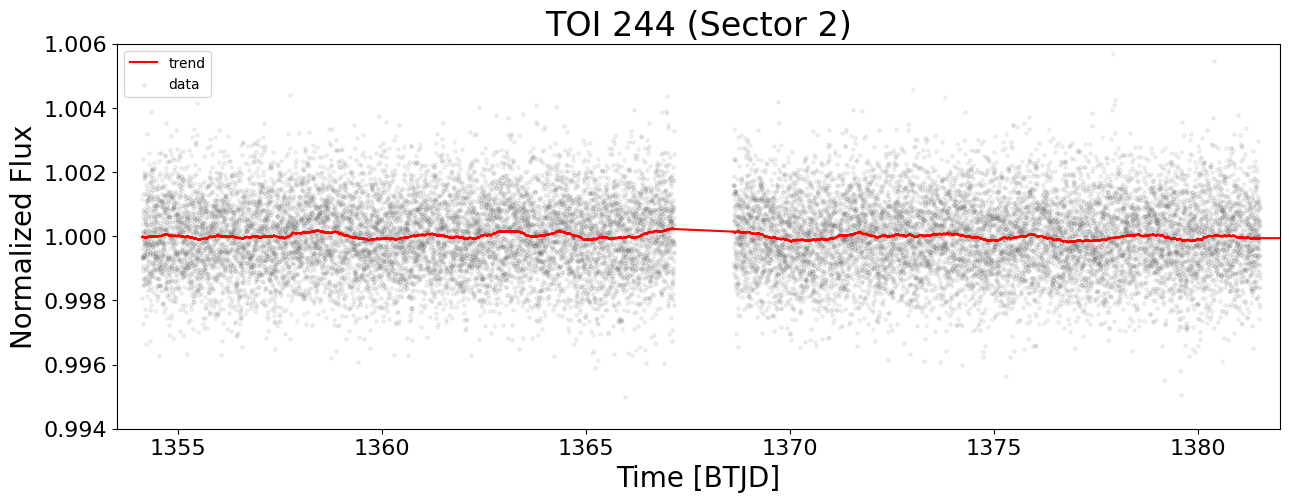}
 \includegraphics[width=.8\linewidth]{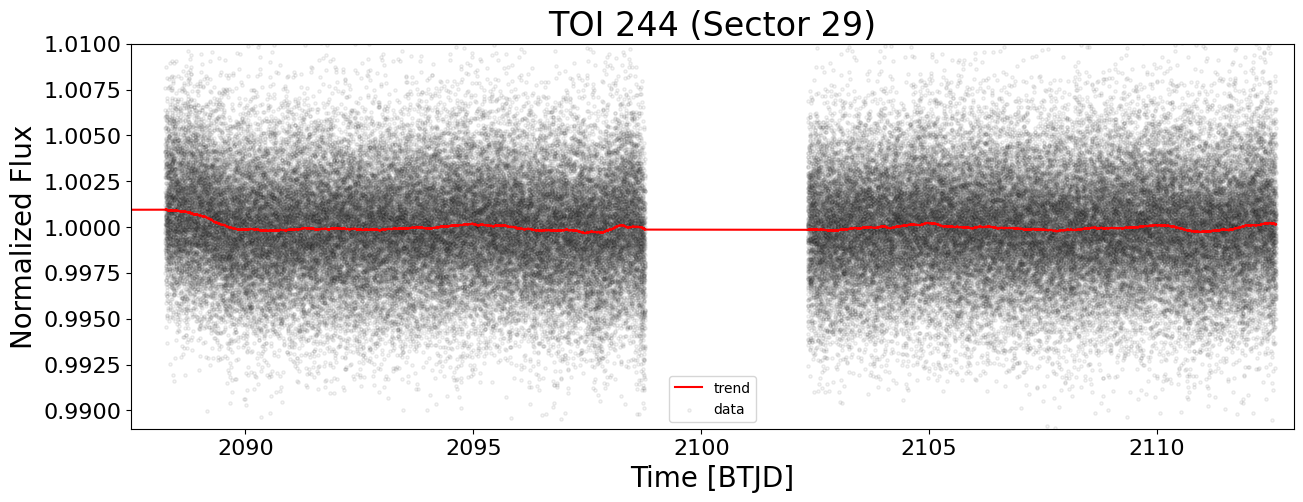}
 \caption{TESS sectors for TOI 244. Red line is the same as Fig. \ref{fig:TOI118_TESS}.}
 \label{fig:TOI244_TESS}
\end{figure*}

\begin{figure*}
\centering
 \includegraphics[width=.8\linewidth]{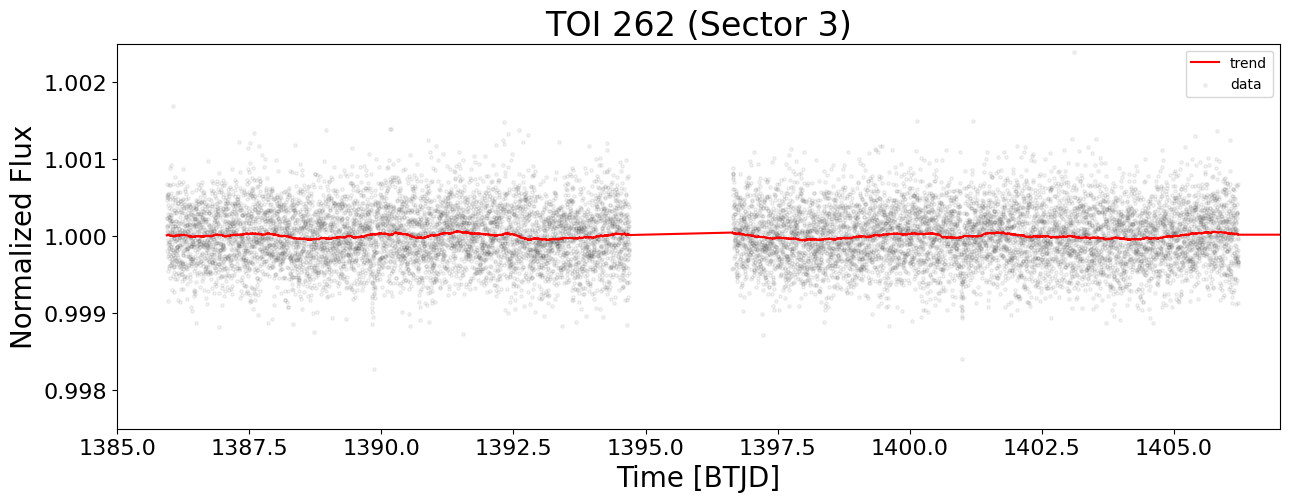}
 \includegraphics[width=.8\linewidth]{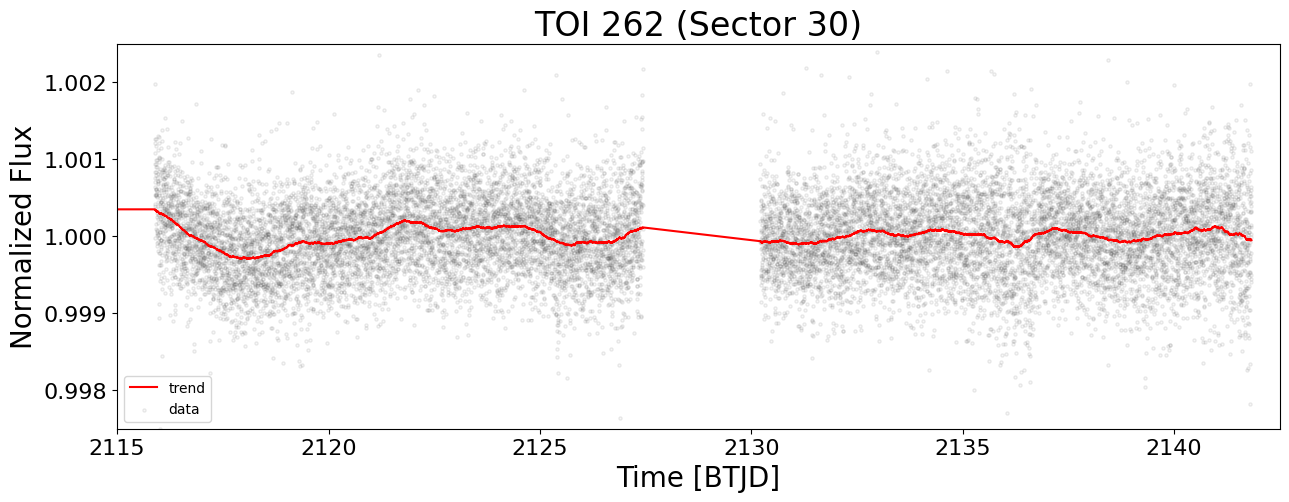}
 \caption{TESS sectors for TOI 262. Red line is the same as Fig. \ref{fig:TOI118_TESS}.}
 \label{fig:TOI262_TESS}
\end{figure*}

\begin{figure*}
\centering
 \includegraphics[width=.8\linewidth]{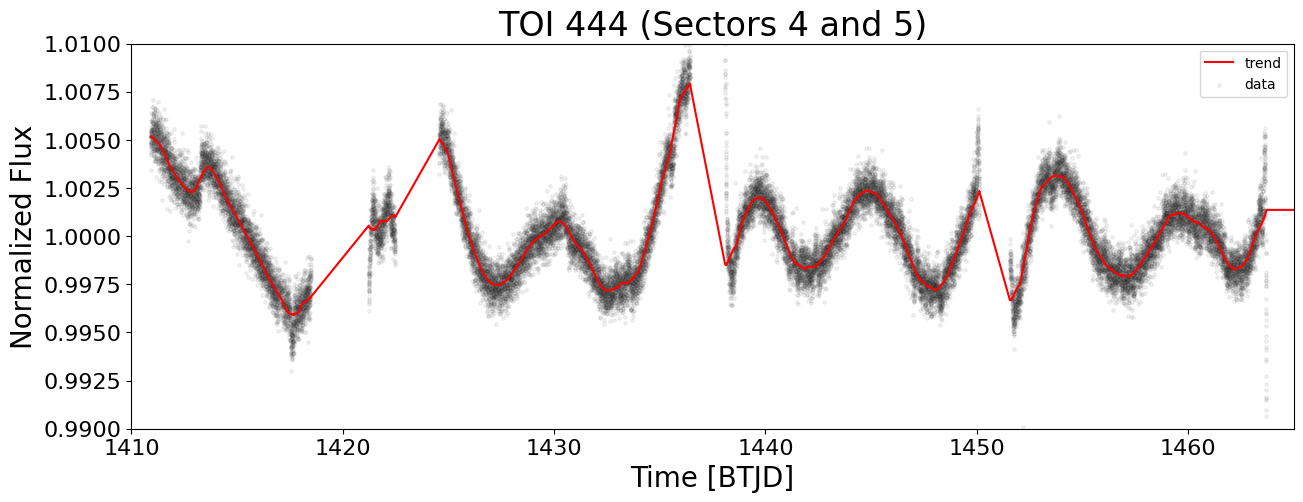}
 \includegraphics[width=.8\linewidth]{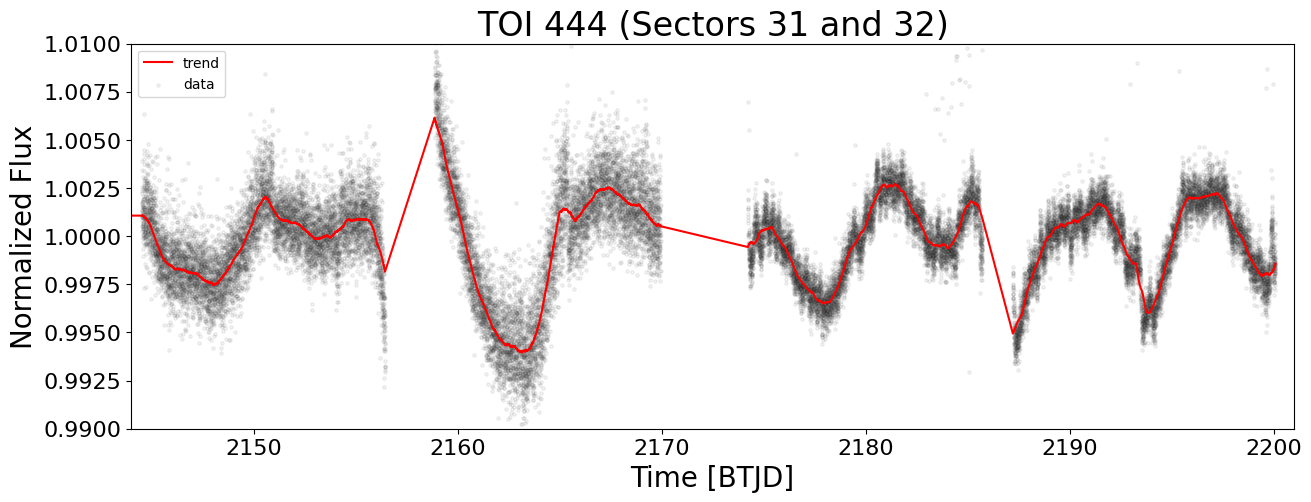}
 \caption{TESS sectors for TOI 444. Red line is the same as Fig. \ref{fig:TOI118_TESS}.}
 \label{fig:TOI444_TESS}
\end{figure*}

\begin{figure*}
\centering
 \includegraphics[width=.8\linewidth]{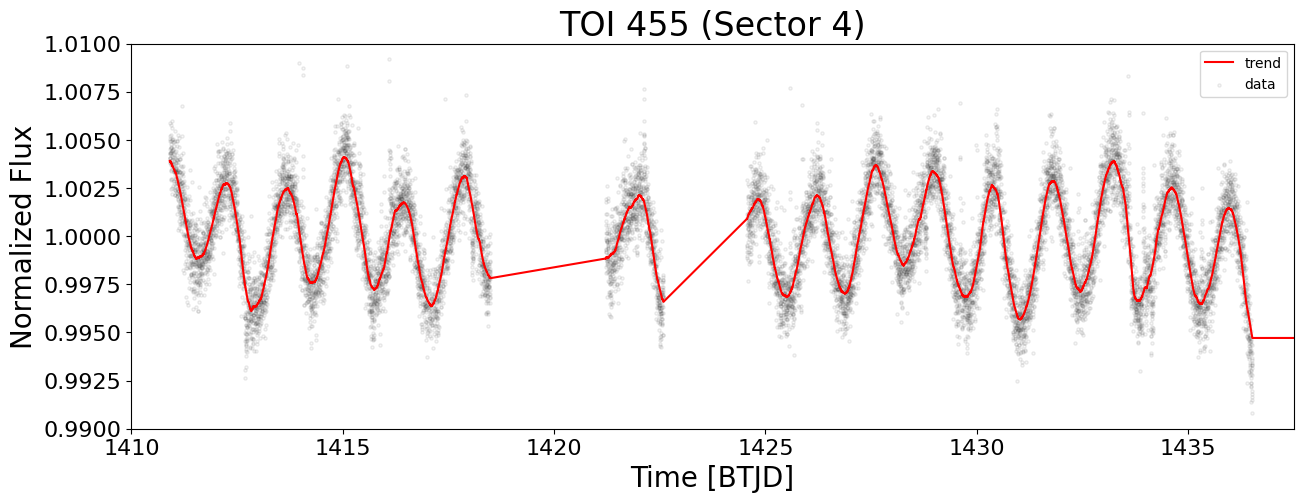}
 \includegraphics[width=.8\linewidth]{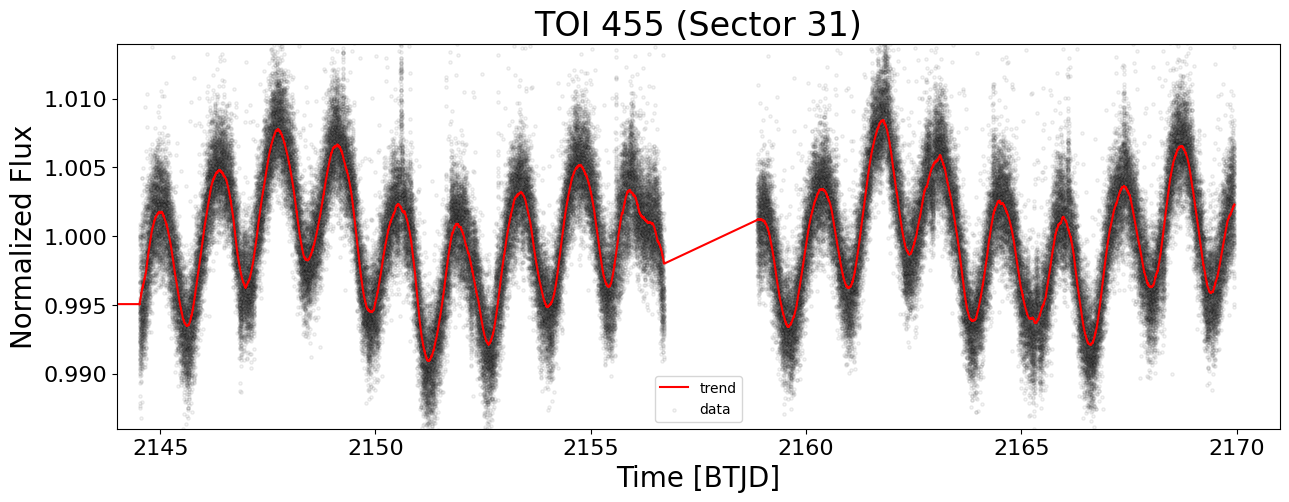}
 \caption{TESS sectors for TOI 455. Red line is the same as Fig. \ref{fig:TOI118_TESS}.}
 \label{fig:TOI455_TESS}
\end{figure*}

\begin{figure*}
\centering
 \includegraphics[width=.8\linewidth]{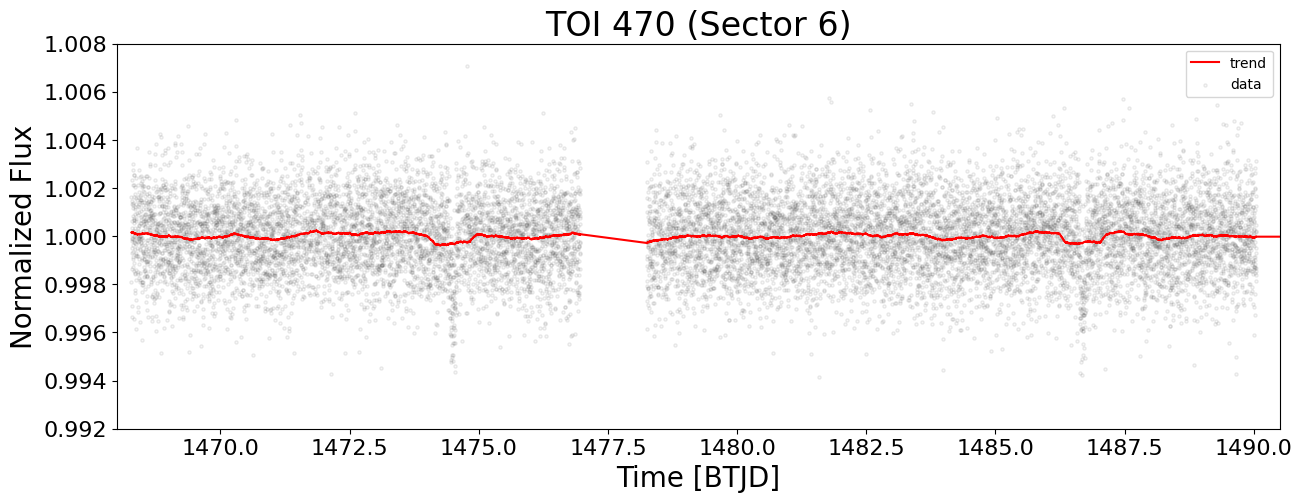}
 \includegraphics[width=.8\linewidth]{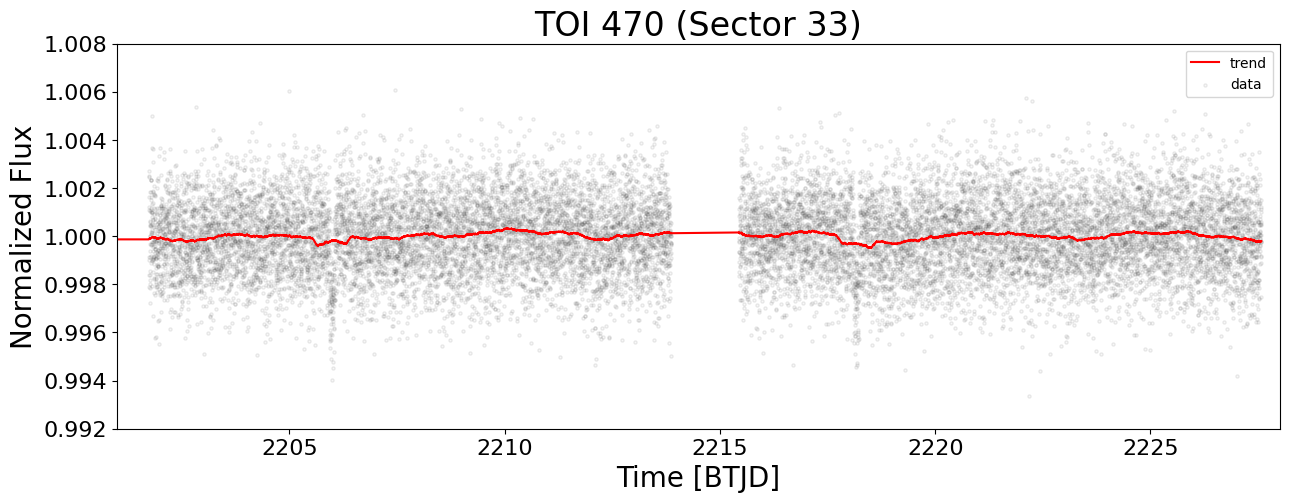}
 \caption{TESS sectors for TOI 470. Red line is the same as Fig. \ref{fig:TOI118_TESS}.}
 \label{fig:TOI470_TESS}
\end{figure*}

\begin{figure*}
\centering
 \includegraphics[width=.8\linewidth]{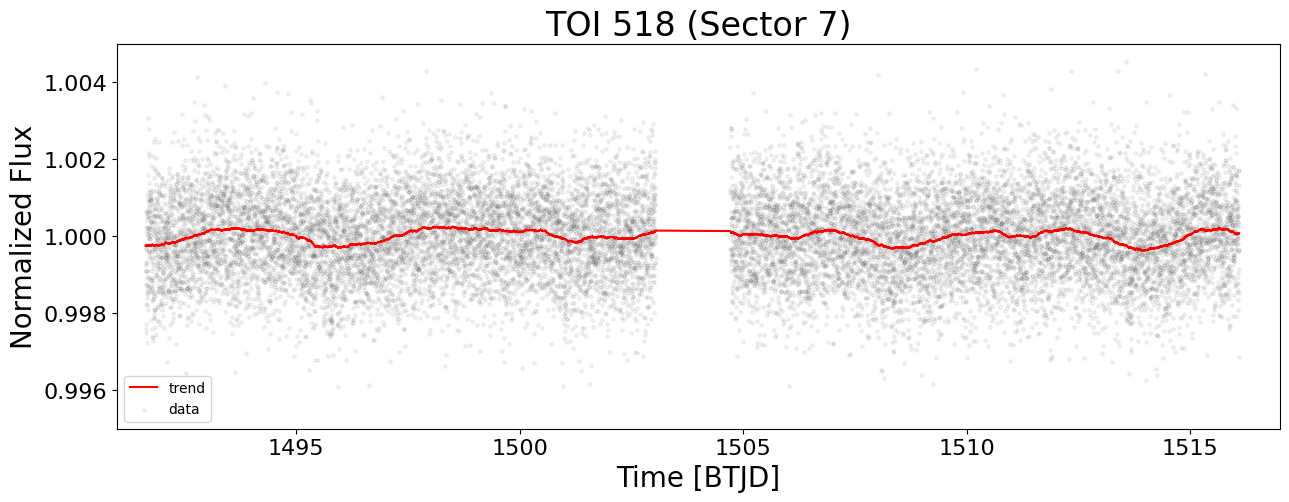}
 \caption{TESS sectors for TOI 518. Red line is the same as Fig. \ref{fig:TOI118_TESS}.}
 \label{fig:TOI518_TESS}
\end{figure*}

\begin{figure*}
\centering
 \includegraphics[width=.8\linewidth]{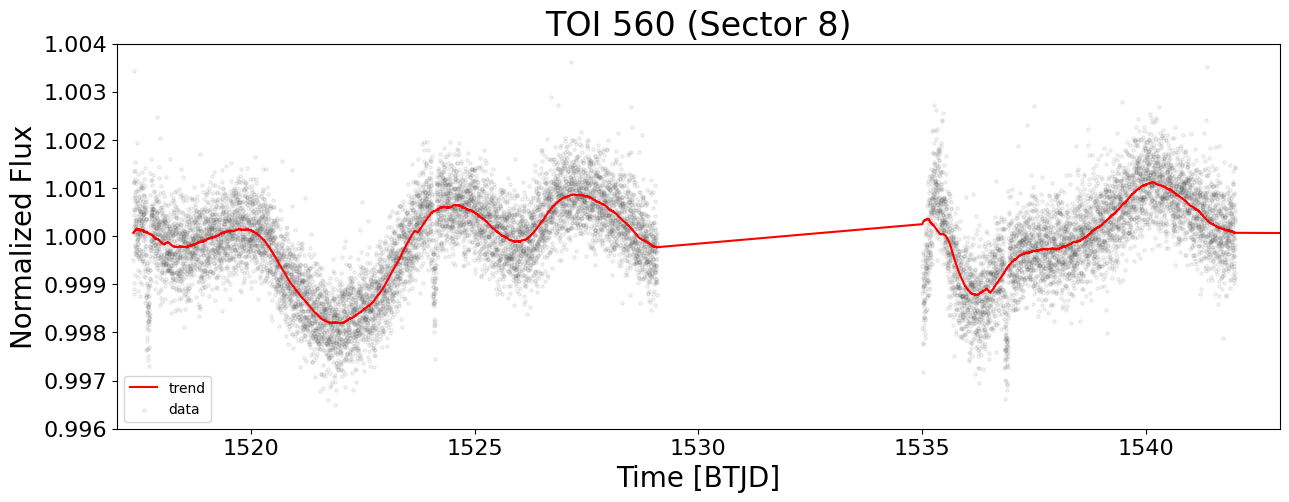}
 \includegraphics[width=.8\linewidth]{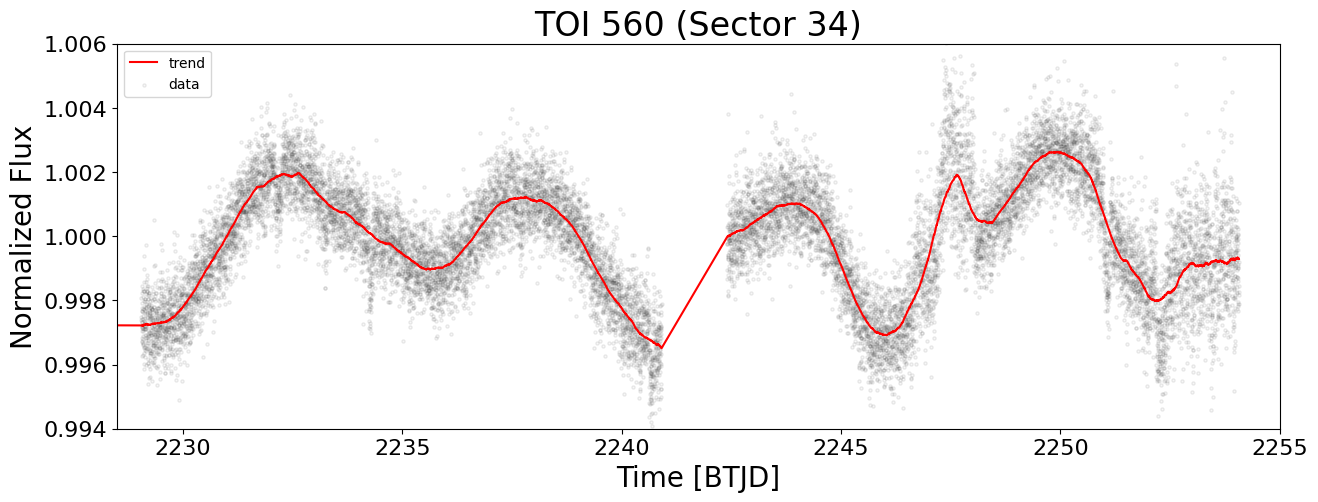}
 \caption{TESS sectors for TOI 560. Red line is the same as Fig. \ref{fig:TOI118_TESS}.}
 \label{fig:TOI560_TESS}
\end{figure*}

\begin{figure*}
\centering
 \includegraphics[width=.8\linewidth]{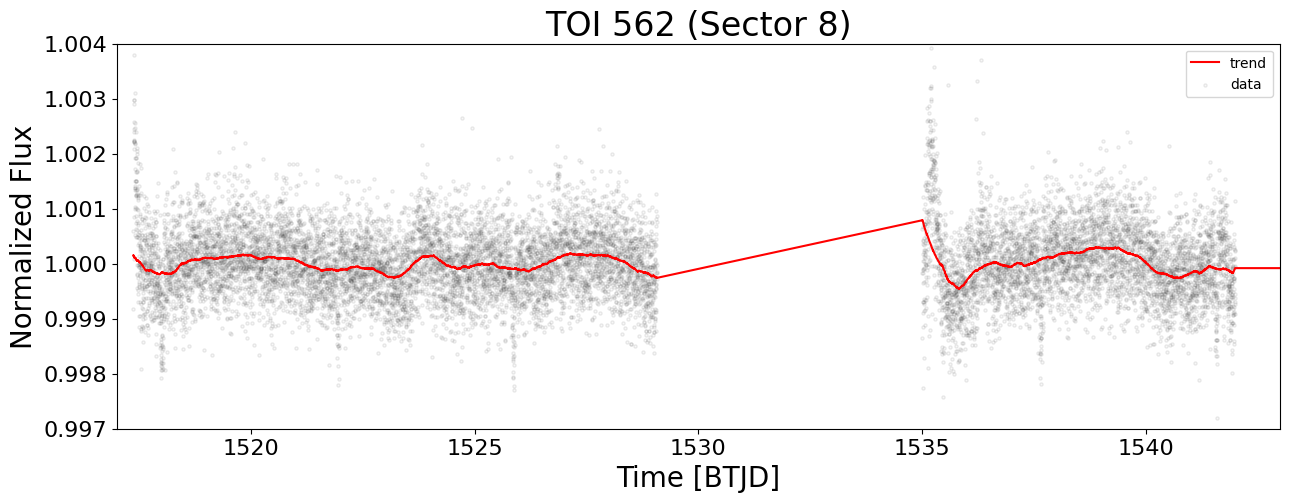}
 \includegraphics[width=.8\linewidth]{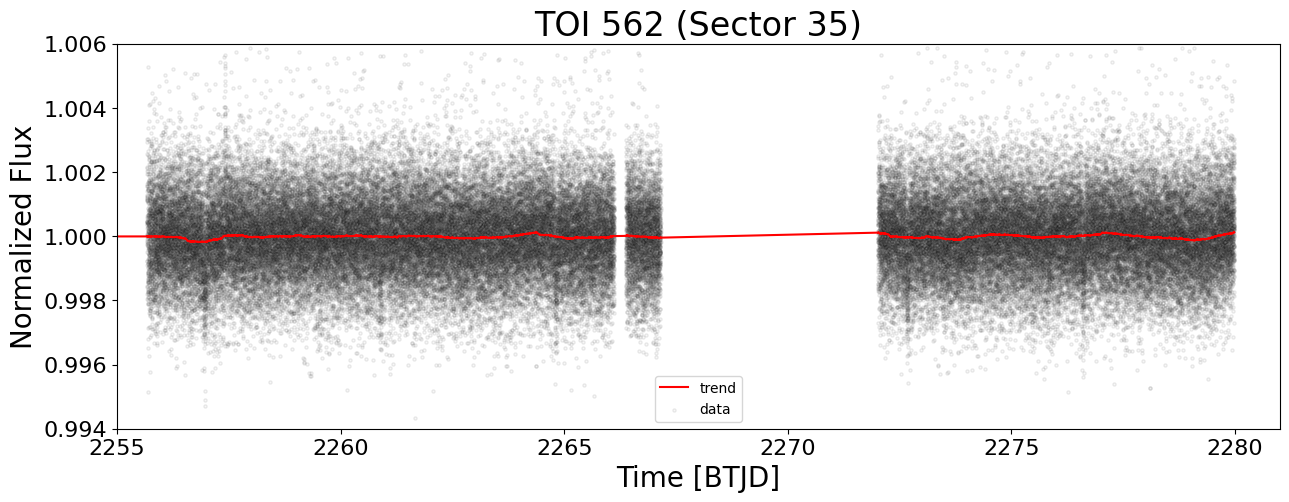}
 \caption{TESS sectors for TOI 562. Red line is the same as Fig. \ref{fig:TOI118_TESS}.}
 \label{fig:TOI562_TESS}
\end{figure*}

\section{\cheops\ light curves}\label{app:ch_LCs}

\begin{figure}
\centering
 \includegraphics[width=\linewidth]{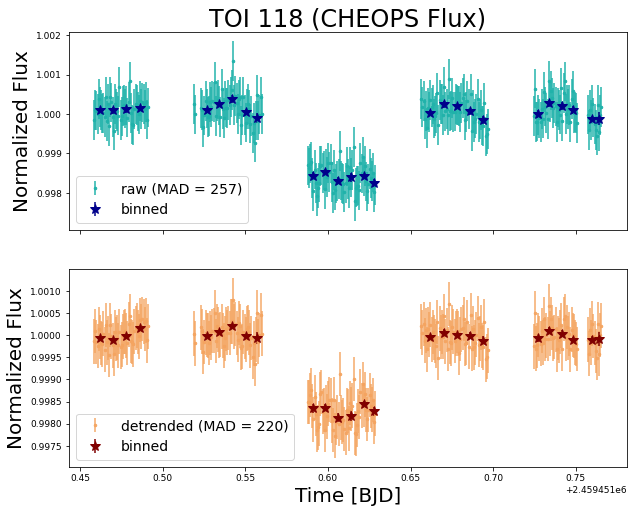}
 \caption{CHEOPS light curve for TOI 118.}
 \label{fig:TOI118_cheops}
\end{figure}

\begin{figure}
\centering
 \includegraphics[width=\linewidth]{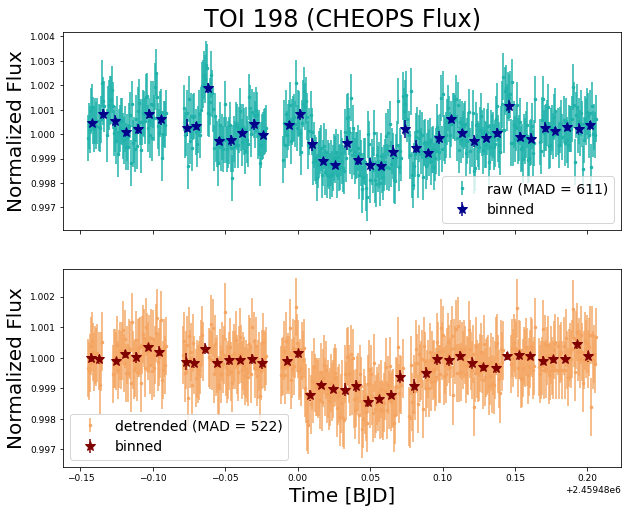}
 \caption{CHEOPS light curve for TOI 198.}
 \label{fig:TOI198_cheops}
\end{figure}

\begin{figure}
\centering
 \includegraphics[width=\linewidth]{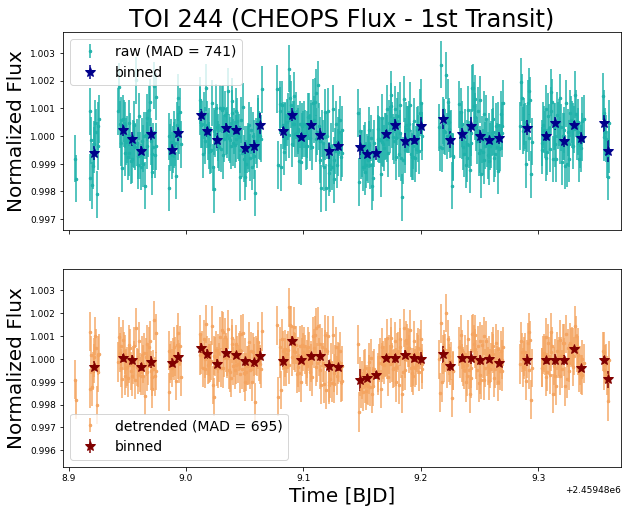}
 \includegraphics[width=\linewidth]{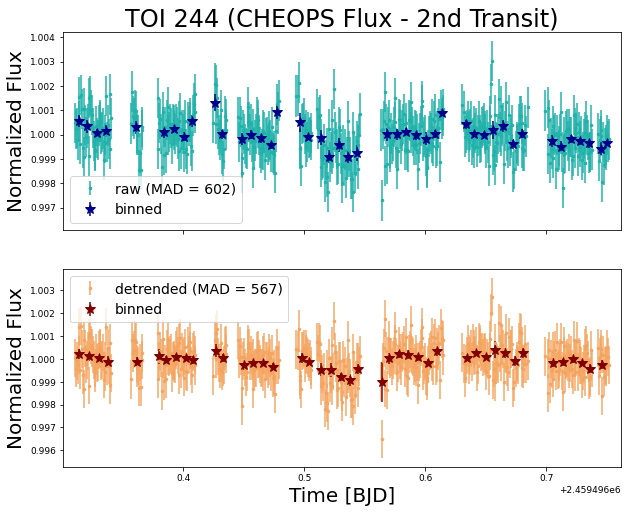}
 \caption{CHEOPS light curves for TOI 244.}
 \label{fig:TOI244_cheops}
\end{figure}

\begin{figure}
\centering
 \includegraphics[width=\linewidth]{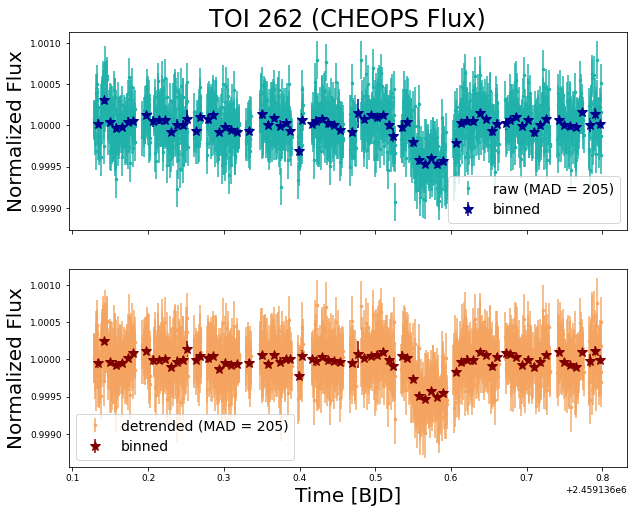}
 \caption{CHEOPS light curve for TOI 262.}
 \label{fig:TOI262_cheops}
\end{figure}

\begin{figure}
\centering
 \includegraphics[width=\linewidth]{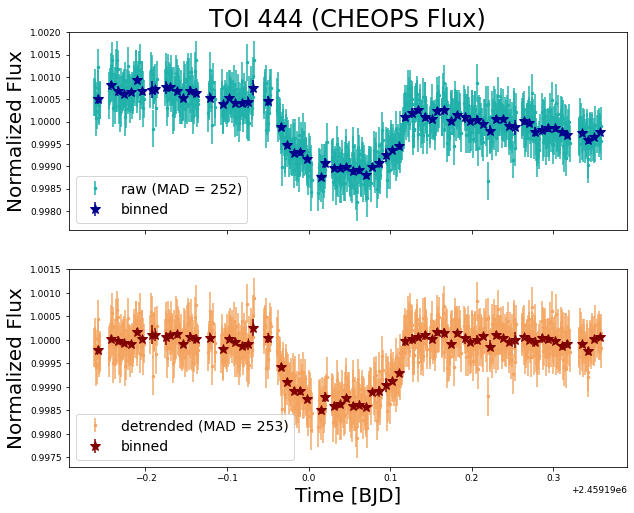}
 \caption{CHEOPS light curve for TOI 444.}
 \label{fig:TOI444_cheops}
\end{figure}

\begin{figure}
\centering
 \includegraphics[width=\linewidth]{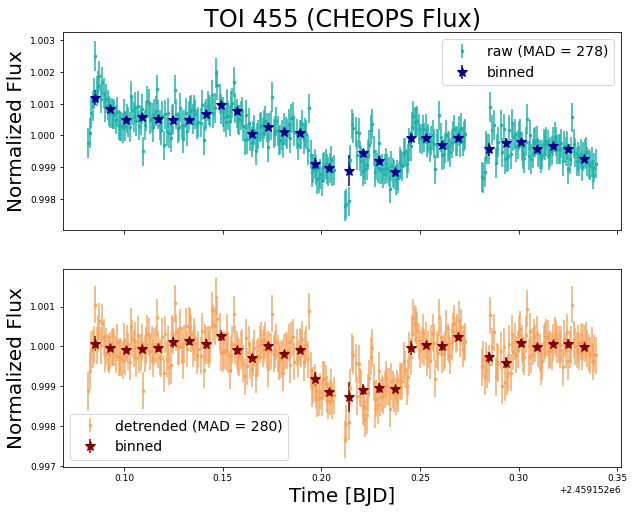}
 \caption{CHEOPS light curve for TOI 455.}
 \label{fig:TOI455_cheops}
\end{figure}

\begin{figure}
\centering
 \includegraphics[width=\linewidth]{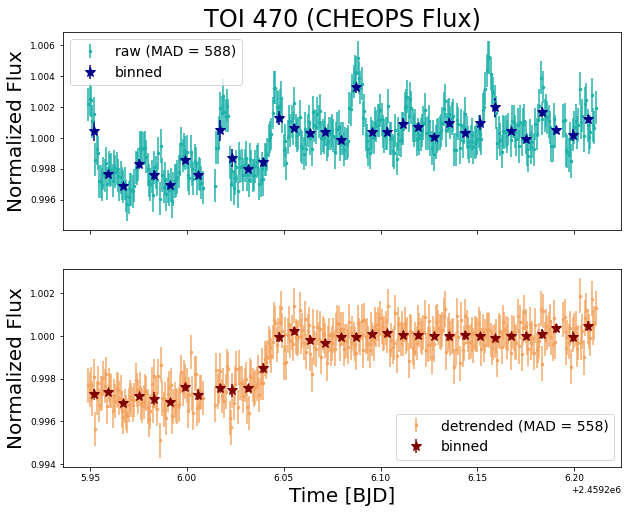}
 \caption{CHEOPS light curve for TOI 470.}
 \label{fig:TOI470_cheops}
\end{figure}

\begin{figure}
\centering
 \includegraphics[width=\linewidth]{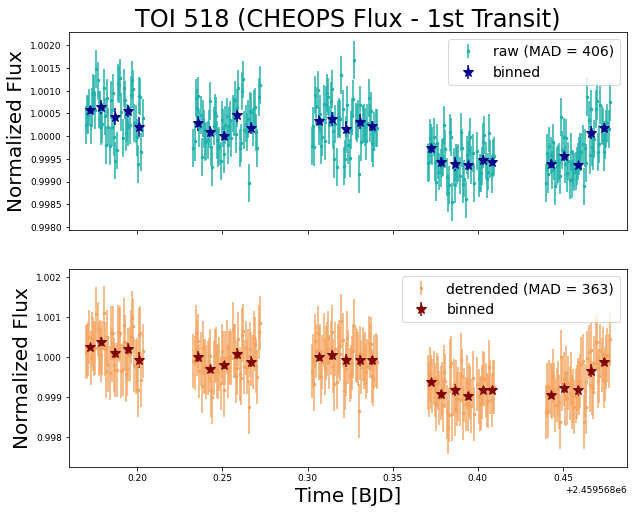}
 \includegraphics[width=\linewidth]{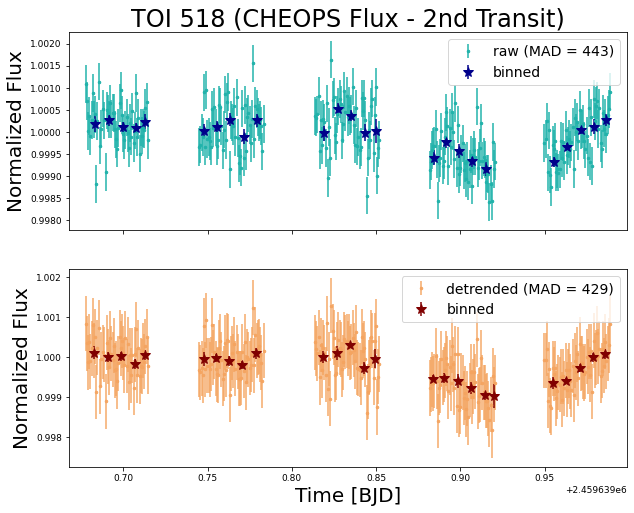}
 \caption{CHEOPS light curve for TOI 518.}
 \label{fig:TOI518_cheops}
\end{figure}

\begin{figure}
\centering
 \includegraphics[width=\linewidth]{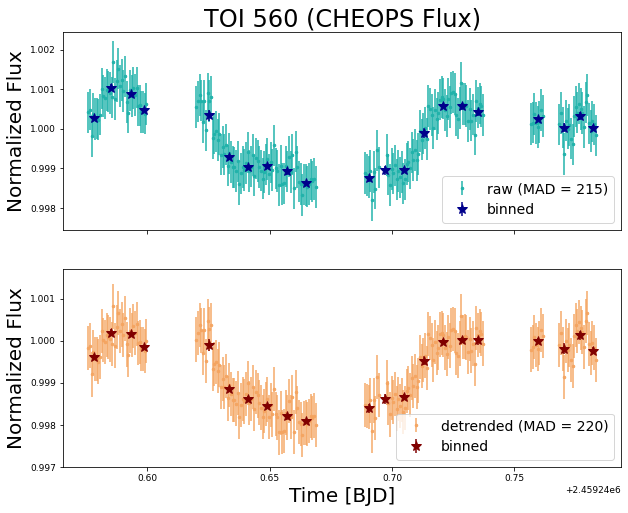}
 \caption{CHEOPS light curve for TOI 560.}
 \label{fig:TOI560_cheops}
\end{figure}

\begin{figure}
\centering
 \includegraphics[width=\linewidth]{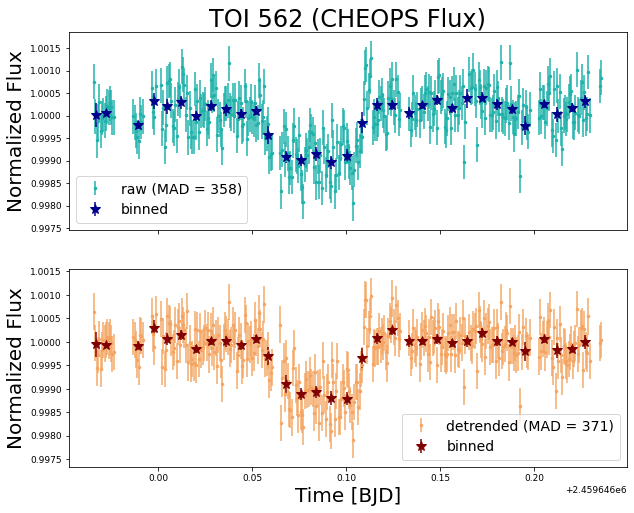}
 \caption{CHEOPS light curve for TOI 562.}
 \label{fig:TOI562_cheops}
\end{figure}

\section{SED Analysis}\label{app:SED}

\begin{figure*}\label{fig:sed}
    \centering
    \includegraphics[width=0.49\linewidth]{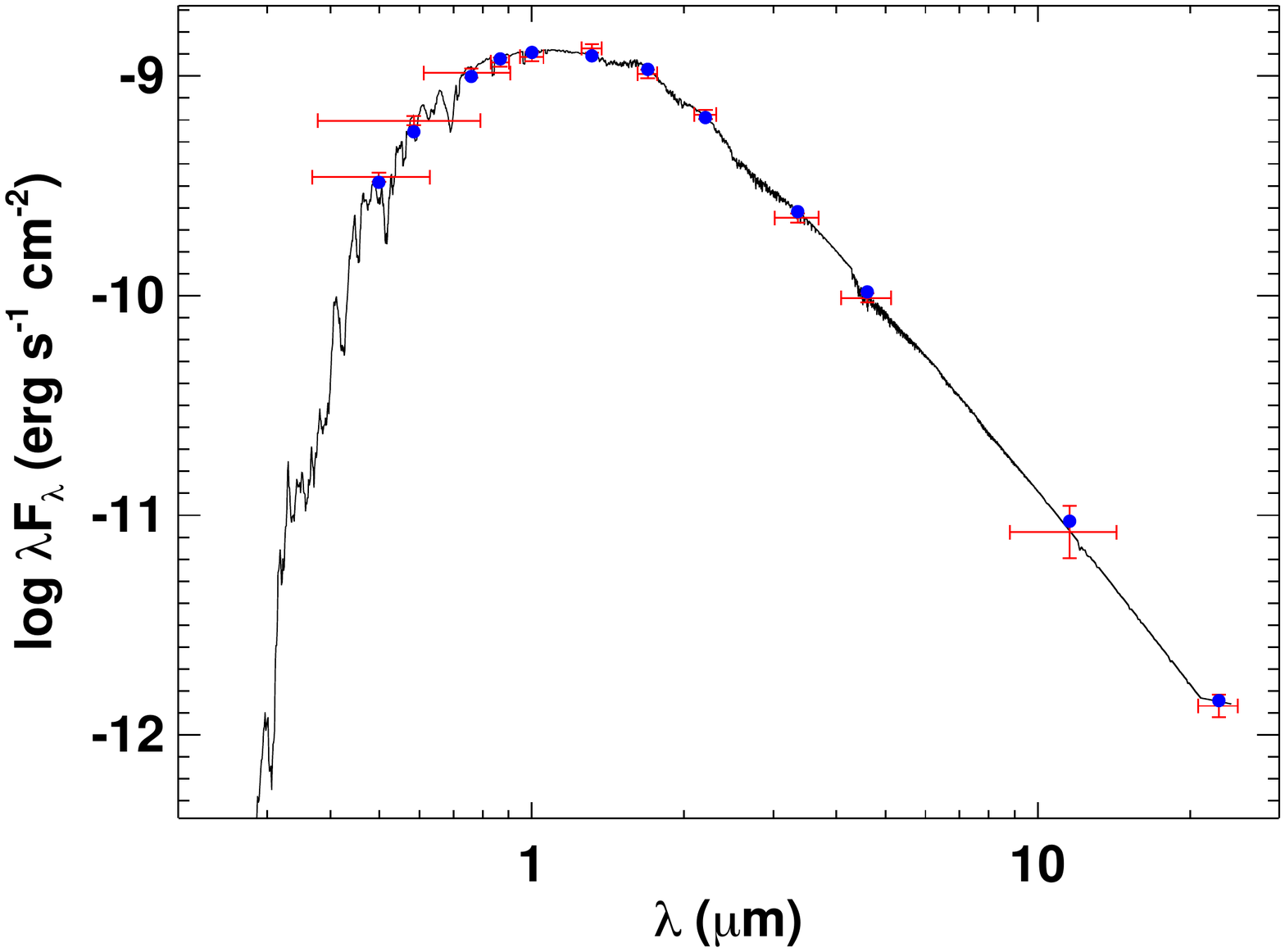}
    \includegraphics[width=0.49\linewidth]{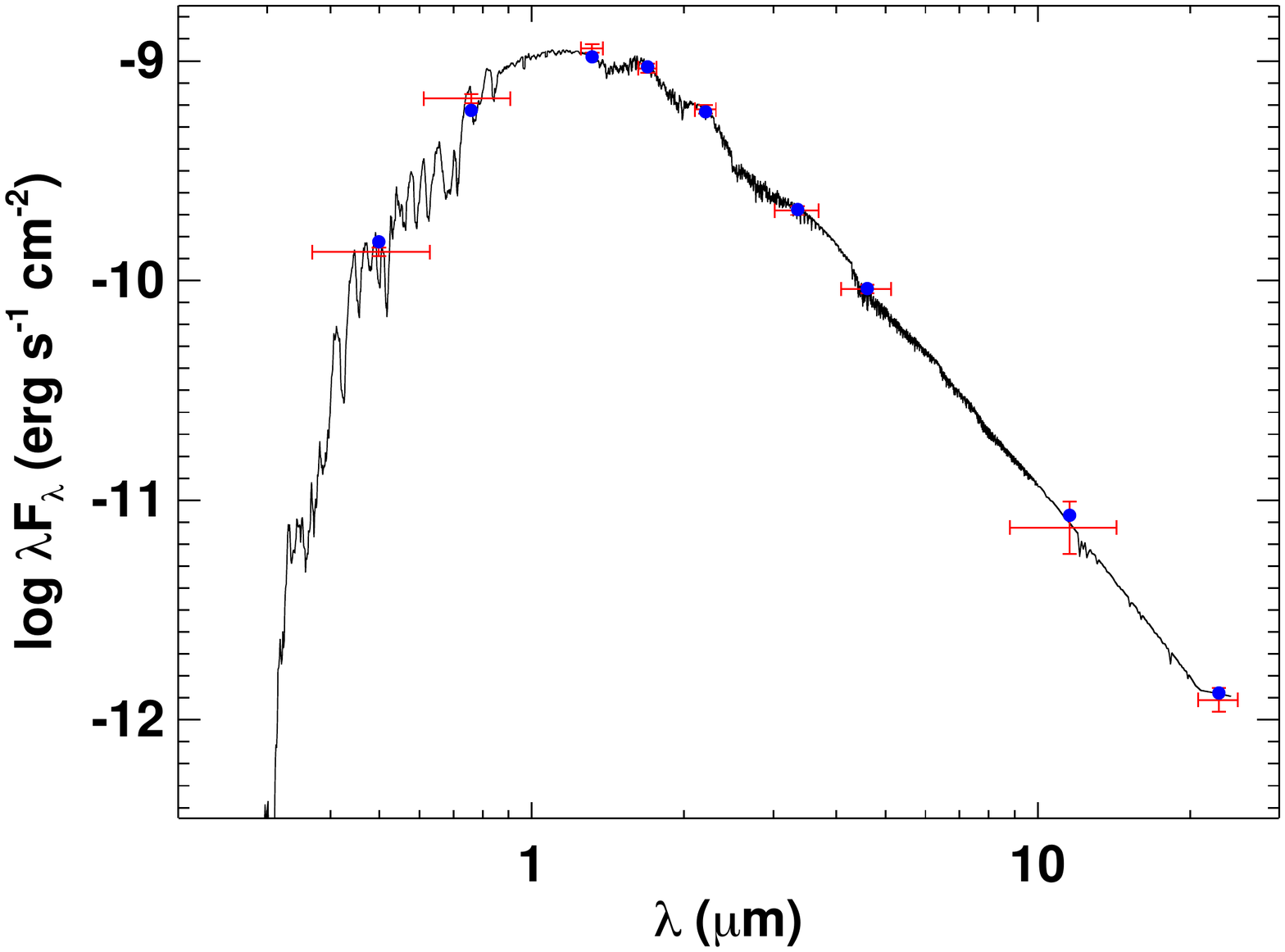}
    \includegraphics[width=0.49\linewidth]{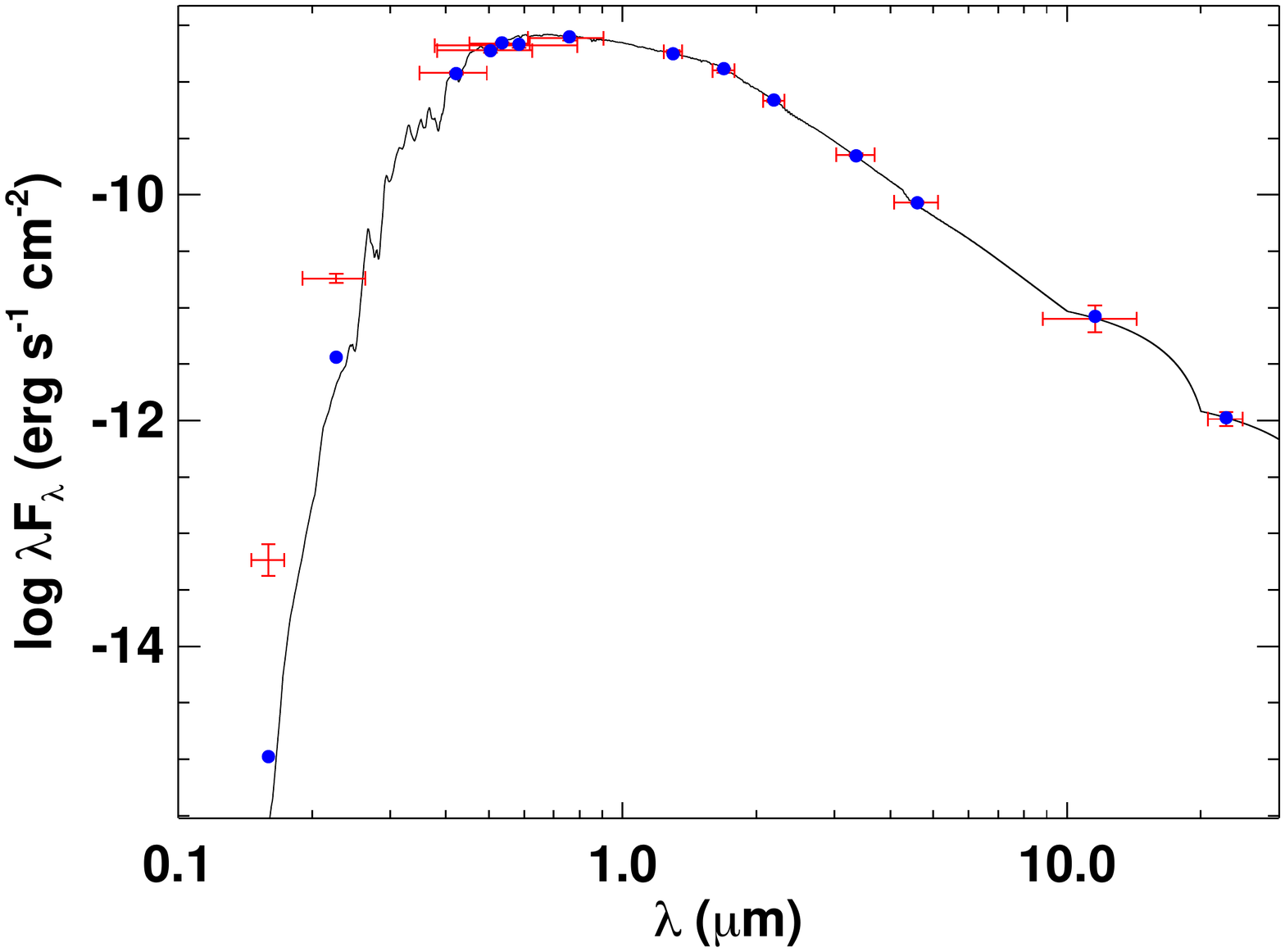}
    \includegraphics[width=0.49\linewidth]{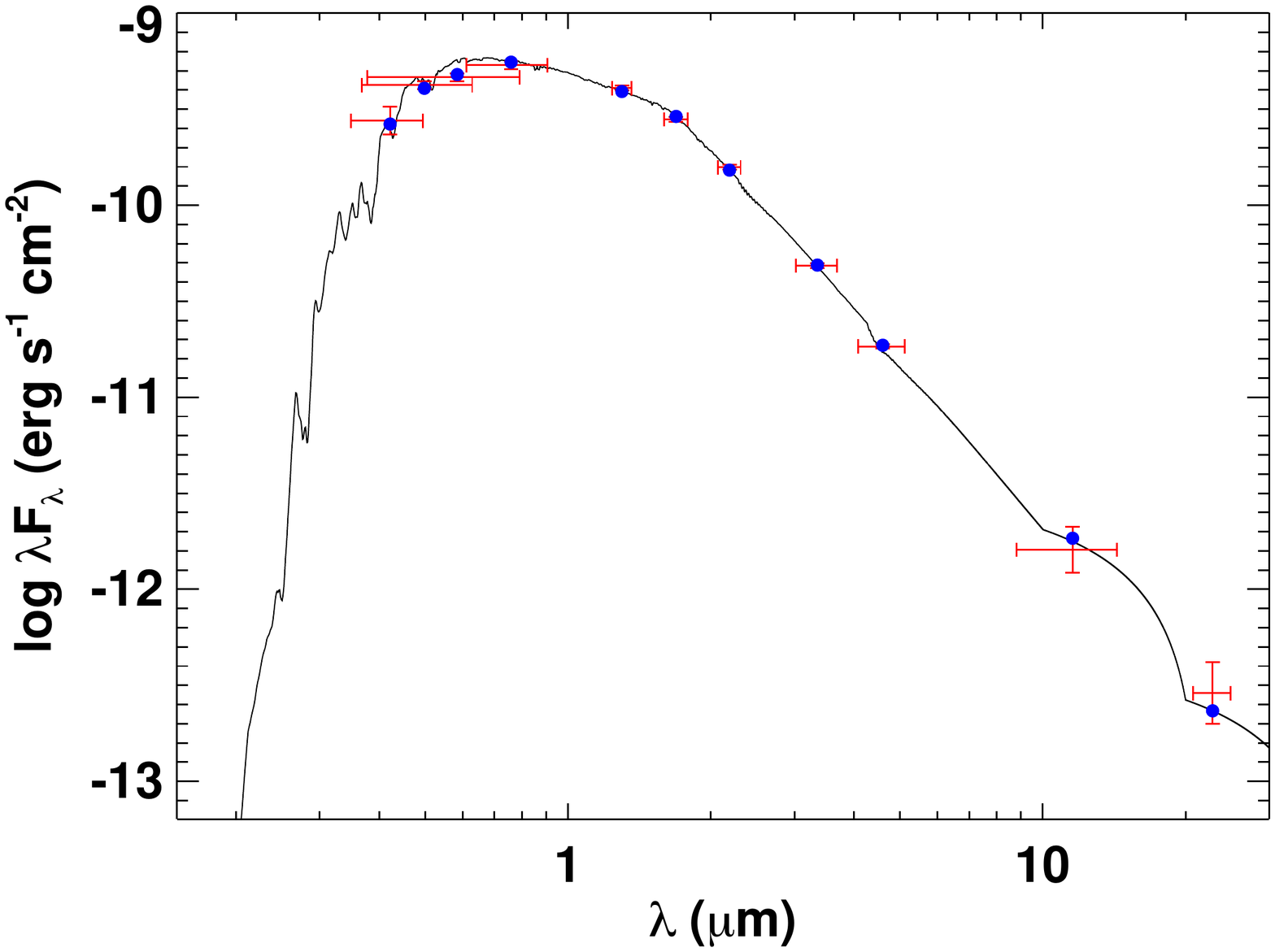}
    \includegraphics[width=0.49\linewidth]{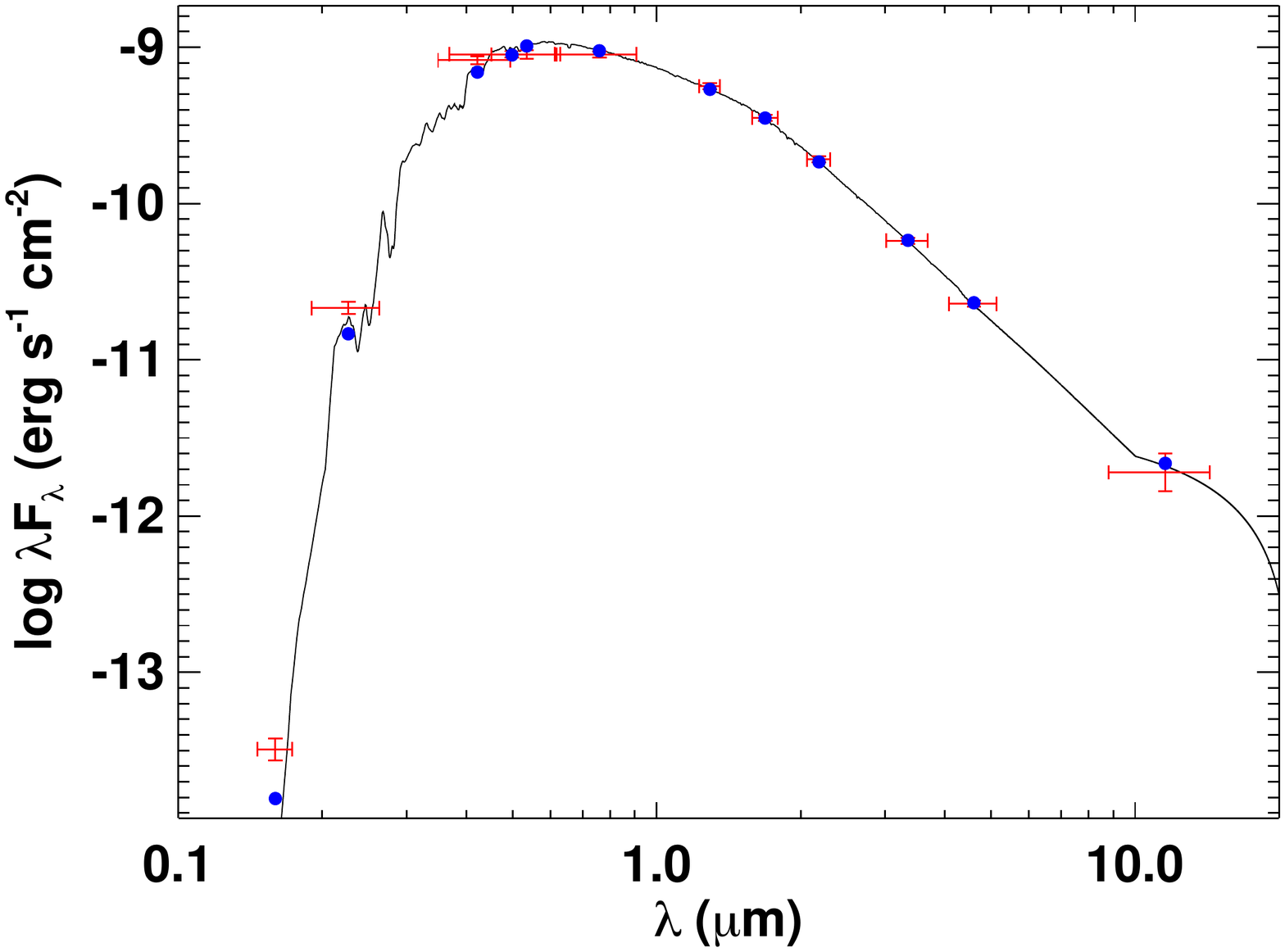}
    \caption{Spectral energy distribution of TOI-198 and TOI-244 (top row), TOI-444 and TOI-470 (middle row), and TOI-518 (bottom row). Red symbols represent the observed photometric measurements, where the horizontal bars represent the effective width of the passband. Blue symbols are the model fluxes from the best-fit Kurucz atmosphere model (black).}
\end{figure*}

\section{Phase-folded data and models}\label{app:pf}

Phase-folded light curves and models centered around the time of mid-transit for both \cheops\ and \tess\ light curves, including residuals around the HOMEBREW joint model.

\begin{figure}
\centering
 \includegraphics[width=\linewidth]{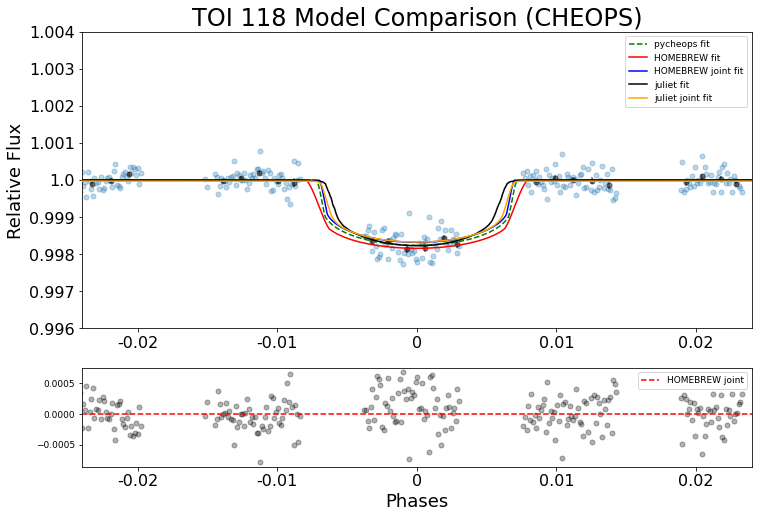}
 \includegraphics[width=\linewidth]{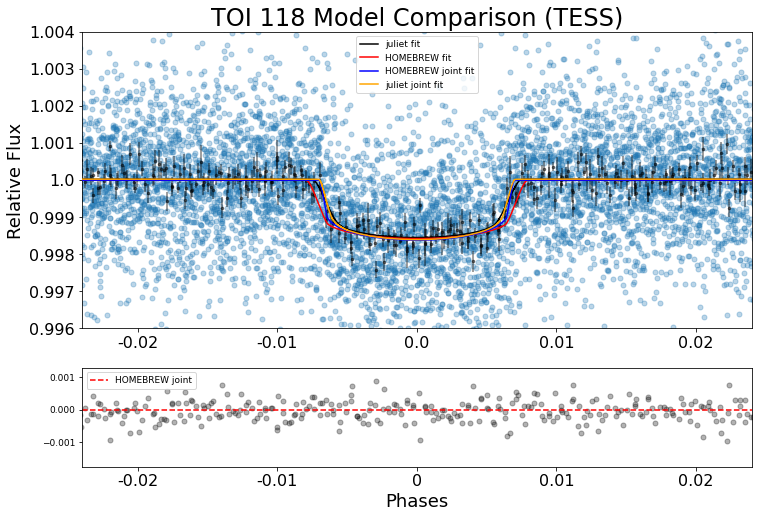}
 \caption{Phase-folded photometry with labeled models calculated with different fit methods. Top: \cheops\ photometric data; Bottom: \tess\ photometric data. }
 \label{fig:TOI118_pfs}
\end{figure}

\begin{figure}
\centering
 \includegraphics[width=\linewidth]{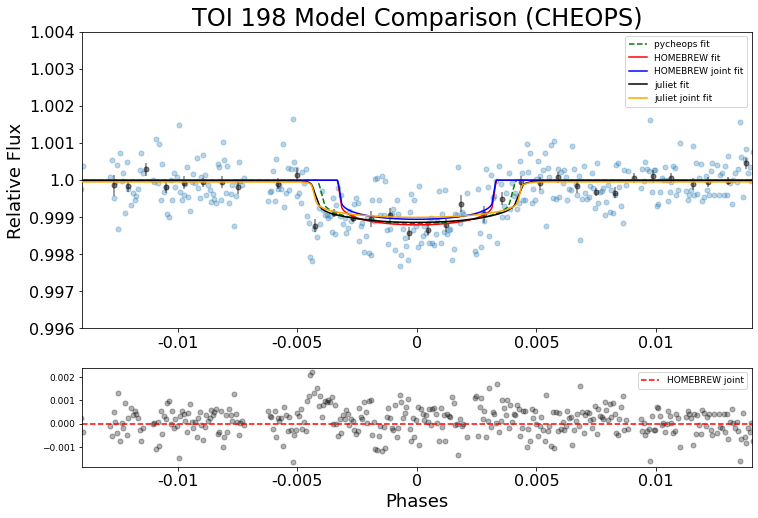}
 \includegraphics[width=\linewidth]{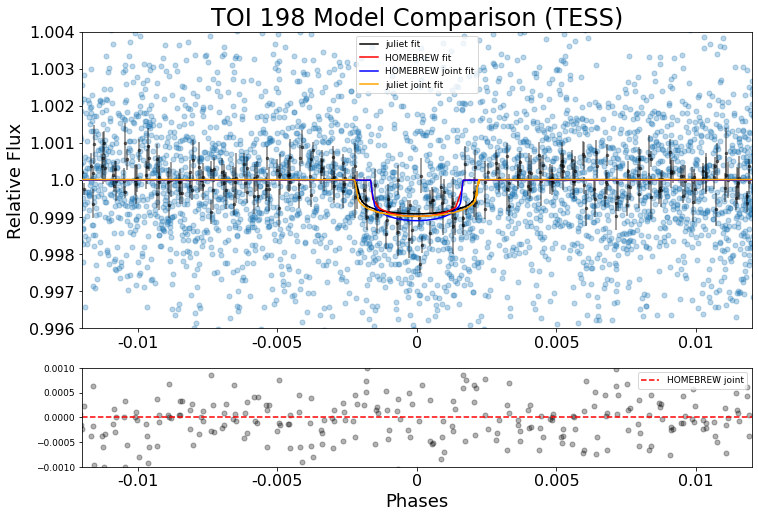}
 \caption{Phase-folded photometry with labeled models calculated with different fit methods. Top: \cheops\ photometric data; Bottom: \tess\ photometric data. }
 \label{fig:TOI198_pfs}
\end{figure}

\begin{figure}
\centering
 \includegraphics[width=\linewidth]{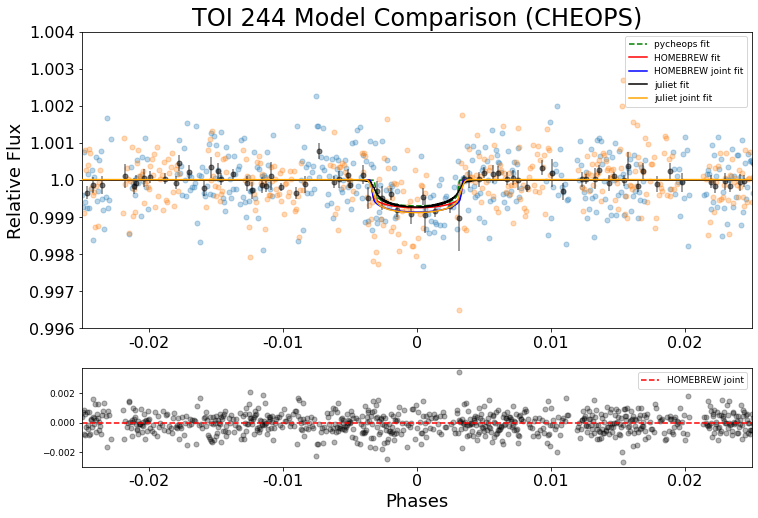}
 \includegraphics[width=\linewidth]{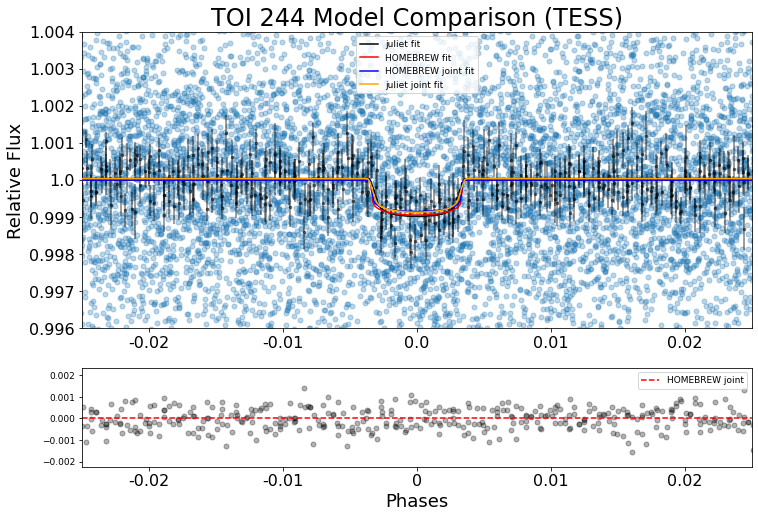}
 \caption{Phase-folded photometry with labeled models calculated with different fit methods. Top: \cheops\ photometric data; Bottom: \tess\ photometric data. }
 \label{fig:TOI244_pfs}
\end{figure}

\begin{figure}
\centering
 \includegraphics[width=\linewidth]{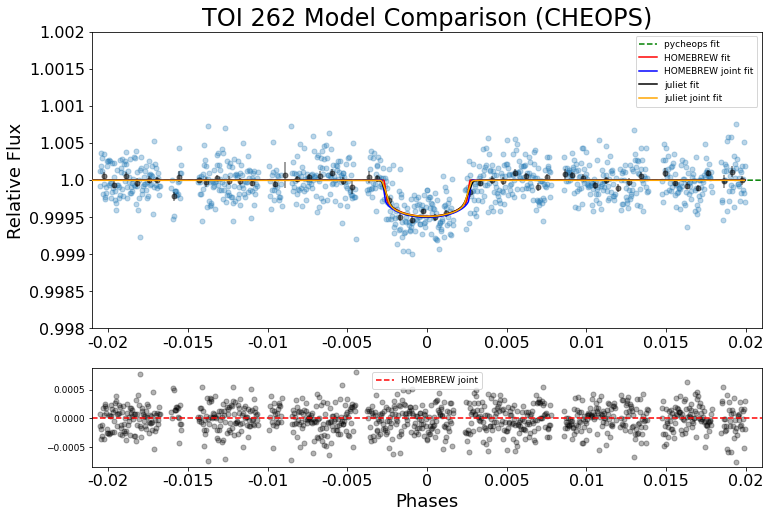}
 \includegraphics[width=\linewidth]{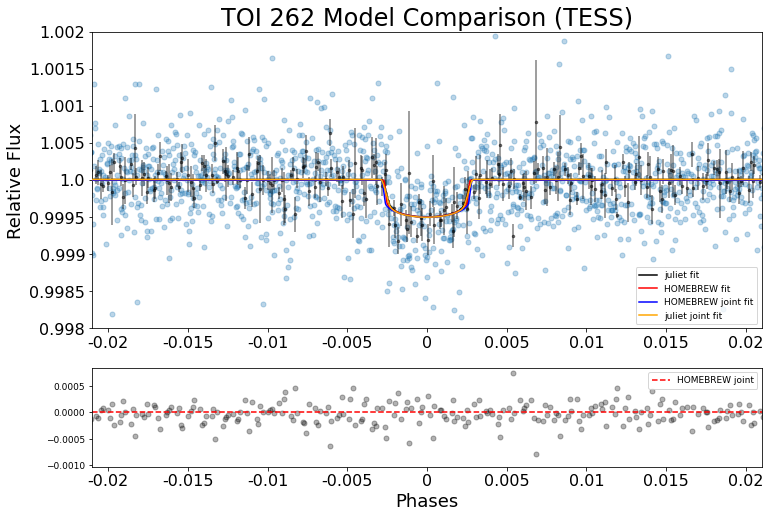}
 \caption{Phase-folded photometry with labeled models calculated with different fit methods. Top: \cheops\ photometric data; Bottom: \tess\ photometric data. }
 \label{fig:TOI262_pfs}
\end{figure}

\begin{figure}
\centering
 \includegraphics[width=\linewidth]{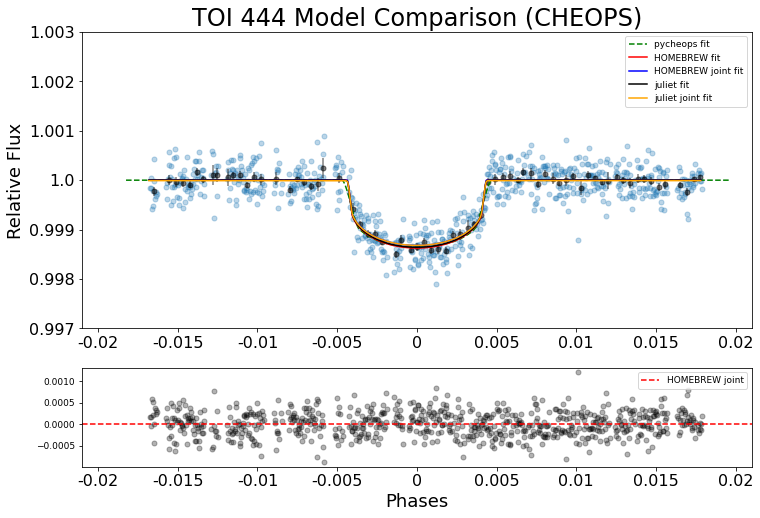}
 \includegraphics[width=\linewidth]{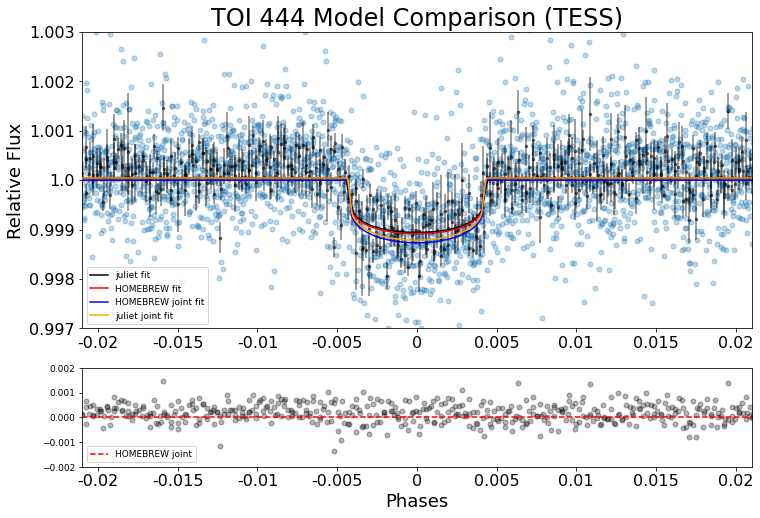}
 \caption{Phase-folded photometry with labeled models calculated with different fit methods. Top: \cheops\ photometric data; Bottom: \tess\ photometric data. }
 \label{fig:TOI444_pfs}
\end{figure}

\begin{figure}
\centering
 \includegraphics[width=\linewidth]{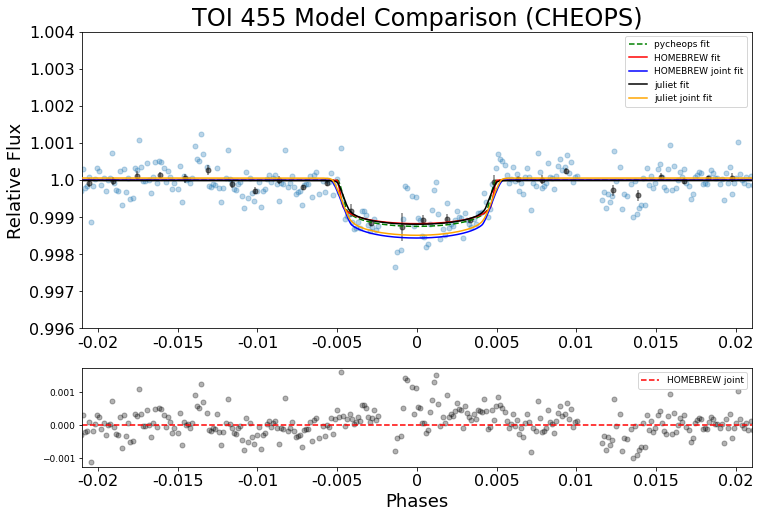}
 \includegraphics[width=\linewidth]{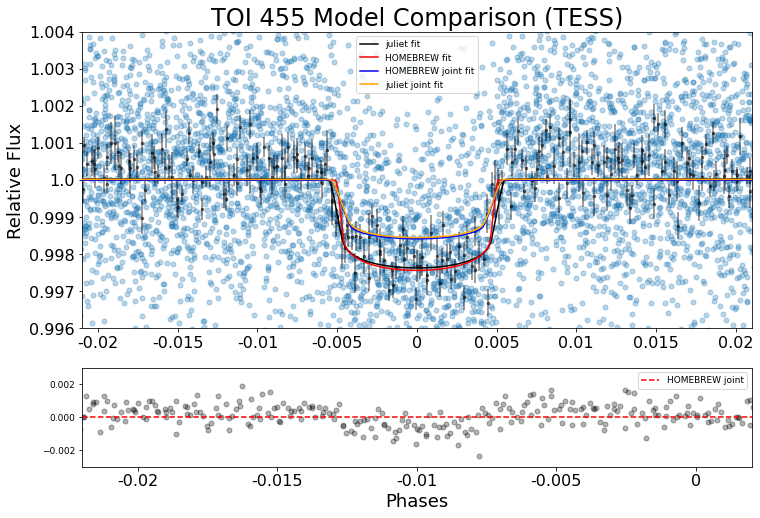}
 \includegraphics[width=\linewidth]{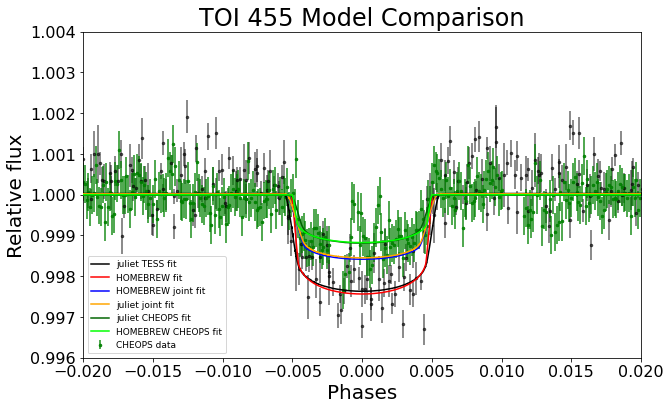}
 \caption{Phase-folded photometry with labeled models calculated with different fit methods. Top: \cheops\ photometric data; Middle: \tess\ photometric data. Bottom: Both photometric datasets overlaid with models.}
 \label{fig:TOI455_pfs}
\end{figure}

\begin{figure}
\centering
 \includegraphics[width=\linewidth]{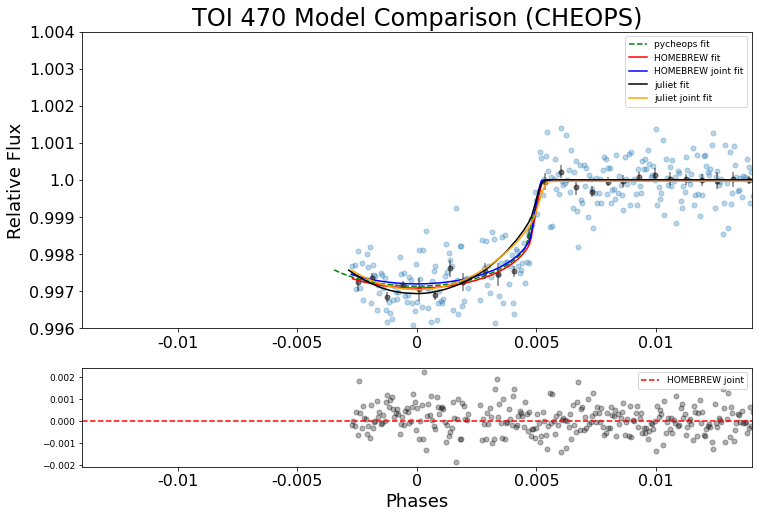}
 \includegraphics[width=\linewidth]{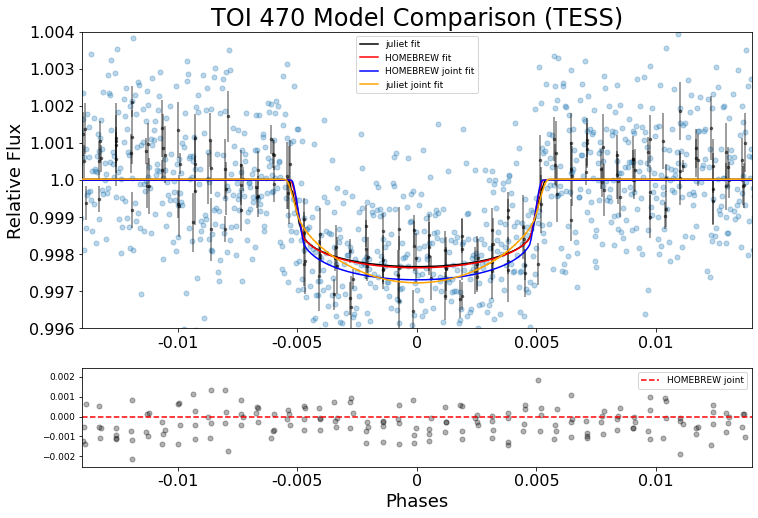}
 \caption{Phase-folded photometry with labeled models calculated with different fit methods. Top: \cheops\ photometric data; Bottom: \tess\ photometric data. }
 \label{fig:TOI470_pfs}
\end{figure}

\begin{figure}
\centering
 \includegraphics[width=\linewidth]{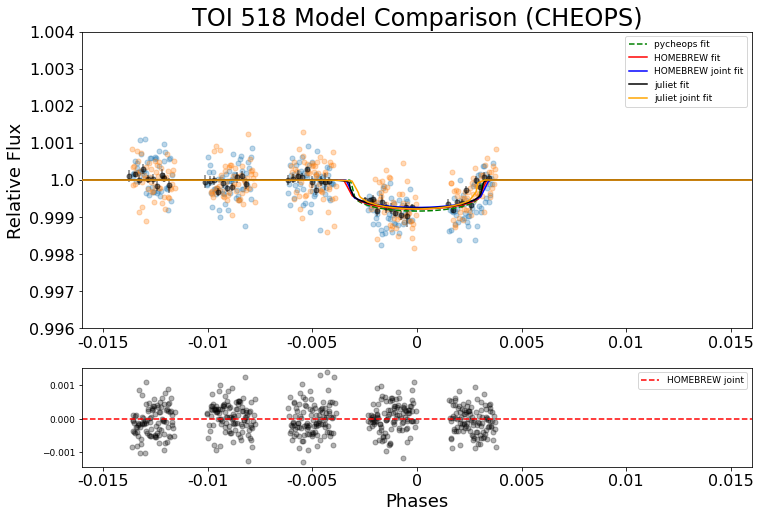}
 \includegraphics[width=\linewidth]{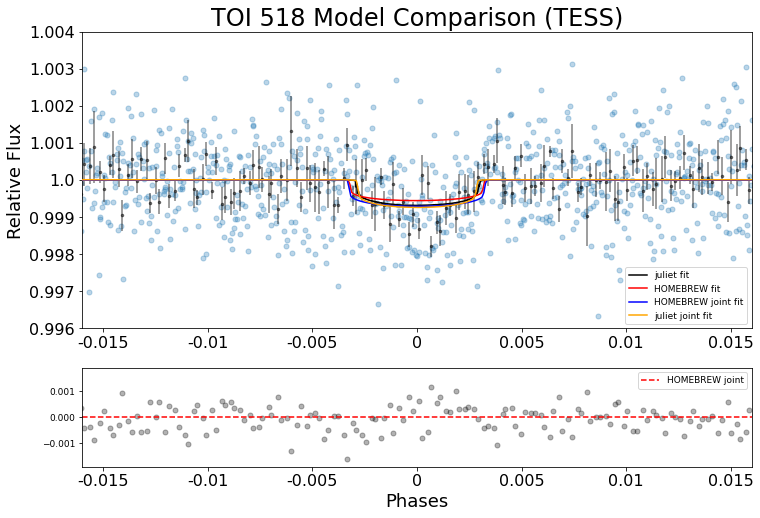}
 \caption{Phase-folded photometry with labeled models calculated with different fit methods. Top: \cheops\ photometric data; Bottom: \tess\ photometric data. }
 \label{fig:TOI518_pfs}
\end{figure}

\begin{figure}
\centering
 \includegraphics[width=\linewidth]{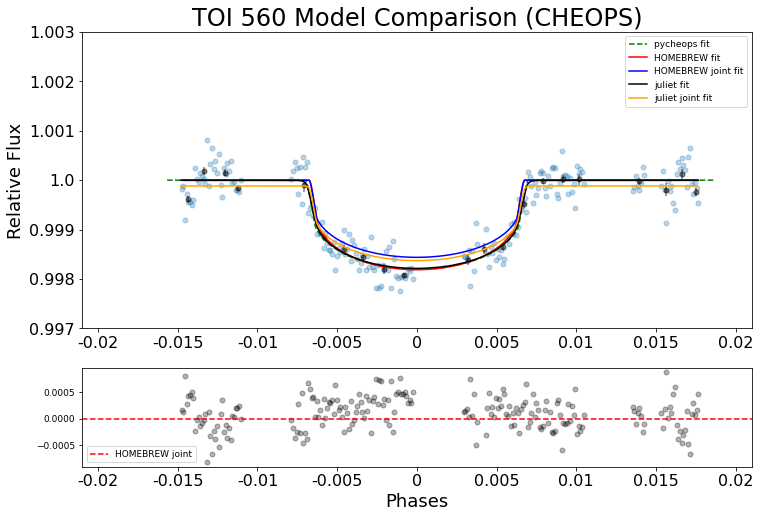}
 \includegraphics[width=\linewidth]{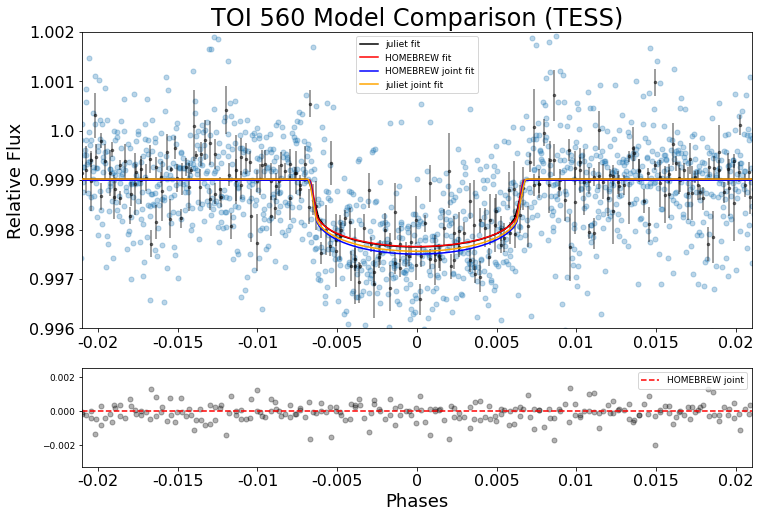}
 \caption{Phase-folded photometry with labeled models calculated with different fit methods. Top: \cheops\ photometric data; Bottom: \tess\ photometric data. }
 \label{fig:TOI560_pfs}
\end{figure}

\begin{figure}
\centering
 \includegraphics[width=\linewidth]{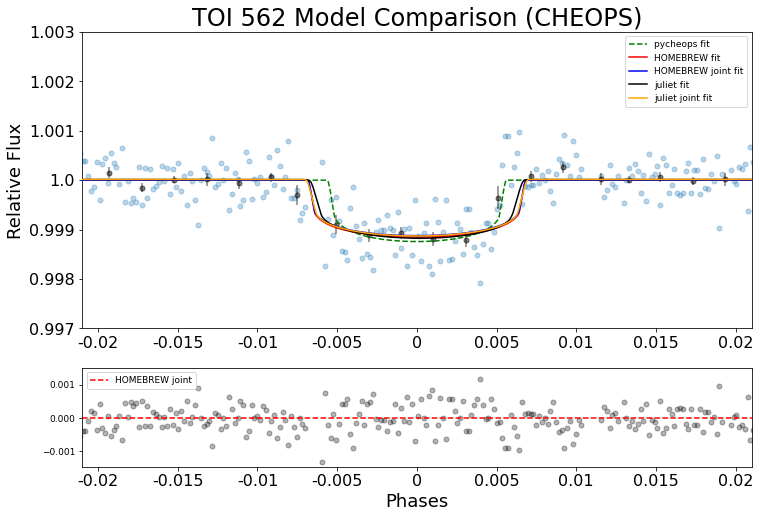}
 \includegraphics[width=\linewidth]{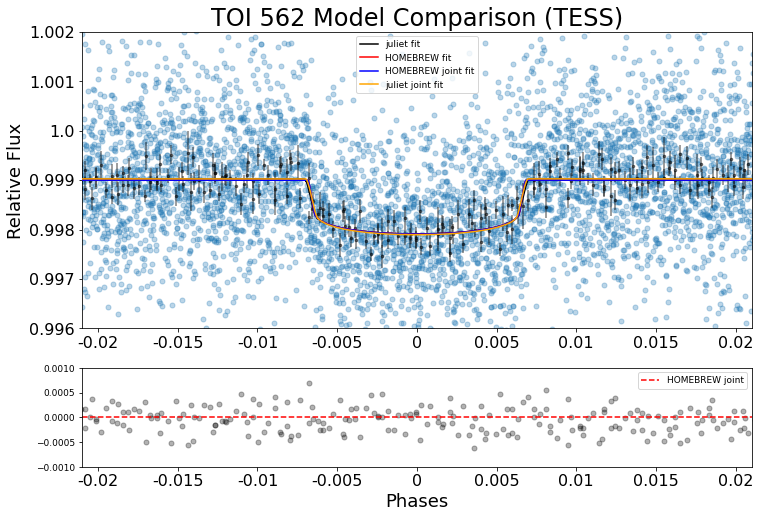}
 \caption{Phase-folded photometry with labeled models calculated with different fit methods. Top: \cheops\ photometric data; Bottom: \tess\ photometric data. }
 \label{fig:TOI562_pfs}
\end{figure}

\section{Corner plots for HOMEBREW joint fits}

\begin{figure*}
    \centering
    \includegraphics[width=.85\linewidth]{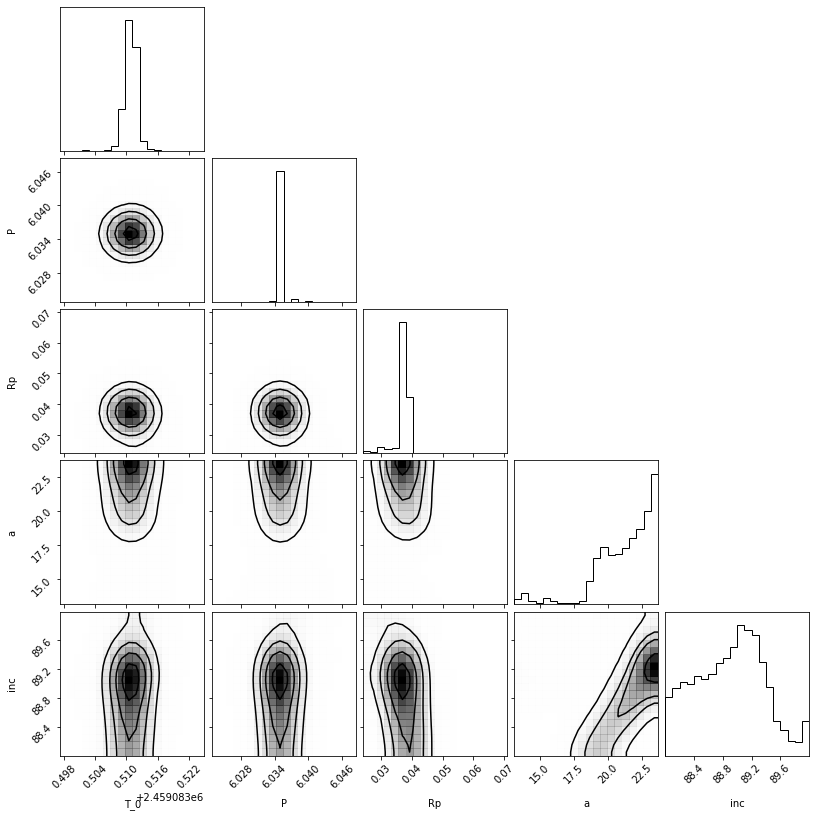}
    \caption{Corner plot for TOI 118 HOMEBREW joint fit.}
    \label{fig:TOI118_corner}
\end{figure*}

\begin{figure*}
    \centering
    \includegraphics[width=.85\linewidth]{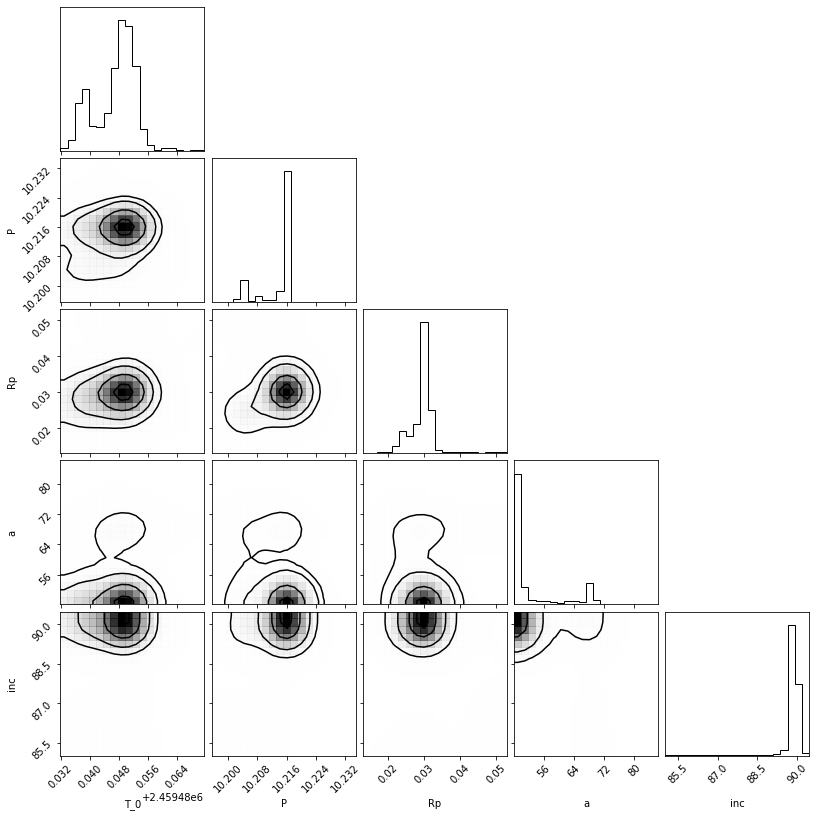}
    \caption{Corner plot for TOI 198 HOMEBREW joint fit.}
    \label{fig:TOI198_corner}
\end{figure*}

\begin{figure*}
    \centering
    \includegraphics[width=.85\linewidth]{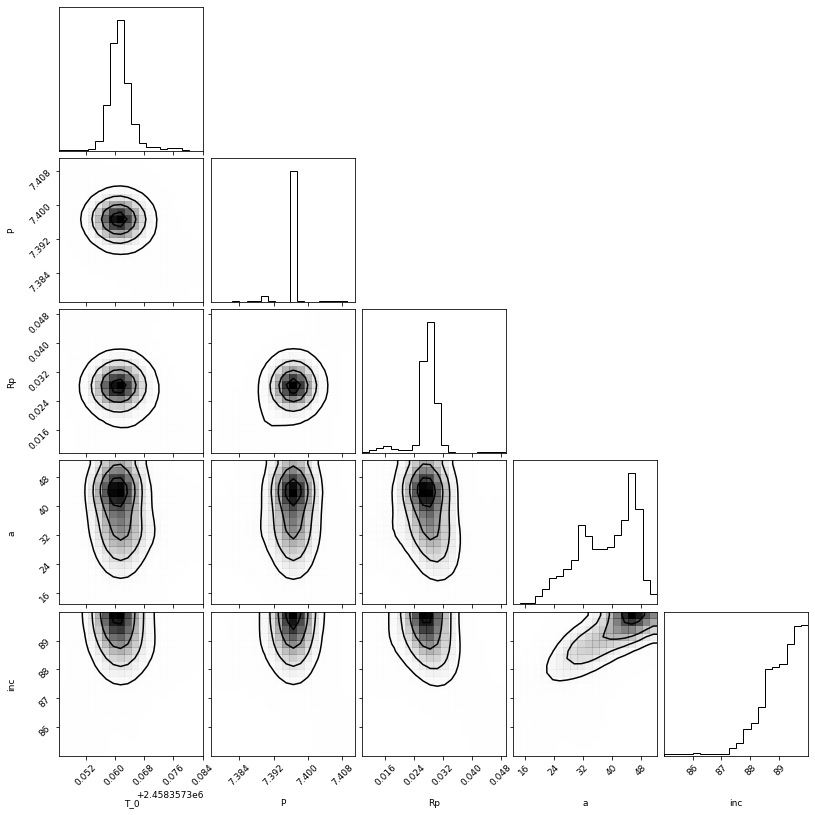}
    \caption{Corner plot for TOI 244 HOMEBREW joint fit.}
    \label{fig:TOI244_corner}
\end{figure*}

\begin{figure*}
    \centering
    \includegraphics[width=.85\linewidth]{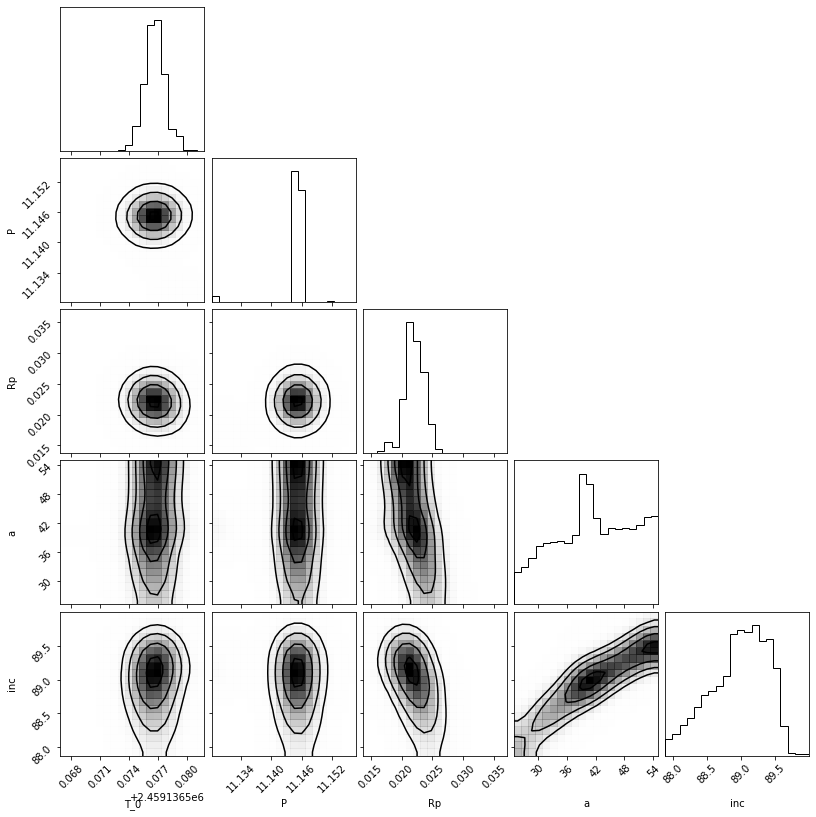}
    \caption{Corner plot for TOI 262 HOMEBREW joint fit.}
    \label{fig:TOI262_corner}
\end{figure*}

\begin{figure*}
    \centering
    \includegraphics[width=.85\linewidth]{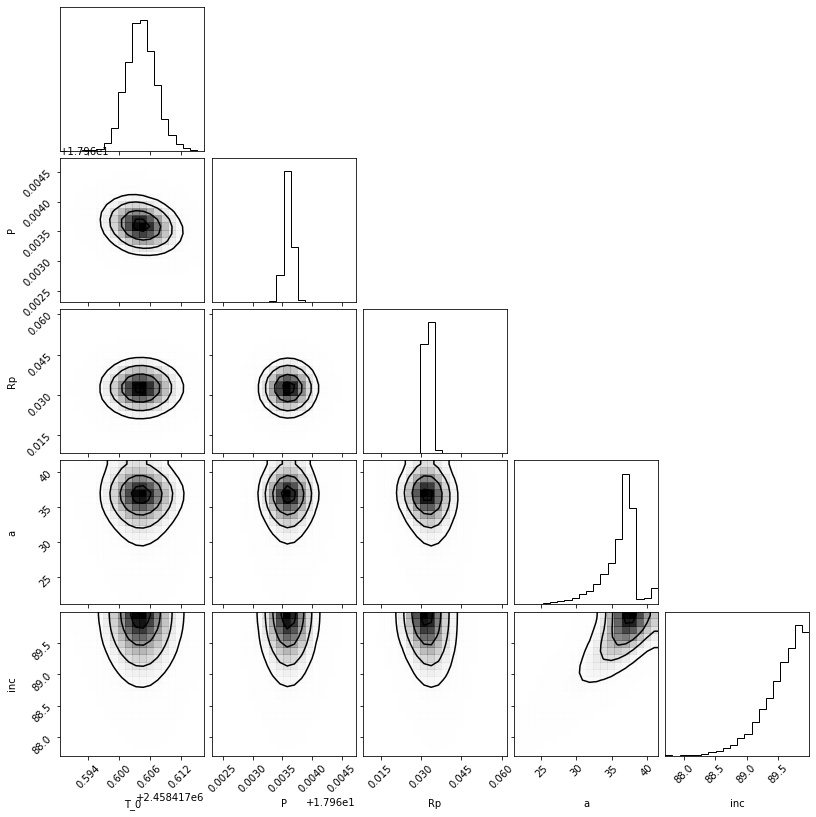}
    \caption{Corner plot for TOI 444 HOMEBREW joint fit.}
    \label{fig:TOI444_corner}
\end{figure*}

\begin{figure*}
    \centering
    \includegraphics[width=.85\linewidth]{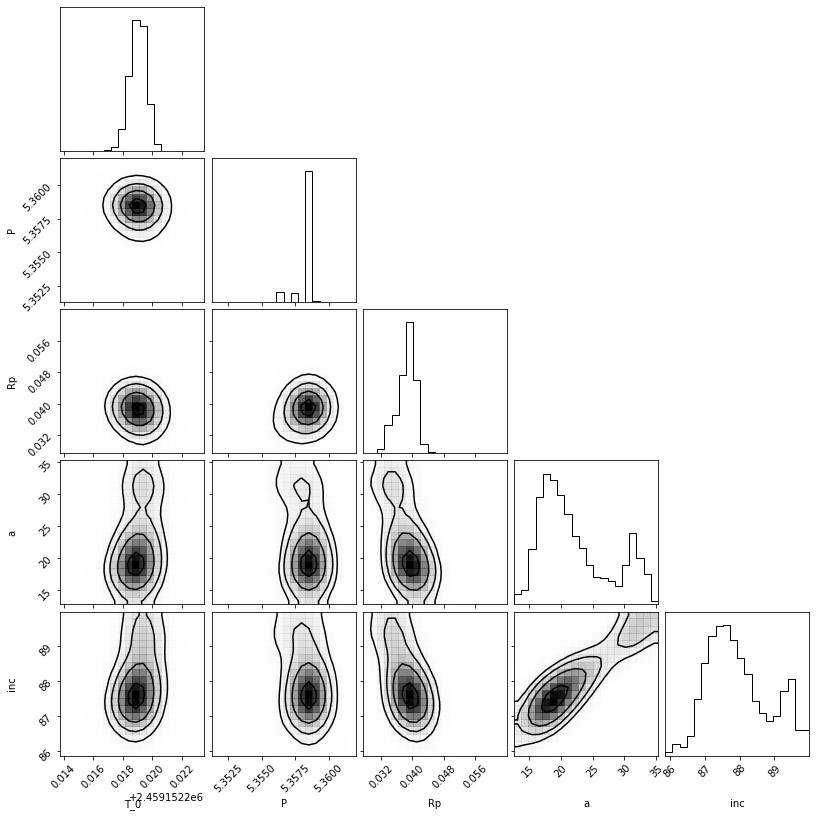}
    \caption{Corner plot for TOI 455 HOMEBREW joint fit.}
    \label{fig:TOI455_corner}
\end{figure*}

\begin{figure*}
    \centering
    \includegraphics[width=.85\linewidth]{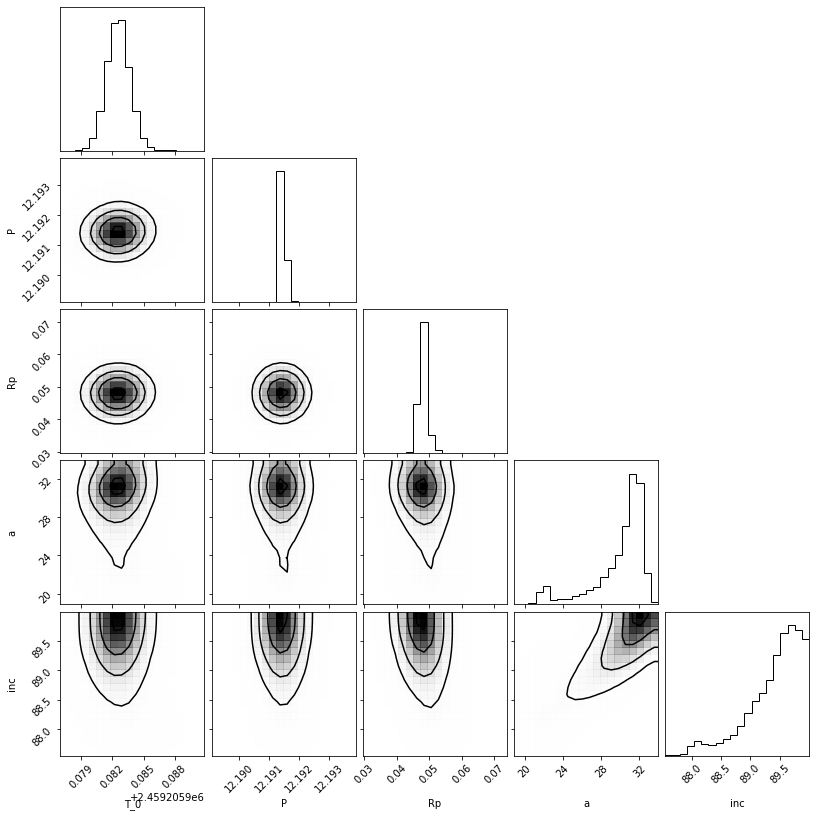}
    \caption{Corner plot for TOI 470 HOMEBREW joint fit.}
    \label{fig:TOI470_corner}
\end{figure*}

\begin{figure*}
    \centering
    \includegraphics[width=.85\linewidth]{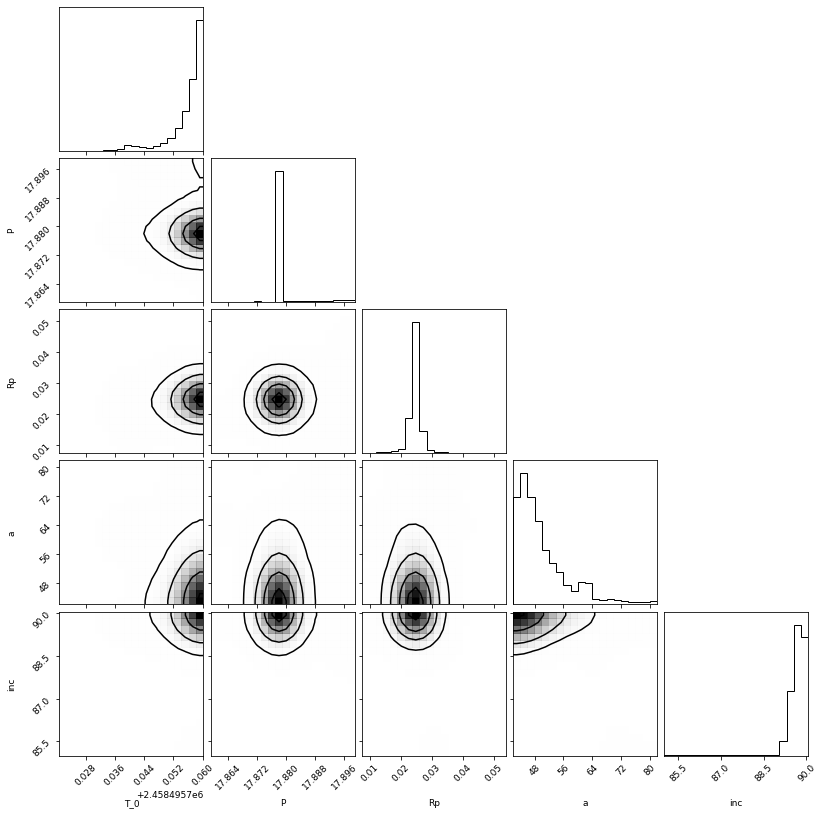}
    \caption{Corner plot for TOI 518 HOMEBREW joint fit.}
    \label{fig:TOI518_corner}
\end{figure*}

\begin{figure*}
    \centering
    \includegraphics[width=.85\linewidth]{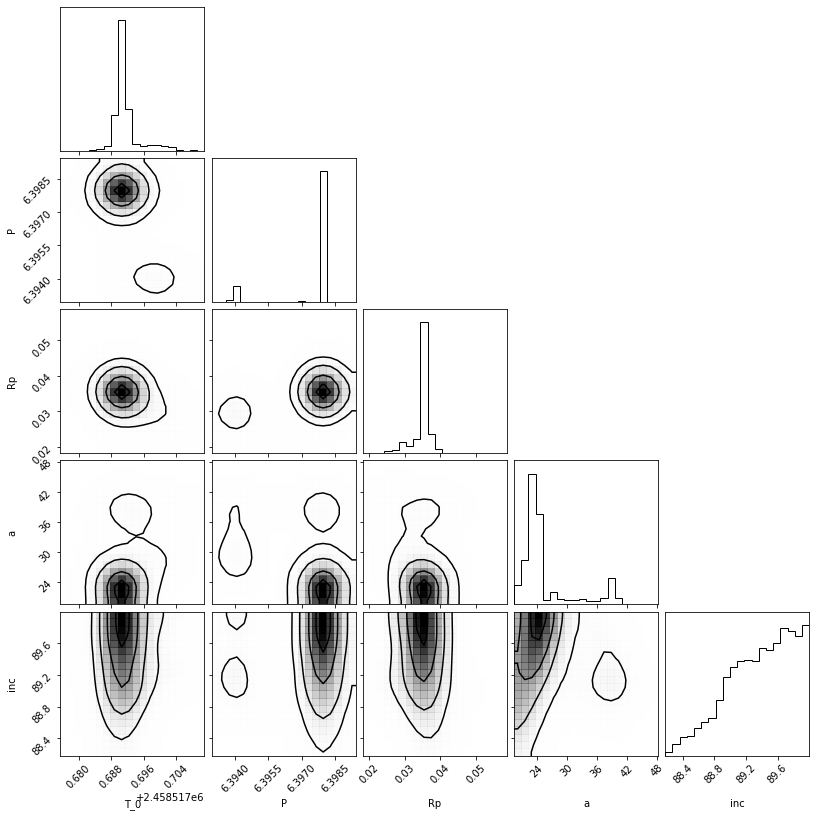}
    \caption{Corner plot for TOI 560 HOMEBREW joint fit.}
    \label{fig:TOI560_corner}
\end{figure*}

\begin{figure*}
    \centering
    \includegraphics[width=.85\linewidth]{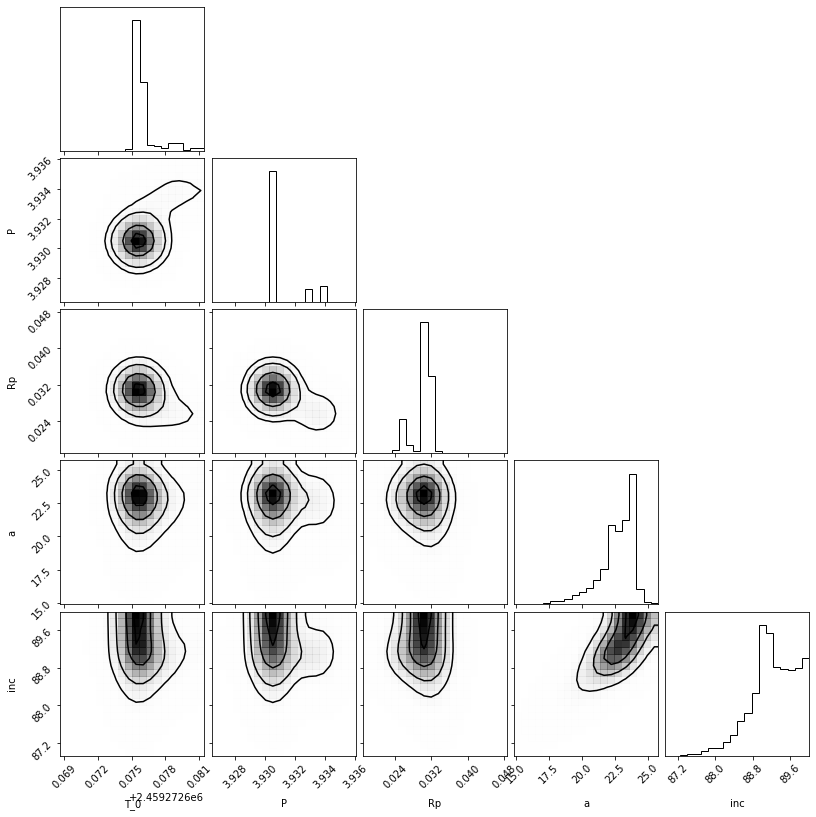}
    \caption{Corner plot for TOI 562 HOMEBREW joint fit.}
    \label{fig:TOI562_corner}
\end{figure*}

\bibliography{cheops_bib}{}
\bibliographystyle{aasjournal}

\end{document}